\documentclass[12pt]{article}

\usepackage{latexsym}
\usepackage{amssymb,amsfonts,amsmath}
\usepackage{graphicx} 
\usepackage{indentfirst}
\usepackage{bbm}
\usepackage{amssymb}
\usepackage{verbatim}
\usepackage{amsmath, amsthm,amssymb}
\usepackage{mathrsfs}
\usepackage{hyperref}
\usepackage{amsfonts}
\usepackage{dsfont}

\parskip=\medskipamount

\newcommand {\cA}{{\cal A}}
\newcommand {\cB}{{\cal B}}

\newcommand {\cD}{{\cal D}}
\newcommand {\cE}{{\cal E}}

\newcommand {\cL}{{\cal L}}

\newcommand {\cN}{{\cal N}}

\newcommand {\cR}{{\cal R}}
\newcommand {\cS}{{\cal S}}

\newcommand {\cV}{{\cal V}}

\def\a{\alpha}

\def\b{\beta}

\def\d{\delta}

\def\f{\phi}
\def\g{\gamma}
\def\G{\Gamma}

\def\j{\psi}

\def\l{\lambda}
\def\m{\mu}

\def\o{\omega}

\def\q{\theta}

\def\s{\sigma}

\def\x{\xi}
\def\z{\zeta}

\def\F{\Phi}
\def\J{\Psi}
\def\L{\Lambda}
\def\O{\Omega}

\def\S{\Sigma}

\def\rd{{\rm d}}
\def\ri{{\rm i}}

\newcommand{\ad}{{\dot{\alpha}}}                           
\newcommand{\bd}{{\dot{\beta}}}                            
\newcommand{\ve}{\varepsilon}                            
\newcommand{\cDB}{{\bar\cD}}                            

\newcommand{\pa}{\partial}                           
\newcommand{\hf}{\frac12}

\newcommand{\vf}{\varphi}

\newcommand{\be}{\begin{equation}}
\newcommand{\ee}{\end{equation}}
\newcommand{\bea}{\begin{eqnarray}}
\newcommand{\eea}{\end{eqnarray}}
\newcommand{\non}{\nonumber}
\newcommand{\bm}[1]{\mbox{\boldmath$#1$}}

\def\double #1{#1{\hbox{\kern-2pt $#1$}}}

\newcommand{\gd}{{\dot\g}}

\newif\ifdtup

\newcommand{\bsubeq}{\begin{subequations}}
\newcommand{\esubeq}{\end{subequations}}

\newcommand{\mub}{{{\bar{\mu}}}}

\numberwithin{equation}{section}

\begin{document}

\begin{titlepage}
\begin{flushright}
May, 2018\\
\end{flushright}
\vspace{5mm}

\begin{center}
{\Large \bf Higher spin 
supercurrents in anti-de Sitter space
}\\ 
\end{center}

\begin{center}

{\bf 
Evgeny I. Buchbinder, Jessica Hutomo and 
Sergei M. Kuzenko
} \\
\vspace{5mm}

\footnotesize{
{\it Department of Physics M013, The University of Western Australia\\
35 Stirling Highway, Crawley W.A. 6009, Australia}}  
\vspace{2mm}
~\\
Email: \texttt{evgeny.buchbinder@uwa.edu.au, 20877155@student.uwa.edu.au,
sergei.kuzenko@uwa.edu.au}\\
\vspace{2mm}

\end{center}

\begin{abstract}
\baselineskip=14pt
We propose higher spin supercurrent multiplets for $\cN=1$ supersymmetric field 
theories in four-dimensional anti-de Sitter space (AdS). Their explicit realisations are 
derived for various supersymmetric theories, including a model of $N$ massive chiral scalar
superfields with an arbitrary mass matrix. 
We also present new off-shell gauge formulations 
for the massless ${\cal N}=1$ supersymmetric 
multiplet of integer superspin $s$ in AdS, where $s =2,3,\dots$, 
as well as for the massless gravitino multiplet (superspin $s=1$)
which requires special consideration.
\end{abstract}

\vfill

\vfill
\end{titlepage}

\newpage
\renewcommand{\thefootnote}{\arabic{footnote}}
\setcounter{footnote}{0}

\tableofcontents{}
\vspace{1cm}
\bigskip\hrule

\allowdisplaybreaks


\section{Introduction} 

There exist only five maximally supersymmetric backgrounds 
in off-shell $\cN=1$  supergravity in four dimensions \cite{FS}.\footnote{The
classification by Festuccia and Seiberg \cite{FS} was given 
purely at the component level.
It was re-derived in \cite{K13} using the superspace formalism
developed in the mid-1990s \cite{BK}. As curved $\cN=1$ superspaces, all maximally  supersymmetric backgrounds were described
in \cite{KT-M}
(see also \cite{K16} for a new derivation of the results in \cite{K13,KT-M}, 
which works equally well  for all known off-shell formulations for $\cN=1$ supergravity).}
Only two of them support maximally symmetric spacetimes.
The latter backgrounds are: Minkowski superspace 
${\mathbb M}^{4|4}$  \cite{AV,SS} and 
anti-de Sitter (AdS) superspace ${\rm AdS}^{4|4}$ \cite{Keck,Zumino77,IS}.

The structure of consistent 
supercurrent multiplets  in $\cN=1$ AdS supersymmetry \cite{BK11,BK12}
considerably differs from that in the $\cN=1$ super-Poincar\'e case, see e.g. 
\cite{MSW,K2010}. 
Specifically, there exist three minimal supercurrents with $12+12$ degrees 
of freedom in ${\mathbb M}^{4|4}$  \cite{K2010}, and only one in ${\rm AdS}^{4|4}$
\cite{BK11},
the latter being the AdS extension of the Ferrara-Zumino supercurrent \cite{FZ}.
Furthermore, the so-called  $\cS$-multiplet 
advocated by Komargodski and Seiberg \cite{KSeiberg} 
does not admit a minimal extension to AdS.\footnote{The consistent 
supergravity extension of the   $\cS$-multiplet was given in \cite{BK11}.}
These differences between the supercurrent multiplets  in
${\mathbb M}^{4|4}$  and  ${\rm AdS}^{4|4}$
have nontrivial dynamical implications. For instance, 
since every $\cN=1$ supersymmetric field theory in AdS should have 
a well-defined Ferrara-Zumino supercurrent \cite{BK11,BK12}, 
the K\"ahler target space of every supersymmetric nonlinear $\s$-model 
in AdS must be non-compact and possess an exact K\"ahler two-form, 
in accordance with the analysis of Komargodski 
and Seiberg \cite{KSeiberg}.\footnote{In the $\cN=2$ extended case, 
AdS supersymmetry imposes nontrivial restrictions on the structure 
of the hyperk\"ahler target spaces of supersymmetric nonlinear 
 $\s$-models  \cite{BKsigma,BKLT-M}.}
The same conclusion was also obtained by direct studies of 
the most general $\cN=1$ supersymmetric nonlinear $\s$-models 
in AdS \cite{FS,AJKL}.
 
It should be pointed out  that  the consistent 
AdS supercurrents  \cite{BK11,BK12} are closely related to two  classes of 
supersymmetric gauge theories: (i) the known off-shell formulations, 
minimal (see, e.g., \cite{BK,GGRS} for reviews) and non-minimal \cite{BK12},
for $\cN=1$ AdS supergravity; and (ii)
the two
dually equivalent series of massless higher spin supermultiplets 
in AdS proposed in \cite{KS94}.
More specifically, as discussed in \cite{BK12},
there are only two
irreducible AdS supercurrents, with $(12+12)$ and $(20+20)$ degrees of 
freedom.\footnote{These supercurrents are related to each other via 
a well-defined improvement transformation \cite{BK12}.}
The former is naturally associated with the so-called longitudinal action
$S^{||}_{(3/2)}$ for a massless superspin-3/2 multiplet in AdS \cite{KS94}, 
which is formulated in terms of  a real vector prepotential
$H_{\alpha \ad}$ and a covariantly chiral superfield $\sigma$. The latter is associated
with a unique dual formulation $S^{\perp}_{(3/2)}$ where the chiral superfield
is replaced by a complex linear superfield $\Gamma$. 
The functional $S^{||}_{(3/2)}$ proves to be the linearised action for
minimal $\cN=1$ AdS supergravity.  The dual action  $S^{\perp}_{(3/2)}$
results from the linearisation around the AdS background of non-minimal 
$\cN=1$ AdS supergravity \cite{BK12}.\footnote{It was believed for almost thirty 
years that there is no off-shell non-minimal formulation for $\cN=1$ AdS supergravity
\cite{GGRS}. However, such a formulation was constructed in \cite{BK12}.}
Both actions
represent the lowest superspin limits of two infinite series of dual models, 
$S^{||}_{(s+\hf)}$ and $S^{\perp}_{(s+\hf)}$,
for off-shell massless gauge  supermultiplets in AdS 
of half-integer superspin $(s+\hf)$, where $s=1,2 \dots$, 
constructed in \cite{KS94}.
Off-shell formulations for massless gauge  supermultiplets in AdS 
of integer superspin $s$, with $s=1,2 \dots$, 
were also constructed in \cite{KS94}.
In the flat-superspace limit, the supersymmetric higher spin theories
of \cite{KS94} reduce to those proposed in \cite{KSP,KS}.

Making use of the gauge  off-shell formulations for massless higher spin 
supermultiplets in AdS \cite{KS94}, one can define consistent higher spin supercurrent 
multiplets in AdS superspace
(i.e. higher spin extensions  of the supercurrent)
 that contain ordinary bosonic and fermionic conserved currents in AdS. 
One can then look for explicit realisations 
of such higher spin supercurrents in concrete supersymmetric theories in AdS, 
for instance  models for massless and massive chiral scalar superfields. 
Such a program is a natural  extension of the flat-space results obtained 
in recent papers \cite{HK1,HK2} in which two of us built on the structure of 
higher spin supercurrent multiplets in models for  superconformal chiral superfields
\cite{KMT}.  In accordance with the standard Noether method
(see, e.g., \cite{VanNieuwenhuizen:1981ae} for a review),
the construction of conserved higher spin supercurrents 
for various supersymmetric theories in AdS is equivalent to generating consistent 
cubic vertices of the type $\int H  J$, 
where $H$ denotes
some off-shell higher spin gauge multiplet \cite{KS94}, 
and $J=\cD^p \F \cD^q\J$ is the higher spin current which is constructed in terms of 
some matter multiplets $\F$ and $\J$ and the 
AdS covariant derivatives $\cD$. 
This is one of the important applications of the results
presented in the present paper.
In the flat-superspace case, several cubic vertices
involving the off-shell higher spin multiplets of  \cite{KSP,KS}
were constructed recently in \cite{BGK1,KKvU,BGK3}, 
as an extension of the superconformal cubic couplings between
 a chiral scalar superfield and an infinite tower of gauge massless multiplets 
of half-integer superspin given in \cite{KMT}.

It should be pointed out that conserved higher spin currents 
for scalar and spinor fields in Minkowski space
have been studied in numerous publications. 
To the best of our knowledge, the spinor case was first described by Migdal \cite{Migdal}
and Makeenko \cite{Makeenko},  while
the conserved higher spin currents for scalar fields were first obtained in 
\cite{Makeenko,CDT,BBvD} (see also \cite{Anselmi,Anselmi2}). 
The conserved higher spin currents for scalar fields
in AdS were studied, e.g., in \cite{MR,MM,FIPT,FTsulaia,BekaertM}. 
Since the curvature of AdS space is non-zero, explicit calculations 
of conserved higher spin currents are much harder than in Minkowski space. 
This is one of the reasons why Refs.   \cite{MR,MM} studied only the conformal scalar, and 
only the first order correction to the flat-space expression was given explicitly. 
The construction presented in \cite{BekaertM} is more complete in the sense that 
all conserved higher spin currents were computed exactly for a free massive scalar field. 
This was achieved by making use of a somewhat unorthodox  formulation in the so-called ambient space.  
All these works dealt with integer spin currents. 
The important feature of supersymmetric theories is that they also possess
half-integer spin currents.  
They belong to the higher spin supercurrent multiplets we construct in this work.
Another nice feature of the supersymmetric case is that the calculation 
of higher spin supercurrent multiplets in AdS superspace is considerably simpler
than the problem of computing the ordinary conserved higher spin currents 
in AdS space. 

Various aspects of supersymmetric field theories on AdS${}_4$ have been studied 
in detail over the last forty years, see, e.g., 
\cite{FS,Zumino77,IS,BKsigma,AJKL,ST,Burgess:1984rz,Burgess:1984ti,Burges:1985qq,Aharony:2010ay,Aharony:2012jf,Aharony:2015hix} and references therein.
The energy-momentum tensor of such a theory belongs to the Ferrara-Zumino 
supercurrent (or, equivalently, to the non-minimal AdS supercurrent which is 
related to the Ferrara-Zumino supermultiplet by 
a well-defined improvement transformation \cite{BK12}.). 
In this paper we present, for the first time, higher spin extensions
of the AdS supercurrents and derive their explicit realisations  for various supersymmetric theories on AdS, 
including a model of $N$ massive chiral scalar
superfields with an arbitrary mass matrix. 
Our results have numerous applications. For instance, 
the conserved higher spin supercurrents computed in section 5 and 6 can 
readily be reduced to component fields. This will give closed-form expressions 
for conserved higher spin bosonic and fermionic currents in models
with massive scalar and spinor fields, thus leading to 
more general results than 
those known in the literature \cite{MR,MM,FIPT,FTsulaia,BekaertM}. 
Another applications of the results obtained are consistent cubic 
coupling between chiral scalar supermultiplets and massless higher spin 
supermultiplets. 
Our results also make it possible to develop an effective action 
approach to massless higher spin supermultiplets
along the lines advocated in 
\cite{Tseytlin2002,Segal,BJM} and more recently in \cite{KMT}.
We also refine some statements given recently 
in the literature, see section 7.

 This paper is organised as follows. Section 2 contains a summary of the results 
 concerning supersymmetric field theory in AdS superspace.
 Section 3 is devoted to a novel formulation for the massless integer superspin 
 multiplets in AdS. This formulation is shown to reduce to that proposed in \cite{KS94} upon  partially fixing the gauge freedom. We also describe off-shell formulations (including a novel one) for the massless gravitino multiplet in AdS. In section 4
we  introduce higher spin supercurrent multiplets in AdS and describe improvement 
transformations for them. Sections 5 and 6 are devoted to the explicit construction
of higher spin supercurrents for $N$ massive chiral multiplets. 
Several nontrivial applications of the results obtained are given in section 7.
The main body of the paper is accompanied by three technical appendices.
Appendix A reviews the irreducible supercurrent multiplets in AdS 
following \cite{BK11,BK12}.
Appendices B and C review the conserved higher spin  currents 
for $N$ scalars and spinors, respectively, with arbitrary mass matrices. 
These results are scattered in the literature, including 
\cite{Migdal,Makeenko,CDT,BBvD}.


\section{Field theory in AdS superspace} \label{section2}

In this section we give a summary of the results which are 
absolutely essential when doing $\cN=1$ supersymmetric field theory in 
AdS in a manifestly  ${\rm OSp(1|4)}$-invariant way.
We mostly follow the presentation in  \cite{KS94}. 
Our notation and two-component spinor conventions agree with \cite{BK}, 
except for the notation for superspace integration measures.

Let $z^M =(x^m , \q^\m, \bar \q_{\dot \m} ) $ be local coordinates
for  $\cN=1$ AdS superspace, ${\rm AdS}^{4|4}$. 
The geometry  of ${\rm AdS}^{4|4}$
may be  described in terms of covariant derivatives
of the form
\bea
\cD_A = (\cD_a , \cD_\a ,\bar \cD^\ad ) = E_A + \O_A~, \qquad
E_A = E_A{}^M \partial_M  ~,
\eea
where $E_A{}^M $ is the inverse superspace vielbein, and 
\bea
\O_A = \frac{1}{2}\,\O_A{}^{bc} M_{bc}
= \O_A{}^{\b \g} M_{\b \g}
+\bar \O_A{}^{\bd \gd} \bar M_{\bd \gd} ~,
\eea	
is the Lorentz connection. 
The Lorentz generators $M_{bc} \Leftrightarrow (M_{\b\g},{\bar M}_{\bd\gd})$ 
act on two-component spinors as follows: 
\begin{subequations} 
\bea
M_{\a\b} \,\j_\g=
\hf(\ve_{\g\a}\j_{\b}+\ve_{\g\b}\j_{\a})
~,\quad&&\qquad M_{\a\b}\, {\bar \j}_{\gd}=0~,\\
{\bar M}_{\ad\bd} \,{\bar \j}_{\gd}=
\hf(\ve_{\gd\ad}{\bar \j}_{\bd}+\ve_{\gd\bd}{\bar \j}_{\ad})
~,\quad&&\qquad {\bar M}_{\ad\bd}\, \j_{\g}=0~.
\eea
\end{subequations} 
The covariant derivatives of ${\rm AdS}^{4|4}$ satisfy the following algebra
\begin{subequations}  \label{1.2}
\bea
&& \qquad \{ \cD_\a , \bar \cD_\ad \} = -2\rm i \cD_{\a \ad} ~, \\
&& \qquad \{\cD_\a, \cD_\b \} = -4\bar \m\, M_{\a \b}~, \qquad
\{ {\bar \cD}_\ad, {\bar \cD}_\bd \} = 4\m\,\bar M_{\ad \bd}~, \\
&& \qquad [ \cD_\a , \cD_{ \b \bd }] 
=\rm i \bar \m\,\ve_{\a \b} \bar \cD_\bd~,  \qquad
\,\,[{ \bar \cD}_{\ad} , \cD_{ \b \bd }] 
=-\rm i \m\,\ve_{\ad \bd} \cD_\b~,    \\
&&\quad \,\,\,\,[ \cD_{\a \ad} , \cD_{ \b \bd } ] = -2 \bar \m \m \Big({\ve}_{\a \b} \bar M_{\ad \bd }+ \ve_{\ad \bd} M_{\a \b}\Big)~,  
\eea
\end{subequations} 
with $\m\neq 0$ being a  complex parameter, which is related to the scalar curvature 
$\cR$ of  AdS space by the rule $\cR = -12 |\m|^2$.

In our calculations, we often make use of the following identities, which can be readily derived from the covariant derivatives algebra \eqref{1.2}:
\begin{subequations} 
\label{1.4}
\bea 
\cD_\a\cD_\b
\!&=&\!\frac{1}{2}\ve_{\a\b}\cD^2-2{\bar \m}\,M_{\a\b}~,
\quad\qquad \,\,\,
{\bar \cD}_\ad{\bar \cD}_\bd
=-\frac{1}{2}\ve_{\ad\bd}{\bar \cD}^2+2\m\,{\bar M}_{\ad\bd}~,  \label{1.4a}\\
\cD_\a\cD^2
\!&=&\!4 \bar \m \,\cD^\b M_{\a\b} + 4{\bar \m}\,\cD_\a~,
\quad\qquad
\cD^2\cD_\a
=-4\bar \m \,\cD^\b M_{\a\b} - 2\bar \m \, \cD_\a~, \label{1.4b} \\
{\bar \cD}_\ad{\bar \cD}^2
\!&=&\!4 \m \,{\bar \cD}^\bd {\bar M}_{\ad\bd}+ 4\m\, \bar \cD_\ad~,
\quad\qquad
{\bar \cD}^2{\bar \cD}_\ad
=-4 \m \,{\bar \cD}^\bd {\bar M}_{\ad\bd}-2\m\, \bar \cD_\ad~,  \label{1.4c}\\
\left[\bar \cD^2, \cD_\a \right]
\!&=&\!4\rm i \cD_{\a\bd} \bar \cD^\bd +4 \m\,\cD_\a = 
4\rm i \bar \cD^\bd \cD_{\a\bd} -4 \m\,\cD_\a~,
 \label{1.4d} \\
\left[\cD^2,{\bar \cD}_\ad\right]
\!&=&\!-4\rm i \cD_{\b\ad}\cD^\b +4\bar \m\,{\bar \cD}_\ad = 
-4\rm i \cD^\b \cD_{\b\ad} -4 \bar \m\,{\bar \cD}_\ad~,
 \label{1.4e}
\eea
\end{subequations} 
where $\cD^2=\cD^\a\cD_\a$, and ${\bar \cD}^2={\bar \cD}_\ad{\bar \cD}^\ad$. 
These relations imply the identity 
\bea
\cD^\a (\cDB^2- 4 \mu) \cD_\a = \cDB_\ad (\cD^2 - 4 \mub) \cDB^\ad ~,
\label{1.44}
\eea
which guarantees the reality of the action functionals considered 
in the next sections.

Complex tensor superfields 
$\G_{\a(m) \ad(n)} :=\G_{\a_1 \dots \a_m \ad_1 \dots \ad_n}
=\G_{(\a_1 \dots \a_m)( \ad_1 \dots \ad_n)}$ and 
$G_{\a(m) \ad(n)}$ 
are referred to as transverse linear and longitudinal linear, respectively, if the constraints
\begin{subequations} \label{1.6}
\bea
&& \bar \cD^\bd \G_{ \a(m) \bd \ad(n - 1) } = 0 ~,  \qquad n \neq 0~,   \label{1.6a}
\\
&& \bar \cD_{(\ad_1} G_{\a(m)\ad_2 \dots \ad_{n+1} )} = 0  \label{1.6b}
\eea
\end{subequations}
are satisfied. 
For $n=0$ the latter constraint 
coincides with  the condition of covariant chirality, 
 $\bar \cD_\bd G_{ \a(m) } = 0$. 
 With the aid of \eqref{1.4a}, the relations \eqref{1.6} lead to the linearity conditions
\begin{subequations}
\bea
(\bar \cD^2-2(n+2)\m)\,\G_{\a(m) \ad(n)} &=& 0~, \label{1.7a}\\
(\bar \cD^2+2n\m)\,G_{\a(m) \ad(n)} &=& 0~. \label{1.7b}
\eea
\end{subequations}
The transverse condition \eqref{1.6a} is not defined for $n=0$. 
However its corollary \eqref{1.7a} remains consistent for the choice $n=0$
and corresponds to complex linear superfields $\G_{\a(m)}$
 constrained by 
\bea
(\bar \cD^2-4\m)\,\G_{\a(m) } = 0~.
\eea
In the family of constrained superfields $\G_{\a(m)}$ introduced, 
the scalar multiplet, $m=0$, is used most often in applications. 
One can define projectors $P^{\perp}_{n}$ and $P^{||}_{n}$
on the spaces of  transverse linear and longitudinal linear superfields
respectively: 
\begin{subequations}
\bea
P^{\perp}_{n}&=& \frac{1}{4 (n+1)\m} (\bar \cD^2+2n\m) ~,\\
P^{||}_{n}&=&- \frac{1}{4 (n+1)\m} (\bar \cD^2-2(n+2)\m ) ~,
\eea
\end{subequations} 
with the properties 
\bea
\big(P^{\perp}_{n}\big)^2 =P^{\perp}_{n} ~, \quad 
\big(P^{||}_{n}\big)^2=P^{||}_{n}~,
\quad P^{\perp}_{n} P^{||}_{n}=P^{||}_{n}P^{\perp}_{n}=0~,
\quad P^{\perp}_{n} +P^{||}_{n} ={\mathbbm 1}~.
\eea
Superfields \eqref{1.6} were introduced and studied by Ivanov and Sorin \cite{IS}
in their analysis of the representations of the AdS supersymmetry.
A nice review of the results of  \cite{IS} is given in the book \cite{West}.

Given a complex tensor superfield $V_{\a(m)  \ad(n)} $ with $n \neq 0$, 
it can be represented
as a sum of transverse linear and longitudinal linear multiplets, 
\bea
V_{\a(m) \ad(n)} = &-& 
\frac{1}{2 \mu (n+2)} \cDB^\gd \cDB_{(\gd} V_{\a(m) \ad_1 \dots  \ad_n)} 
- \frac{1}{2 \mu (n+1)} \cDB_{(\ad_1} \cDB^{|\gd|} V_{\a(m) \ad_2 \dots \ad_{n} ) \gd} 
~ . ~~~
\eea
Choosing $V_{\a(m) \ad(n)} $ to be  transverse linear ($\G_{\a(m) \ad(n)} $)
or longitudinal linear ($G_{\a(m) \ad(n)} $), the above relation
gives
\begin{subequations}
\bea
 \G_{\a(m) \ad(n)}&=& \bar \cD^\bd 
{ \Phi}_{\a(m)\,(\bd \ad_1 \cdots \ad_{n}) } ~,
 \\
 G_{\a(m) \ad(n)} &=& {\bar \cD}_{( \ad_1 }
 \Psi_{ \a(m) \, \ad_2 \cdots \ad_{n}) } ~,
\eea
\end{subequations}
for some 
prepotentials $ \F_{\a(m) \ad(n+1)}$ and $  \J_{\a(m) \ad(n-1)}$.
The constraints \eqref{1.6} hold for unconstrained $ \F_{\a(m) \ad(n+1)}$ and $  \J_{\a(m) \ad(n-1)}$.
These prepotentials  are defined modulo gauge transformations of the form:
\begin{subequations}
\bea
\d_\x \Phi_{\a(m)\, \ad (n+1)} 
&=&  \bar \cD^\bd 
{ \x}_{\a(m)\, (\bd \ad_1 \cdots \ad_{n+1}) } ~,
\\
\d_\z  \Psi_{ \a(m) \, \ad {(n-1}) } &=&  {\bar \cD}_{( \ad_1 }
 \z_{ \a(m) \, \ad_2 \cdots \ad_{n-1}) } ~,
\eea
\end{subequations}
with the gauge parameters $ { \x}_{\a(m)\,  \ad (n+2) } $
and $ \z_{ \a(m) \, \ad (n-2)}$ being unconstrained.

The isometry group of $\cN=1$ AdS superspace is ${\rm OSp(1|4)}$.
The  isometries transformations of AdS$^{4|4}$ are generated by the Killing vector fields
$\L^A E_A$ which are defined to solve the Killing equation
\bea
\big[\L+\hf \o^{bc}M_{bc},\cD_{A} \big]=0~,\qquad
\L:=\L^B \cD_B =
\l^{b} \cD_{b}+\l^\b \cD_\b+{\bar \l}_\bd {\bar \cD}^\bd~, 
\label{N=1-killings-0}
\eea
for some Lorentz superfield parameter   $\o^{bc}= -\o^{cb}$. 
As shown in \cite{BK}, 
the equations in (\ref{N=1-killings-0}) are equivalent to 
\begin{subequations}
\bea
\cD_{(\a}\l_{\b)\bd}&=&0~, \qquad  {\bar \cD}^\bd\l_{\a\bd}   + 8\ri\l_\a=0~,\\
\cD_\a\l^\a&=&0~,
\qquad
{\bar \cD}_\ad\l_\a  + \frac{\ri}{ 2}{\mu}\l_{\a\ad}  =0~, \\
\o_{\a\b}&=&\cD_\a\l_\b~.
\eea
\end{subequations}
The solution to these equations is given in \cite{BK}.
If $T$ is a tensor superfield (with  suppressed indices), 
its infinitesimal ${\rm OSp(1|4)}$ transformation is 
\bea
\d T = \big( \L+\hf \o^{bc}M_{bc} \big) T ~.
\eea

In Minkowski space, there are two ways to generate supersymmetric invariants, 
one of which corresponds to the 
integration over the full superspace and the other over its chiral subspace. 
In AdS superspace, every chiral integral can be always recast as 
a full superspace integral.
Associated with a scalar superfield $\cL$ is the following ${\rm OSp(1|4)}$ invariant
\bea
\int \rd^4x \rd^2 \q  \rd^2 \bar \q
\,E\,{\cal L} &=& 
-\frac{1}{4} \int
\rd^4x \rd^2 \q  
\, \cE\, {({\bar \cD}^2 - 4 \m)} {\cal L} ~, \qquad
E^{-1}= {\rm Ber}\, (E_{\rm A}{}^M)~,
\eea
where 
$\cE$ denotes the chiral integration measure.\footnote{In the chiral 
representation \cite{BK,GGRS}, the chiral measure
is $\cE= \vf^3$, where $\vf$ is the chiral compensator of old minimal 
supergravity \cite{Siegel78}.} 
Let $\cL_{\rm c}$ be a chiral scalar, $\bar \cD_\ad \cL_{\rm c} =0$. 
It generates the supersymmetric invariant 
$
\int \rd^4x \rd^2 \q  \, \cE \,{\cal L}_{\rm c}. 
$
The specific feature
of AdS superspace is that the chiral action can equivalently
be written as an integral over the full superspace  \cite{Siegel78,Zumino}
\bea
\int \rd^4x \rd^2 \q  \, \cE \,{\cal L}_{\rm c} 
= \frac{1}{\m} \int \rd^4x \rd^2 \q  \rd^2 \bar \q
\,{E}\, {\cal L}_{\rm c} ~.
\eea
Unlike the flat superspace case, the integral on the right does not vanish in AdS.


\section{Massless integer superspin multiplets} \label{section3}

Let $s$ be a positive integer. The longitudinal formulation for the massless
superspin-$s$ multiplet in AdS was realised in \cite{KS94} in terms
of the following dynamical variables 
\bea
v^{||}_{(s)} &=& \left\{ H_{\a(s-1) \ad(s-1)}(z), G_{\a(s)\ad(s)}(z), \bar{G}_{\a(s)\ad(s)}(z) \right\} \ . 
\eea
Here, $H_{\a(s-1)  \ad(s-1)}$ is an unconstrained real superfield,  
and $G_{\a(s)\ad(s)}$ is a longitudinal linear superfield.
The latter is a field strength associated with a complex unconstrained 
prepotential $\J_{\a(s) \ad(s-1)}$,
\bea
G_{\a_1 \dots \a_s \ad_1 \dots \ad_s} := 
\bar \cD_{(\ad_1} \J_{\a_1 \dots \a_s \ad_2 \dots \ad_s)}
\quad \Longrightarrow \quad
\bar \cD_{(\ad_1} G_{\a_1 \dots \a_s \ad_2 \dots \ad_{s+1})}=0~.
\label{g2.5}
\eea
The gauge freedom postulated in \cite{KS94} is given by 
\begin{subequations} \label{oldgaugefreedom}
\bea
\d H_{\a(s-1)\ad(s-1)} &=& \cD^\b L_{\b \a(s-1) \ad(s-1)} - \cDB^\bd \bar{L}_{\a(s-1)\bd\ad(s-1)} \ ,
\\
\d G_{\b \a(s-1) \bd \ad(s-1)} &=& \hf \cDB_{(\bd} \cD_{(\b} \cD^{|\g|} L_{\a(s-1))\g\ad(s-1))} \ , 
\eea
\end{subequations}
where the gauge parameter is $L_{\a(s)\ad(s-1)}$ is unconstrained.

In this section we propose a reformulation of the longitudinal theory 
that is obtained by enlarging the gauge freedom \eqref{oldgaugefreedom}
at the cost of 
introducing a new purely gauge superfield variables in addition to 
$H_{\a(s-1)  \ad(s-1)}$, $\J_{\a(s)  \ad(s-1)}$ and $\bar\J_{\a(s-1)  \ad(s)}$.
In such a setting, the  gauge freedom of  $\J_{\a(s)  \ad(s-1)}$ 
coincides with that of  a superconformal multiplet of superspin $s$ \cite{KMT}.
The new formulation will be an extension of the one given in \cite{HK2}
in the flat-superspace case.

\subsection{New formulation}

Given a positive integer $s \geq 2$, a massless superspin-$s$ multiplet can be described in ${\rm AdS}^{4|4}$
by using the following superfield variables: 
(i) an unconstrained prepotential
 $\J_{\a(s)\ad(s-1)}  $ 
and its complex conjugate $\bar \J_{\a(s-1)\ad(s)}$; 
(ii) a real  superfield 
 $H_{\a(s-1)\ad(s-1)}  =\bar H_{\a(s-1)\ad(s-1)}  $; and (iii)
a complex superfield $\S_{\a(s-1) \ad (s-2) }$ 
and its conjugate $\bar \S_{\a (s-2) \ad(s-1)}$, 
where $\S_{\a(s-1) \ad (s-2) }$ is constrained 
to be transverse linear,
\bea
\bar \cD^\bd \S_{\a(s-1) \bd \ad(s-3)} =0~.
\label{2.1}
\eea
The constraint \eqref{2.1} is solved in terms of a complex unconstrained prepotential $Z_{\a(s-1) \ad (s-1)}$ by the rule
\bea
\S_{\a(s-1) \ad (s-2)} = \bar \cD^\bd Z_{\a(s-1) (\bd \ad_1 \dots \ad_{s-2} )} ~.
\label{2.2}
\eea
This prepotential is defined modulo gauge transformations 
\bea
\d_\x Z_{\a(s-1) \ad (s-1)}=  \bar \cD^\bd \x_{\a(s-1) (\bd \ad_1 \dots \ad_{s-1} )} ~,
\label{2.3}
\eea
with the gauge parameter  $\x_{\a(s-1) \ad (s) }$ being  unconstrained.

The gauge freedom of $\J_{\a_1 \dots \a_s \ad_1 \dots \ad_{s-1}} $ is
chosen to coincide with that of the superconformal superspin-$s$ multiplet 
\cite{KMT}, which is
\begin{subequations} \label{2.4}
\bea
 \d_{ {\mathfrak V} ,\z} \J_{\a_1 \dots \a_s \ad_1 \dots \ad_{s-1}} 
 &=& \hf \cD_{(\a_1}  {\mathfrak V}_{\a_2 \dots \a_s)\ad_1 \dots \ad_{s-1}}
+  \bar \cD_{(\ad_1} \z_{\a_1 \dots \a_s \ad_2 \dots \ad_{s-1} )}  ~ , \label{2.4a}
\eea
with unconstrained gauge parameters ${\mathfrak V}_{\a(s-1) \ad(s-1)}$ 
and $\z_{\a(s) \ad(s-2)}$. 
The $\mathfrak V$-transformation is defined to act on the superfields $H_{\a(s-1) \ad(s-1)}$
and $\S_{\a(s-1) \ad(s-2) }$ as follows
\bea
\d_{\mathfrak V} H_{\a(s-1) \ad(s-1)}&=& {\mathfrak V}_{\a(s-1) \ad(s-1)} +\bar {\mathfrak V}_{\a(s-1) \ad(s-1)}
~, \label{2.4b}\\
\d_{\mathfrak V} \S_{\a(s-1) \ad(s-2) }&=&  \bar \cD^\bd \bar {\mathfrak V}_{\a(s-1) \bd \ad(s-2)}
\quad \Longrightarrow \quad \d_{\mathfrak V} Z_{\a(s-1) \ad (s-1)}
=\bar  {\mathfrak V}_{\a(s-1) \ad (s-1)}~.~~~
\label{2.4c}
\eea
\end{subequations}
The longitudinal linear superfield defined by \eqref{g2.5}
is invariant under the $\z$-transformation \eqref{2.4a} 
and  varies under the $\mathfrak V$-transformation as 
\bea
 \d_{ {\mathfrak V} } G_{\a_1 \dots \a_s \ad_1 \dots \ad_{s}} 
 &=& \hf \bar \cD_{(\ad_1} \cD_{(\a_1}  {\mathfrak V}_{\a_2 \dots \a_s)\ad_2 \dots \ad_{s})}~.
\eea

Our next task is to derive an AdS extension of the gauge-invariant action in Minkowski superspace (given by eq. (2.8) in \cite{HK2}). The geometry of AdS superspace is completely determined by the algebra \eqref{1.2}. We start with the following action functional in AdS superspace, which is a minimal AdS extension of the action constructed in \cite{HK2}.
\bea
S^{\|}_{(s)} &=&
\Big( - \frac{1}{2}\Big)^s  \int 
 \rd^4x \rd^2 \q  \rd^2 \bar \q
\, E
\left\{ \frac{1}{8} H^{ \a (s-1) \ad (s-1) }  \cD^\b {\bar \cD}^2 \cD_\b 
H_{\a (s-1) \ad (s-1)} \right. \non \\
&&+ \frac{s}{s+1}H^{ \a(s-1) \ad(s-1) }
\Big( \cD^{\b}  {\bar \cD}^{\bd} G_{\b\a(s-1) \bd\ad(s-1) }
- {\bar \cD}^{\bd}  \cD^{\b} 
{\bar G}_{\b \a (s-1) \bd \ad (s-1) } \Big) \non \\
&&+ 2 \bar G^{ \a (s) \ad (s) } G_{ \a (s) \ad (s) } 
+ \frac{s}{s+1}\Big( G^{ \a (s) \ad (s) } G_{ \a (s) \ad (s) } 
+ \bar G^{ \a (s) \ad (s) }  \bar G_{ \a (s) \ad (s) } 
 \Big) \non \\
 &&+ \frac{s-1}{4s}H^{ \a(s-1) \ad(s-1) }
\Big( \cD_{\a_1} \bar \cD^2 \bar \S_{\a_2 \dots \a_{s-1}\ad(s-1)}
 - {\bar \cD}_{\ad_1}  \cD^2 \S_{\a(s-1) \ad_2 \dots \ad_{s-1} } \Big)  \non \\
&&+\frac{1}{s} \J^{\a(s) \ad(s-1)} \Big( 
\cD_{\a_1} \bar \cD_{\ad_1} -2\ri (s-1) \cD_{\a_1 \ad_1} \Big)
\S_{\a_2 \dots \a_s \ad_2 \dots \ad_{s-1} }\non  \\
&&+\frac{1}{s} \bar \J^{\a(s-1) \ad(s)} \Big( 
 \bar \cD_{\ad_1} \cD_{\a_1}-2\ri (s-1) \cD_{\a_1 \ad_1} \Big)
\bar \S_{\a_2 \dots \a_{s-1} \ad_2 \dots \ad_{s} }\non \\
&&+ \frac{s-1}{8s} \Big( \S^{\a(s-1) \ad(s-2) } \cD^2 \S_{\a(s-1) \ad(s-2)} 
- \bar \S^{\a(s-2) \ad(s-1) }\bar \cD^2 \bar \S_{\a(s-2) \ad(s-1)} \Big) \non \\
&&- \frac{1}{s^2}\bar \S^{\a(s-2) \ad(s-2)\bd } \Big( \hf (s^2 +1) \cD^\b \bar \cD_\bd 
+\ri  {(s-1)^2} \cD^\b{}_\bd \Big) \S_{\b \a(s-2) \ad(s-2)}\Big\}
+\dots~~~
\label{2.7}
\eea

The gauge-invariant action in AdS is expected to differ from \eqref{2.7} by some $\m$-dependent terms, which are required to ensure invariance under the gauge transformations \eqref{2.4} and, by construction, \eqref{2.3}. We compute the variation of \eqref{2.7} under \eqref{2.4} and then add certain $\m$-dependent terms to achieve an invariant action. The identities \eqref{1.4} prove to be useful in carrying out such calculations. 

The above procedure leads to the following action in AdS, which is invariant under the gauge transformations \eqref{2.4} and \eqref{2.3}
\bea
S^{\|}_{(s)} &=&
\Big( - \frac{1}{2}\Big)^s  \int 
 \rd^4x \rd^2 \q  \rd^2 \bar \q
\, E
\left\{ \frac{1}{8} H^{ \a (s-1) \ad (s-1) }  \cD^\b ({\bar \cD}^2- 4\mu) \cD_\b 
H_{\a (s-1) \ad (s-1)} \right. \non \\
&&+ \frac{s}{s+1}H^{ \a(s-1) \ad(s-1) }
\Big( \cD^{\b}  {\bar \cD}^{\bd} G_{\b\a(s-1) \bd\ad(s-1) }
- {\bar \cD}^{\bd}  \cD^{\b} 
{\bar G}_{\b \a (s-1) \bd \ad (s-1) } \Big) \non \\
&&+ \frac{(s+1)^2}{2} \bar \mu \mu 
H^{\a (s-1) \ad (s-1)} H_{\a (s-1) \ad (s-1)} \non \\
&&+ 2 \bar G^{ \a (s) \ad (s) } G_{ \a (s) \ad (s) } 
+ \frac{s}{s+1}\Big( G^{ \a (s) \ad (s) } G_{ \a (s) \ad (s) } 
+ \bar G^{ \a (s) \ad (s) }  \bar G_{ \a (s) \ad (s) } 
 \Big) \non \\
 &&+ \frac{s-1}{4s}H^{ \a(s-1) \ad(s-1) }
\Big( \cD_{\a_1} \bar \cD^2 \bar \S_{\a_2 \dots \a_{s-1}\ad(s-1)}
 - {\bar \cD}_{\ad_1}  \cD^2 \S_{\a(s-1) \ad_2 \dots \ad_{s-1} } \Big)  \non \\
&&+\frac{1}{s} \J^{\a(s) \ad(s-1)} \Big( 
\cD_{\a_1} \bar \cD_{\ad_1} -2\ri (s-1) \cD_{\a_1 \ad_1} \Big)
\S_{\a_2 \dots \a_s \ad_2 \dots \ad_{s-1} }\non  \\
&&+\frac{1}{s} \bar \J^{\a(s-1) \ad(s)} \Big( 
 \bar \cD_{\ad_1} \cD_{\a_1}-2\ri (s-1) \cD_{\a_1 \ad_1} \Big)
\bar \S_{\a_2 \dots \a_{s-1} \ad_2 \dots \ad_{s} }\non \\
&&- \mu \frac{s^2+4s-1}{2s} H^{\a (s-1) \ad (s-1)} \cD_{\a_1} 
\bar \S_{\a_2 \dots \a_{s-1} \ad(s-1)} \non \\
&&+ \bar \mu \frac{s^2+4s-1}{2s} H^{\a (s-1) \ad (s-1)} \bar \cD_{\ad_1} 
 \S_{ \a(s-1) \ad_2 \dots \ad_{s-1}} \non \\
&&+ \frac{s-1}{8s} \Big( \S^{\a(s-1) \ad(s-2) } \cD^2 \S_{\a(s-1) \ad(s-2)} 
- \bar \S^{\a(s-2) \ad(s-1) }\bar \cD^2 \bar \S_{\a(s-2) \ad(s-1)} \Big) \non \\
&&- \frac{1}{s^2}\bar \S^{\a(s-2) \ad(s-2)\bd } \Big( \hf (s^2 +1) \cD^\b \bar \cD_\bd 
+\ri  {(s-1)^2} \cD^\b{}_\bd \Big) \S_{\b \a(s-2) \ad(s-2)} \non\\
&&+ \mu \frac{s^2+4s-1}{4s} \bar \S^{\a(s-2) \ad(s-1)} \bar \S_{\a(s-2) \ad(s-1)} \non \\
&&+ \bar \mu \frac{s^2+4s-1}{4s} \S_{\a(s-1) \ad(s-2)} \S^{\a(s-1) \ad(s-2)} \Big\}
~.~~~
\label{action}
\eea
The above action is real due to the identity \eqref{1.44}.
In the limit of vanishing curvature of the AdS superspace ($\m\rightarrow 0$), \eqref{action} reduces to the action constructed in \cite{HK2}. 

The $\mathfrak V$-gauge freedom \eqref{2.4} allows us to gauge away $\S_{\a(s-1) \ad(s-2) }$,
\bea
 \S_{\a(s-1) \ad(s-2) }=0~.
\label{2.10}
\eea
In this gauge, the action \eqref{action} reduces to 
that describing the longitudinal formulation for the massless superspin-$s$ multiplet  \cite{KS94}. The gauge condition \eqref{2.10} does not 
fix completely the $\mathfrak V$-gauge freedom. The residual gauge transformations  
are generated by 
\bea
{\mathfrak V}_{\a(s-1) \ad(s-1)} = \cD^\b L_{(\b \a_1 \dots \a_{s-1}) \ad(s-1)}~,
\label{2.11}
\eea
with $L_{\a(s) \ad(s-1)}$ being an unconstrained superfield. 
With this expression for ${\mathfrak V}_{\a(s-1) \ad(s-1)}$, the gauge transformations \eqref{2.4a}  and \eqref{2.4b} coincide with 
 \eqref{oldgaugefreedom}.
Thus, the action \eqref{action} indeed provides an off-shell formulation for the massless superspin-$s$ multiplet in the AdS superspace.

One can impose an alternative gauge fixing 
\bea
H_{\a(s-1) \ad(s-1)} =0~.
\label{2.12}
\eea
In accordance with \eqref{2.4b}, in this gauge
the residual gauge freedom is described by 
\bea
{\mathfrak V}_{\a(s-1) \ad(s-1)} = \ri {\mathfrak R}_{\a(s-1) \ad(s-1)}~, \qquad 
\bar{\mathfrak R}_{\a(s-1) \ad(s-1)}={\mathfrak R}_{\a(s-1) \ad(s-1)}~.
\eea

The action \eqref{action} includes a single term which involves the `naked' 
gauge field $\J_{\a(s)\ad(s-1)} $ and not the field strength $G_{\a(s)\ad(s)} $, 
the latter being 
defined by \eqref{g2.5} and invariant under the $\z$-transformation \eqref{2.4a}.
This is actually a BF term, for it can be written in two different forms
\bea
\frac{1}{s}  \int  \rd^4x \rd^2 \q  \rd^2 \bar \q \, E
 \,
 \J^{\a(s) \ad(s-1)} \Big( 
\cD_{\a_1} \bar \cD_{\ad_1} &-&2\ri (s-1) \cD_{\a_1 \ad_1} \Big)
\S_{\a_2 \dots \a_s \ad_2 \dots \ad_{s-1} } \non \\
=- \frac{1}{s+1}  \int 
 \rd^4x \rd^2 \q  \rd^2 \bar \q \, E
\,
 G^{\a(s) \ad(s)} \Big( \bar \cD_{\ad_1} \cD_{\a_1}  
&+&2\ri (s+1) \cD_{\a_1 \ad_1} \Big)
Z_{\a_2 \dots \a_s \ad_2 \dots \ad_{s} }~.
\label{2.14}
\eea
The former makes the gauge symmetry \eqref{2.3} manifestly realised, 
while the latter
turns the $\z$-transformation \eqref{2.4a} into a manifest symmetry.

Making use of \eqref{2.14} leads to a different representation 
for the action \eqref{action}. It is 
\bea
S^{\|}_{(s)} &=&
\Big( - \frac{1}{2}\Big)^s  \int 
 \rd^4x \rd^2 \q  \rd^2 \bar \q
\, E
\left\{ \frac{1}{8} H^{ \a (s-1) \ad (s-1) }  \cD^\b ({\bar \cD}^2- 4\mu) \cD_\b 
H_{\a (s-1) \ad (s-1)} \right. \non \\
&&+ \frac{s}{s+1}H^{ \a(s-1) \ad(s-1) }
\Big( \cD^{\b}  {\bar \cD}^{\bd} G_{\b\a(s-1) \bd\ad(s-1) }
- {\bar \cD}^{\bd}  \cD^{\b} 
{\bar G}_{\b \a (s-1) \bd \ad (s-1) } \Big) \non \\
&&+ \frac{(s+1)^2}{2} \bar \mu \mu 
H^{\a (s-1) \ad (s-1)} H_{\a (s-1) \ad (s-1)} \non \\
&&+ 2 \bar G^{ \a (s) \ad (s) } G_{ \a (s) \ad (s) } 
+ \frac{s}{s+1}\Big( G^{ \a (s) \ad (s) } G_{ \a (s) \ad (s) } 
+ \bar G^{ \a (s) \ad (s) }  \bar G_{ \a (s) \ad (s) } 
 \Big) \non \\
 &&+ \frac{s-1}{4s}H^{ \a(s-1) \ad(s-1) }
\Big( \cD_{\a_1} \bar \cD^2 \bar \S_{\a_2 \dots \a_{s-1}\ad(s-1)}
 - {\bar \cD}_{\ad_1}  \cD^2 \S_{\a(s-1) \ad_2 \dots \ad_{s-1} } \Big)  \non \\
&&-\frac{1}{s+1} G^{\a(s) \ad(s-1)} \Big( 
\bar \cD_{\ad_1} \cD_{\a_1} +2\ri (s+1) \cD_{\a_1 \ad_1} \Big)
Z_{\a_2 \dots \a_s \ad_2 \dots \ad_s }\non  \\
&&+ \frac{1}{s+1} G^{\a(s) \ad(s-1)} \Big( 
\cD_{\a_1} \bar \cD_{\ad_1} +2\ri (s+1) \cD_{\a_1 \ad_1} \Big)
\bar Z_{\a_2 \dots \a_s \ad_2 \dots \ad_s }\non \\
&&- \mu \frac{s^2+4s-1}{2s} H^{\a (s-1) \ad (s-1)} \cD_{\a_1} 
\bar \S_{\a_2 \dots \a_{s-1} \ad(s-1)} \non \\
&&+ \bar \mu \frac{s^2+4s-1}{2s} H^{\a (s-1) \ad (s-1)} \bar \cD_{\ad_1} 
 \S_{ \a(s-1) \ad_2 \dots \ad_{s-1}} \non \\
&&+ \frac{s-1}{8s} \Big( \S^{\a(s-1) \ad(s-2) } \cD^2 \S_{\a(s-1) \ad(s-2)} 
- \bar \S^{\a(s-2) \ad(s-1) }\bar \cD^2 \bar \S_{\a(s-2) \ad(s-1)} \Big) \non \\
&&- \frac{1}{s^2}\bar \S^{\a(s-2) \ad(s-2)\bd } \Big( \hf (s^2 +1) \cD^\b \bar \cD_\bd 
+\ri  {(s-1)^2} \cD^\b{}_\bd \Big) \S_{\b \a(s-2) \ad(s-2)} \non\\
&&+ \mu \frac{s^2+4s-1}{4s} \bar \S^{\a(s-2) \ad(s-1)} \bar \S_{\a(s-2) \ad(s-1)} \non \\
&&+ \bar \mu \frac{s^2+4s-1}{4s} \S_{\a(s-1) \ad(s-2)} \S^{\a(s-1) \ad(s-2)} \Big\}
~.~~~
\label{action2}
\eea


\subsection{Dual formulation}

As in the case of the flat superspace \cite{HK2}, the theory with action \eqref{action2} can be reformulated in terms of a transverse linear superfield by applying the duality transformation introduced in \cite{KS94}.

We now associate with our theory \eqref{action2} 
the following first-order action\footnote{The specific normalisation of the Lagrange multiplier in \eqref{action3} is chosen to match that of \cite{KS94}.} 
\bea
S_{\text{first-order} }&=&S^{\|}_{(s)}[U, \bar U, H , Z, \bar Z]  \non \\
&&+ \Big( \frac{-1}{2} \Big)^s\int  \rd^4x \rd^2 \q  \rd^2 \bar \q \,E \,
 \Big(\frac{2}{s+1} \G^{\a(s) \ad(s)} U_{\a(s) \ad(s)} 
  +{\rm c.c.} \Big)~,~~
\label{action3}
\eea
where $S^{\|}_{(s)}[U, \bar U, H , Z, \bar Z] $ is obtained from the action
\eqref{action2} by replacing $G_{\a(s) \ad(s)} $ with an unconstrained 
complex superfield $U_{\a(s) \ad(s)} $, and the Lagrange multiplier $\G_{\a(s) \ad(s)} $
is transverse linear, 
\bea \label{2.17}
\bar \cD^\bd \G_{\a(s) \bd \ad_1 \dots \ad_{s-1}} =0 ~.
\eea
Varying \eqref{action3} with respect to the Lagrange multiplier and taking into account the constraint \eqref{2.17} yields $U_{\a(s) \ad(s)} = G_{\a(s) \ad(s)} $; then, $S_{\text{first-order} }$ turns into the original action \eqref{action2}. On the other hand, we can eliminate the auxiliary superfields  $U_{\a(s) \ad(s)}$ and  $\bar U_{\a(s) \ad(s)}$ from \eqref{action3} using their equations of motion. This leads to the dual action
\bea
S^{\perp}_{(s)} &=& - \Big( - \hf \Big)^s 
  \int  \rd^4x \rd^2 \q  \rd^2 \bar \q \, E \,
\Bigg\{ - \frac{1}{8} H^{\a(s-1) \ad(s-1)} \cD^\b ({\bar \cD}^2- 4\mu) \cD_\b 
H_{\a (s-1) \ad (s-1)} \non\\
&&+ \frac{1}{8} \frac{s^2}{(s+1)(2s+1)} [\cD^\b, \bar \cD^\bd] H^{\a(s-1)\ad(s-1)} 
[\cD_{(\b}, \bar \cD_{(\bd}] H_{\a(s-1))\ad(s-1))} \non\\
&&+\hf \frac{s^2}{s+1} \cD^{\b\bd} H^{\a(s-1)\ad(s-1)} \cD_{(\b(\bd} H_{\a(s-1))\ad(s-1))} \non \\
&&- \frac{(s+1)^2}{2} \bar \mu \mu 
H^{\a (s-1) \ad (s-1)} H_{\a (s-1) \ad (s-1)} \non \\
&&+ \frac{2 \ri s}{2s+1} H^{\a(s-1)\ad(s-1)} \cD^{\b\bd} 
\Big({\bm \G}_{\b\a(s-1)\bd\ad(s-1)} 
- \bar{\bm \G}_{\b\a(s-1)\bd\ad(s-1)}\Big) \non\\
&& 
+ \frac{2}{2s+1} \bar{\bm \G}^{ \a (s) \ad (s) } {\bm \G}_{ \a (s) \ad (s) } 
- \frac{s}{(s+1)(2s+1)} \Big({\bm \G}^{ \a (s) \ad (s) }  {\bm \G}_{ \a (s) \ad (s) } 
+ \bar{\bm \G}^{ \a (s) \ad (s) } \bar{\bm \G}_{ \a (s) \ad (s) }\Big) \non\\
 &&- \frac{s-1}{2(2s+1)} H^{ \a(s-1) \ad(s-1) }
\Big( \cD_{\a_1} \bar \cD^2 \bar \S_{\a_2 \dots \a_{s-1}\ad(s-1)}
 - {\bar \cD}_{\ad_1}  \cD^2 \S_{\a(s-1) \ad_2 \dots \ad_{s-1} } \Big)  \non \\
&&+ \frac{1}{2(2s+1)} H^{ \a(s-1) \ad(s-1) }
\Big( \cD^2 {\bar \cD}_{\ad_1} \S_{\a(s-1) \ad_2 \dots \ad_{s-1} }
 - \bar \cD^2 \cD_{\a_1} \bar \S_{\a_2 \dots \a_{s-1}\ad(s-1)} \Big)  \non \\
&&- \ri \frac{(s-1)^2}{s(2s+1)} H^{ \a(s-1) \ad(s-1) }
 \cD_{\a_1 \ad_1} \Big( \cD^\b \S_{ \b \a_2 \dots \a_{s-1} \ad_2 \dots \ad_{s-1}}
+ \bar \cD^\bd \bar \S_{ \a_2 \dots \a_{s-1} \bd \ad_2 \dots \ad_{s-1}} \Big)  \non \\
&&+  \mu \frac{(s+2)(s+1)}{2s+1} H^{\a (s-1) \ad (s-1)}  \cD_{\a_1} 
\bar \S_{\a_2 \dots \a_{s-1} \ad(s-1)} \non \\
&&- \bar \mu \frac{(s+2)(s+1)}{2s+1} H^{\a (s-1) \ad (s-1)} \bar \cD_{\ad_1} 
 \S_{ \a(s-1) \ad_2 \dots \ad_{s-1}} \non \\
&&- \frac{s-1}{8s} \Big( \S^{\a(s-1) \ad(s-2) } \cD^2 \S_{\a(s-1) \ad(s-2)} 
- \bar \S^{\a(s-2) \ad(s-1) }\bar \cD^2 \bar \S_{\a(s-2) \ad(s-1)} \Big) \non \\
&&+ \frac{1}{s^2}\bar \S^{\a(s-2) \ad(s-2)\bd } \Big( \hf (s^2 +1) \cD^\b \bar \cD_\bd 
+\ri  {(s-1)^2} \cD^\b{}_\bd \Big) \S_{\b \a(s-2) \ad(s-2)} \non\\
&&- \mu \frac{s^2+4s-1}{4s} \bar \S^{\a(s-2) \ad(s-1)} \bar \S_{\a(s-2) \ad(s-1)} \non \\
&&- \bar \mu \frac{s^2+4s-1}{4s} \S_{\a(s-1) \ad(s-2)} \S^{\a(s-1) \ad(s-2)}
\Bigg\} ~,
\label{action4}
\eea
 where we have defined
\bea
{\bm \G}_{ \a (s) \ad (s) } = \G_{ \a (s) \ad (s) }
-\hf \bar \cD_{(\ad_1} \cD_{(\a_1} Z_{\a_2 \dots \a_s) \ad_2 \dots \ad_s) } 
-\ri (s+1)\cD_{(\a_1 (\ad_1 } Z_{\a_2 \dots \a_s) \ad_2 \dots \ad_s) } ~.~~~
\label{shifted}
\eea

The first-order model introduced is equivalent to the original theory \eqref{action2}. The action \eqref{action3} is invariant under the gauge $\x$-transformation 
 \eqref{2.3} which acts on $U_{\a (s) \ad (s)}$ and
 $\G_{\a(s) \ad(s)}$ by the rule
\begin{subequations}
 \bea
 \d_\x U_{\a (s) \ad (s)} &=&0~,\\
 \d_\x \G_{\a(s) \ad(s)} &=& \bar \cD^\bd \Big\{  \frac{s+1}{2(s+2)}
\bar \cD_{(\bd} \cD_{\a_1} \x_{\a_2 \dots \a_s \ad_1 \dots \ad_s) } 
+\ri (s+1)\cD_{\a_1 (\bd } \x_{\a_2 \dots \a_s \ad_1 \dots \ad_s) } \Big\}~.~~~ \label{2.20}
\eea
\end{subequations}
${\bm \G}_{ \a (s) \ad (s) }$ is invariant under the gauge transformations \eqref{2.3} and \eqref{2.20}.

The first-order action  \eqref{action3} is also invariant under the gauge $\mathfrak V$-transformation \eqref{2.4b} and \eqref{2.4c}, which acts on $U_{\a (s) \ad (s)}$ and
$\G_{\a(s) \ad(s)} $ as
\begin{subequations}
\bea
 \d_{ {\mathfrak V} } U_{\a (s) \ad (s)} 
 &=& \hf \bar \cD_{(\ad_1} \cD_{(\a_1}  {\mathfrak V}_{\a_2 \dots \a_s)\ad_2 \dots \ad_{s})}~, \\
\d_{\mathfrak V} \G_{\a(s) \ad(s)} &=&0~.
\eea 
\end{subequations}
 
In accordance with \eqref{2.4c}, the $\mathfrak V$-gauge freedom may be used to impose the condition 
\bea
Z_{\a(s-1) \ad(s-1)} =0~.
\label{3.11}
\eea
In this gauge the action \eqref{action4} reduces to the one defining 
the transverse formulation for the massless superspin-$s$ multiplet \cite{KS94}.
The gauge condition \eqref{3.11} is preserved by residual local 
$\mathfrak V$- and $\x$-transformations of the form 
\bea
  \bar \cD^\bd \x_{\a(s-1) \bd \ad (s-1 )}  +
  \bar {\mathfrak V}_{\a(s-1)\ad (s-1 )} =0~.
\eea
Making use of the parametrisation \eqref{2.11}, the residual gauge freedom is
\begin{subequations}
\bea
\d H_{\a(s-1)\ad(s-1)} &=& \cD^\b L_{\b \a(s-1) \ad(s-1)} - \bar \cD^\bd \bar{L}_{\a(s-1)\bd\ad(s-1)} \ ,\\
\d \G_{\a(s) \ad(s)} &=& \frac{s+1}{2(s+2)} \bar \cD^\bd \Big\{ 
\bar \cD_{(\bd} \cD_{(\a_1} 
+ 2\ri (s+2) \cD_{(\a_1(\bd}\Big\}
 \bar{L}_{\a_2 \dots \a_{s})\ad_1 \dots \ad_{s})} \ ,
\eea
\end{subequations}
which is exactly the gauge symmetry of the transverse formulation for the massless superspin-$s$ multiplet \cite{KS94}.

\subsection{Models for the massless gravitino multiplet in AdS}

The massless gravitino multiplet (i.e., the massless superspin-1 multiplet)
was excluded from the above consideration. Here we will fill the gap. 
 
The (generalised) longitudinal formulation for the gravitino multiplet is described by the action
\begin{subequations}\label{3.26}
\bea
S^{\|}_{\rm GM} &=&
 - \int  \rd^4x \rd^2 \q  \rd^2 \bar \q\, E
\left\{ \frac{1}{16} H  \cD^\a ({\bar \cD}^2- 4\mu) \cD_\a H 
+ \frac{1}{4} H
\big( \cD^{\a}  {\bar \cD}^{\ad} G_{\a \ad}
- {\bar \cD}^{\ad}  \cD^{\a} {\bar G}_{\a \ad} \big) 
\right. 
\non \\
&& \qquad +  \bar G^{ \a  \ad  } G_{ \a  \ad } 
+ \frac 14 \big( G^{ \a  \ad } G_{ \a  \ad } 
+ \bar G^{ \a  \ad  }  \bar G_{ \a  \ad  } 
\big)   \non\\
&& 
\qquad
 \left. 
+|\m|^2 \Big( H - \frac{\F}{\m} -\frac{\bar \F}{\bar \m} \Big)^2
+\Big( \frac{\F}{\m} +\frac{\bar \F}{\bar \m}\Big)
\Big( \m \cD^\a \J_\a + \bar \m \bar \cD_\ad \bar \J^\ad \Big) 
\right\}~,
\eea
where $\F$ is a chiral scalar superfield, $\bar \cD_\ad \F=0$, and
\bea
G_{\a\ad} = \bar \cD_\ad \J_\a ~, \qquad \bar G_{\a\ad} = - \cD_\a \bar \J_\ad~.
\eea
\end{subequations}
This action is invariant under gauge transformations of the form 
\begin{subequations}
\bea
\d H&=& {\mathfrak V} +\bar {\mathfrak V}~,  \label{3.28a} \\
\d \J_\a &=& = \hf \cD_\a  {\mathfrak V}+ \eta_\a~, \qquad \bar \cD_\ad \eta_\a =0~,
\label{3.28b} \\
\d \F &=& -\frac 14 (\bar \cD^2 -4\m) \bar  {\mathfrak V}~.
\label{3.28c} 
\eea
\end{subequations}
This is one of the two models for the massless gravitino multiplet in AdS introduced in 
\cite{BK11}.  In a flat-superspace limit, the action reduces to that given in \cite{GS}.
Imposing the gauge condition $\F=0$ reduces the action \eqref{3.26}
to the original longitudinal formulation for the massless gravitino multiplet in AdS \cite{KS94}.

The action \eqref{3.26} involves the chiral scalar $\F$ and its conjugate only in the combination $(\vf + \bar \vf)$, where $\vf = \F /\m$. This means that the model \eqref{3.26} 
possesses a dual formulation realised in terms of a real linear superfield $L$, 
\bea
\big( \bar \cD^2 -4\m\big) L =0~, \qquad \bar L =L~.
\eea
The dual model is described by the action \cite{BK11}
\bea
S_{\rm GM} &=&
 - \int  \rd^4x \rd^2 \q  \rd^2 \bar \q\, E
\left\{ \frac{1}{16} H  \cD^\a ({\bar \cD}^2- 4\mu) \cD_\a H 
+ \frac{1}{4} H
\big( \cD^{\a}  {\bar \cD}^{\ad} G_{\a \ad}
- {\bar \cD}^{\ad}  \cD^{\a} {\bar G}_{\a \ad} \big) 
\right. 
\non \\
&& \qquad +  \bar G^{ \a  \ad  } G_{ \a  \ad } 
+ \frac 14 \big( G^{ \a  \ad } G_{ \a  \ad } 
+ \bar G^{ \a  \ad  }  \bar G_{ \a  \ad  } 
\big)  + |\m|^2H^2 \non\\
&& 
\qquad
 \left. 
-\frac 14  \Big( 2 |\m|H + L- \frac{\m}{|\m|} \cD^\a \J_\a - \frac{\bar \m}{|\m|} 
\bar \cD_\ad \bar \J^\ad \Big)^2
\right\}~.
\label{3.299}
\eea
This action is invariant under the gauge transformations \eqref{3.28a}, \eqref{3.28b}
and
\bea
\d L =\frac{1}{|\m|} \big( {\m} \cD^\a \eta_\a +{\bar \m} \bar \cD_\ad \bar \eta^\ad \big)~.
\eea
In a flat-superspace limit, the action \eqref{3.299} reduces to that given in \cite{LR2}.

In Minkowski superspace, there exists one more dual realisation for the massless gravitino 
multiplet model \cite{HK2} which is obtained by performing a Legendre transformation 
converting $\F$ into a complex linear superfield. This formulation cannot be lifted to 
the AdS case, the reason being the fact that 
the action \eqref{3.26} involves the chiral scalar $\F$ and its conjugate only in the combination $(\vf + \bar \vf)$, where $\vf = \F /\m$. 

The dependence on $\J_\a $ and $\bar \J_\ad$
in  the last term of \eqref{3.26} 
can be expressed in terms of $G_{\a\ad}$ and $\bar G_{\a\ad}$
if we introduce a complex unconstrained prepotential $U$
for $\F$ in the standard way
\bea
\F = -\frac 14 (\bar \cD^2 -4\m) U~.
\eea
Then making use of \eqref{1.4d} gives
\bea
 \int  \rd^4x \rd^2 \q  \rd^2 \bar \q\, E \, \F \cD^\a \J_\a 
 = - \int  \rd^4x \rd^2 \q  \rd^2 \bar \q\, E\,G^{\a\ad}
 \Big( \frac 14 \bar \cD_\ad \cD_\a  +\ri \cD_{\a\ad}  \Big)U~.
 \eea
 Since the resulting action depends on $G_{\a\ad}$ and $\bar G_{\a\ad}$, 
we can introduce a dual formulation for the theory that is obtained
turning $G_{\a\ad}$ and $\bar G_{\a\ad}$ into a transverse linear superfield
\bea
\G_{\a\ad} = \bar \cD^\bd 
{ \Phi}_{\a\,\ad \bd } ~, \qquad { \Phi}_{\a\,\bd \ad } ={ \Phi}_{\a\,\ad \bd } 
\eea
and its conjugate using the scheme described in \cite{KS94}.
The resulting action is 
\bea
S^{\perp}_{\rm GM} &=&
\int  \rd^4x \rd^2 \q  \rd^2 \bar \q\, E
\left\{ -\frac{1}{16} H  \cD^\a ({\bar \cD}^2- 4\mu) \cD_\a H \right.
\non \\
&&+ \frac{1}{96} [\cD^\a, \bar \cD^\ad] H \, [\cD_\a, \bar \cD_\ad] H + \frac{1}{8} {\cD}^{\a \ad} H\, {\cD}_{\a \ad}H 
\non\\
&&+ \frac{1}{3} \bar {\bm \G}^{\a \ad} {\bm \G}_{\a \ad} - \frac{1}{12} \Big( {\bm \G}^{\a \ad} {\bm \G}_{\a \ad} + \bar {\bm \G}^{\a \ad} \bar {\bm \G}_{\a \ad} \Big)
+ \frac{\ri}{3}   \Big(\bar {\bm \G}^{\a \ad} - {\bm \G}^{\a \ad} \Big)
{\cD}_{\a \ad} H
\non\\
&&- \frac{1}{6} \F {\cD}^2 H- \frac{1}{6} {\bar \F} {\bar \cD}^2 H -|\m|^2 \Big( H - \frac{\F}{\m} - \left. \frac{\bar \F}{\bar \m} \Big)^2 \right\}~,
\label{Tspin1}
\eea
where we have defined
\bea
{\bm \G}_{\a \ad} := \G_{\a \ad} -\hf {\bar \cD}_\ad {\cD}_\a U -2\ri\, {\cD}_{\a \ad} U ~.
\label{shifted.s1}
\eea
The action \eqref{Tspin1} is invariant under the following gauge transformations 
\begin{subequations}
\bea
\d_\x U &=& {\bar \cD}_\ad \bar \x^\ad ~,\\
\d_\x \G_{\a \ad} &=& -\frac{1}{3} \bar \cD^\bd \Big\{\bar \cD_{(\bd} \cD_{\a} \bar \x_{\ad) } 
+6 \ri \, \cD_{\a (\bd } \bar \x_{\ad) } \Big\}~.~~~ \label{xi-spin1}
\eea
\end{subequations}
Both $\F$ and ${\bm \G}_{\a \ad}$ are invariant under $\x$-gauge transformations.
The action \eqref{Tspin1} is also invariant under the gauge transformations \eqref{3.28a}, \eqref{3.28c} and
\begin{subequations}
\bea
\d_{\mathfrak V} U &=& \bar {\mathfrak V}~, \\
\d_{\mathfrak V} \G_{\a \ad} &=& 0~.
\eea 
\end{subequations}
Imposing the gauge condition $U=0$ reduces the action \eqref{Tspin1}
to the original  transverse formulation for the massless gravitino multiplet in AdS \cite{KS94}.


\section{Higher spin supercurrents}

In this section we introduce higher spin supercurrent multiplets in AdS. 
First of all, we recall the structure of the gauge superfields in terms of which
the massless half-integer superspin multiplets are described
\cite{KS94}.

\subsection{Massless half-integer superspin multiplets} \label{subsection4.1}

For a  massless multiplet of half-integer superspin $s+1/2$, with $s=2, 3, \ldots$,
there exist two off-shell formulations \cite{KS94}
which are referred to as transverse and longitudinal. 
They are 
described in terms of 
the following dynamical variables:
\begin{subequations}
\bea
\cV^\bot_{s+1/2}& = &\Big\{H_{\a(s)\ad(s)}~, ~
\G_{\a(s-1) \ad(s-1)}~,
~ \bar{\G}_{\a(s-1) \ad(s-1)} \Big\} ~,    \\
\label{10}
\cV^{\|}_{s+1/2} &=& 
\Big\{H_{\a(s)\ad(s)}~, ~
G_{\a(s-1) \ad(s-1)}~,
~ \bar{G}_{\a(s-1) \ad(s-1)} \Big\}
~.
\eea
\end{subequations}
Here $H_{\a(s) \ad (s)}$ is a real unconstrained superfield.
The complex superfields 
$\G_{\a (s-1) \ad (s-1)} $ and 
$G_{\a (s-1) \ad (s-1)}$ are  transverse linear and 
longitudinal linear, respectively,
\begin{subequations}
\bea
{\bar \cD}^\bd \,\G_{\a(s-1) \bd \ad(s-2)} &=&  0 ~,
\label{transverse}
\\
{\bar \cD}_{ (\ad_1} \,G_{\a(s-1) \ad_2 \dots \ad_{s})}&=&0 ~.
\label{longitudinal}
\eea
\end{subequations}
These constraints are solved in terms of unconstrained prepotentials
as follows:
\begin{subequations}
\bea
 \G_{\a(s-1) \ad(s-1)}&=& \bar \cD^\bd 
{ \Phi}_{\a(s-1)\,(\bd \ad_1 \cdots \ad_{s-1}) } ~,
\label{6.3a}
 \\
 G_{\a(s-1) \ad(s-1)} &=& {\bar \cD}_{( \ad_1 }
 \Psi_{ \a(s-1) \, \ad_2 \cdots \ad_{s-1}) } ~.
\label{6.3b}
\eea
\end{subequations}
The prepotentials  are defined modulo gauge transformations of the form:
\begin{subequations}
\bea
\d_\x \Phi_{\a(s-1)\, \ad (s)} 
&=&  \bar \cD^\bd 
{ \x}_{\a(s-1)\, (\bd \ad_1 \cdots \ad_{s}) } ~,
\label{tr-prep-gauge}
\\
\d_\z  \Psi_{ \a(s-1) \, \ad {(s-2}) } &=&  {\bar \cD}_{( \ad_1 }
 \z_{ \a(s-1) \, \ad_2 \cdots \ad_{s-2}) } ~,
\label{lon-prep-gauge}
\eea
\end{subequations}
with the gauge parameters $ { \x}_{\a(s-1)\,  \ad (s+1) } $
and $ \z_{ \a(s-1) \, \ad (s-3)}$ being unconstrained.

The gauge transformations of the superfields $H$, $\G$ and $G$ are 
\begin{subequations} \label{6.5}
\bea 
\d_\L H_{\a_1 \dots \a_s \ad_1  \dots \ad_s} 
&= &\bar \cD_{(\ad_1} \L_{\a_1\dots  \a_s \ad_2 \dots \ad_s )} 
- \cD_{(\a_1} \bar{\L}_{\a_2 \dots \a_s)\ad_1  \dots \ad_s} \ , \label{6.5a} \\
\d_\L \G_{\a_1 \dots \a_{s-1} \ad_1  \dots \ad_{s-1}} 
&= & -\frac{s}{2(s+1)} {\bar \cD}^\bd {\cD}^\b {\cD}_{(\b} {\bar \L}_{\a(s-1)) \bd \ad(s-1)} \non\\
&=& -\frac{1}{4} \bar \cD^\bd \cD^2 \bar{\L}_{\a_1 \dots \a_{s-1}\bd \ad_1 
\dots \ad_{s-1}} \non\\
&&-\hf {\bar \m} (s-1) {\bar \cD}^\bd \bar{\L}_{\a_1 \dots \a_{s-1}\bd \ad_1 
\dots \ad_{s-1}} \ ,\label{6.5b} \\
\d_\L G_{\a_1 \dots \a_{s-1}\ad_1 \dots \ad_{s-1}} &= & - \hf \bar \cD_{(\ad_1} 
\bar \cD^{|\bd|} \cD^\b \L_{\b\a_1 \dots \a_{s-1} \ad_2 \dots \ad_{s-1}) \bd} \non\\
&&+ \ri (s-1) \bar \cD_{(\ad_1} \cD^{\b |\bd|} 
\L_{\b \a_1 \dots \a_{s-1} \ad_2 \dots \ad_{s-1} ) \bd} \ . \label{6.5c}
\eea
\end{subequations}
Here the gauge parameter $\L_{\a_1 \dots \a_s \ad_1 \dots \ad_{s-1}}
=\L_{(\a_1 \dots \a_s )(\ad_1 \dots \ad_{s-1})}$ 
is unconstrained. 
The symmetrisation in \eqref{6.5c} is extended only to the indices 
$\ad_1, \ad_2, \dots,  \ad_{s-1}$. 
It follows from \eqref{6.5b} and \eqref{6.5c} that the transformation laws of the prepotentials $\Phi_{\a(s-1) \ad(s)}$ and $\J_{\a(s-1) \ad(s-2)}$ are
\begin{subequations} \label{6.6}
\bea
\d_\L \Phi_{\a_1 \dots \a_{s-1}\ad_1 \dots \ad_{s}} &= &  -\frac{1}{4} \cD^2 \bar{\L}_{\a_1 \dots \a_{s-1} \ad_1 
\dots \ad_{s}} 
-\hf {\bar \m} (s-1) \bar{\L}_{\a_1 \dots \a_{s-1} \ad_1 
\dots \ad_{s}}\ , \label{6.6a}\\
\d_\L \J_{\a_1 \dots \a_{s-1}\ad_1 \dots \ad_{s-2}} &= & - \hf 
\Big( \bar \cD^{\bd} \cD^\b -2\ri (s-1) \cD^{\b \bd} \Big)
\L_{\b\a_1 \dots \a_{s-1} \bd \ad_1 \dots \ad_{s-2}} \ . \label{6.6b}
\eea
\end{subequations}


\subsection{Non-conformal supercurrents: Half-integer superspin} 

In the framework of the longitudinal formulation, 
let us couple the prepotentials 
$H_{ \a (s) \ad (s) } $, $\J_{ \a (s-1) \ad (s-2) }$ and $\bar \J_{ \a (s-2) \ad (s-1) }  $,
to external sources
\bea
S^{(s+\hf)}_{\rm source}=\int \rd^4x \rd^2 \q  \rd^2 \bar \q \, E\, \Big\{ 
H^{ \a (s) \ad (s) } J_{ \a (s) \ad (s) }
&+& \J^{ \a (s-1) \ad (s-2) } T_{ \a (s-1) \ad (s-2) } \non \\
&+& \bar \J_{ \a (s-2) \ad (s-1) } \bar T^{ \a (s-2) \ad (s-1) } \Big\}~.
\label{7.1}
\eea
Requiring $S^{(s+\hf)}_{\rm source}$ to be invariant under 
\eqref{lon-prep-gauge} gives
\begin{subequations} \label{7.3}
\bea
\bar \cD^\bd T_{\a(s-1) \bd \ad_1 \dots \ad_{s-3}} =0~,
\label{7.2}
\eea
and therefore $T_{ \a (s-1) \ad (s-2) } $ is a transverse linear superfield. 
Requiring $S^{(s+\hf)}_{\rm source}$ to be invariant under the gauge transformations
(\ref{6.5a}) and (\ref{6.6b}) gives the following conservation equation:
\bea
\bar \cD^\bd J_{\a_1 \dots \a_s \bd \ad_1 \dots \ad_{s-1}} 
+\hf \Big( \cD_{(\a_1} \bar \cD_{(\ad_1}
-2\ri (s-1) \cD_{ (\a_1 (\ad_1 } \Big)  T_{\a_2\dots \a_s) \ad_2 \dots \ad_{s-1})} =0~.
\label{7.3a}
\eea
For completeness, we also give the conjugate equation
\bea
\cD^\b J_{\b \a_1 \dots \a_{s-1}  \ad_1 \dots \ad_{s}} 
-\hf \Big( \bar \cD_{(\ad_1}  \cD_{(\a_1}
-2\ri (s-1) \cD_{ (\a_1 (\ad_1 } \Big)  \bar T_{\a_2\dots \a_{s-1}) \ad_2 \dots \ad_{s})} 
=0~. \label{7.3c}
\eea
\end{subequations}

Similar considerations for the transverse formulation lead to the following non-conformal supercurrent multiplet
\begin{subequations}\label{TSupercurrent}
\bea
\bar \cD^\bd {\mathbb J}_{\a_1 \dots \a_s \bd \ad_1 \dots \ad_{s-1}} 
-\frac{1}{4} (\bar \cD^2 + 2 \m (s-1)) {\mathbb F}_{\a_1 \dots \a_s \ad_1 \dots \ad_{s-1}} &=&0~,\\
\cD_{(\a_1 } {\mathbb F}_{\a_2 \dots \a_{s+1} )\ad_1 \dots \ad_{s-1}}&=&0  ~.
\eea
\end{subequations}
Thus,  the trace multiplet $\bar {\mathbb F}_{\a(s-1) \ad(s)}$ is longitudinal linear.

In the flat-superspace limit, the higher spin supercurrent multiplets  
\eqref{7.3} and \eqref{TSupercurrent}
reduce to those described in \cite{HK1}.

As in \cite{HK1}, it is useful to introduce 
auxiliary complex variables $\z^\a \in {\mathbb C}^2$ and their conjugates 
$\bar \z^\ad$. Given a tensor superfield $U_{\a(m) \ad(n)}$, we associate with it 
the following  field on ${\mathbb C}^2$ 
\bea
U_{(m,n)} (\z, \bar \z):= \z^{\a_1} \dots \z^{\a_m} \bar \z^{\ad_1} \dots \bar \z^{\ad_n}
U_{\a_1 \dots \a_m \ad_1 \dots \ad_n}~,
\label{4.100}
\eea
which is homogeneous of degree $(m,n)$ in the variables $\z^\a$ and $\bar \z^\ad$.
We introduce operators that  increase the degree 
of homogeneity in the variables $\z^\a$ and $\bar \z^\ad$, 
\begin{subequations}
\bea
{\cD}_{(1,0)} &:=& \z^\a \cD_\a~,\\
{\bar \cD}_{(0,1)} &:=& \bar \z^\ad \bar \cD_\ad~, \\
{\cD}_{(1,1)} &:=& 2\ri \z^\a \bar \z^\ad \cD_{\a\ad}
= -\big\{ {\cD}_{(1,0)} , \bar {\cD}_{(0,1)} \big\}
~.
\eea
\end{subequations}
We also introduce two operators that decrease the degree 
of homogeneity in the variables $\z^\a$ and $\bar \z^\ad$, specifically
\begin{subequations}
\bea
\cD_{(-1,0)} &:=& \cD^\a \frac{\pa}{\pa \z^\a}~,\\
\bar \cD_{(0,-1)}& :=& \bar \cD^\ad \frac{\pa}{\pa \bar \z^\ad}~ 
~.
\eea
\end{subequations}

Making use of the above notation, the transverse linear condition \eqref{7.2} and its conjugate become
\begin{subequations}
\bea
\bar \cD_{(0,-1)} T_{(s-1,s-2)} &=&0~,  \label{7.8a}\\
\cD_{(-1,0)} \bar T_{(s-2,s-1)} &=&0~.  \label{7.8b}
\eea
\end{subequations}
The conservation equations \eqref{7.3a} and \eqref{7.3c} turn into 
\begin{subequations}
\bea
\frac{1}{s}\bar \cD_{(0,-1)} J_{(s,s)} -\hf A_{(1,1)} T_{(s-1, s-2)}&=&0~, \label{7.9a}\\
\frac{1}{s}\cD_{(-1,0)} J_{(s,s)} -\hf \bar A_{(1,1)} \bar T_{(s-2, s-1)}&=&0~. \label{7.9b}
\eea
\end{subequations}
where 
\bea
A_{(1,1)} := -\cD_{(1,0)} \bar \cD_{(0,1)} +(s-1) \cD_{(1,1)} ~, \quad
\bar A_{(1,1)} := \bar \cD_{(0,1)}  \cD_{(1,0)} -(s-1) \cD_{(1,1)} ~. 
\eea
Since 
$\bar \cD_{(0,-1)} ^2 J_{(s,s)} =0$,
the conservation equation \eqref{7.9a} is consistent provided
\bea
\bar \cD_{(0,-1)}  A_{(1,1)} T_{(s-1, s-2)}=0~.
\eea
This is indeed true, as a consequence of the transverse linear condition
\eqref{7.8a}. 


\subsection{Improvement transformations}

The conservation equations \eqref{7.3} and \eqref{TSupercurrent}
define two consistent higher spin supercurrents in AdS. 
Similar to the two 
irreducible AdS supercurrents \cite{BK12}, with $(12+12)$ and $(20+20)$ degrees of 
freedom, the  higher spin supercurrents \eqref{7.3} and \eqref{TSupercurrent}
are equivalent in the sense that there always exists a well defined improvement 
transformation that converts \eqref{7.3} into \eqref{TSupercurrent}.
Such an improvement transformation is constructed below.

Since the trace multiplet $T_{\a(s-1) \ad (s-2)}$ is transverse, eq. \eqref{7.2}, 
there exists a well-defined complex tensor operator  $X_{\a(s-1) \ad (s-1)}$ such that 
\bea
T_{\a(s-1) \ad (s-2)}
= \bar \cD^\bd X_{\a(s-1) (\bd \ad_1 \dots \ad_{s-2})} ~.
\eea
Let us introduce the real $U_{\a(s-1) \ad (s-1)}$ and imaginary $V_{\a(s-1) \ad (s-1)}$
parts of $X_{\a(s-1) \ad (s-1)}$, 
\bea
X_{\a(s-1) \ad (s-1)} = U_{\a(s-1) \ad (s-1)} + \ri V_{\a(s-1) \ad (s-1)}~.
\eea
Then it may be checked that the operators
\begin{subequations}
\bea
{\mathbb J}_{\a(s) \ad (s)} &:=& J_{\a(s) \ad (s)}
+\frac{s}{2} \big[ \cD_{(\a_1}, \bar \cD_{(\ad_1} \big]
U_{\a_2 \dots \a_s) \ad_2 \dots \ad_s ) }
+ s \cD_{(\a_1 (\ad_1 } V_{\a_2 \dots \a_s) \ad_2 \dots \ad_s)}~, ~~~\\
{\mathbb F}_{\a(s) \ad(s-1)} &:=& 
\cD_{(\a_1} \Big\{ (2s+1) U_{\a_2 \dots \a_s) \ad(s-1)}
- \ri V_{\a_2 \dots \a_s) \ad(s-1)}\Big\} \eea
\end{subequations}
enjoy the conservation equation \eqref{TSupercurrent}.

In accordance with the result obtained, for all applications it suffices to work with the 
longitudinal supercurrent \eqref{7.3}. This is why in the integer superspin case, 
which will be studied in section \ref{subsection4.4}, we will introduce only 
the longitudinal  supercurrent.

There exists an improvement transformation for the supercurrent multiplet \eqref{7.3}. 
Given a chiral scalar superfield $\O$, introduce
\begin{subequations} \label{improvement4.20}
\bea
\widetilde{J}_{(s, s)} &:=& J_{(s,s)} 
+ \cD^s_{(1,1)} \big( \O +(-1)^s \bar \O\big)~, \qquad \bar \cD_\ad \O=0~,\\
\widetilde{T}_{(s-1,s-2)} &:=& T_{(s-1,s-2)} 
+ \frac{2  (-1)^s }{s(s-1)} \bar \cD_{(0,-1)} \cD^{s-1}_{(1,1)} \bar \O \non \\
&& \phantom{T_{\a(s-1) \ad (s-2)}}
+ \frac{4(s+1)}{s} \m \cD^{s-2}_{(1,1)} \cD_{(1,0)} \O~. 
\eea
\end{subequations}
The operators $\widetilde{J}_{(s,s)} $ and 
$\widetilde{T}_{(s-1,s-2)} $ prove to obey the conservation 
equation \eqref{7.3}. 


\subsection{Non-conformal supercurrents: Integer superspin} \label{subsection4.4}

We now make use of the new gauge formulation \eqref{action}, 
or  equivalently \eqref{action2}, for the integer superspin-$s$ multiplet to derive the AdS analogue of the non-conformal higher spin supercurrents constructed in \cite{HK2}.

Let us couple the prepotentials 
$H_{ \a (s-1) \ad (s-1) } $, $Z_{ \a (s-1) \ad (s-1) }$ and $\Psi_{ \a (s) \ad (s-1) } $ to external sources
\bea
S^{(s)}_{\rm source} &=& \int \rd^4x \rd^2 \q  \rd^2 \bar \q \, E\, \Big\{ 
\Psi^{ \a (s) \ad (s-1) } J_{ \a (s) \ad (s-1) }
-\bar \Psi^{ \a (s-1) \ad (s) } \bar J_{ \a (s-1) \ad (s) }
\non \\
&&+H^{ \a (s-1) \ad (s-1) } S_{ \a (s-1) \ad (s-1) } \non \\
&&+ Z^{ \a (s-1) \ad (s-1) } T_{ \a (s-1) \ad (s-1) } 
+ \bar Z^{ \a (s-1) \ad (s-1) } \bar T_{ \a (s-1) \ad (s-1) }
 \Big\}~.
\label{4.1}
\eea
In order for $S^{(s)}_{\rm source}$ to be invariant under the $\z$-transformation 
in \eqref{2.4a}, the source  $J_{ \a (s) \ad (s-1) }$ must satisfy
\bea
\bar \cD^\bd J_{\a(s) \bd \ad(s-2)} =0 \quad \Longleftrightarrow \quad
\cD^\b \bar J_{\b \a(s-2)  \ad(s)} =0 ~.
\label{4.2a}
\eea
Next, requiring $S^{(s)}_{\rm source}$ to be invariant under the transformation \eqref{2.3} leads to
\bea
\bar \cD_{(\ad_1} T_{\a(s-1) \ad_2 \dots \ad_{s})} =0
 \quad \Longleftrightarrow \quad
 \cD_{(\a_1} \bar T_{\a_2 \dots \a_{s})  \ad (s-1)} =0~.
\label{4.2b}
\eea
We see that  the superfields $J_{ \a (s) \ad (s-1) }$ and $T_{ \a (s-1) \ad (s-1) } $ are transverse linear and longitudinal linear, respectively.
Finally, requiring $S^{(s)}_{\rm source}$ to be invariant under the 
$\mathfrak V$-transformation 
\eqref{2.4} gives the following conservation equation
\begin{subequations} \label{3.4}
\bea
-\hf \cD^\b J_{\b \a(s-1) \ad(s-1)} 
+S_{\a(s-1) \ad(s-1)} + \bar T_{\a(s-1) \ad(s-1)} =0
\label{4.2c}
\eea
as well as its conjugate
\bea
\hf \bar \cD^\bd \bar J_{ \a(s-1) \bd \ad(s-1)} 
+S_{\a(s-1) \ad(s-1)} + T_{\a(s-1) \ad(s-1)} =0~. 
\label{3.4b}
\eea
\end{subequations}

As a consequence of  \eqref{4.2b}, from \eqref{4.2c} we deduce
\bea
\frac{1}{4} \cD^2 J_{ \a(s) \ad(s-1)} -\hf \bar \m(s+2)J_{\a(s) \ad(s-1)} + \cD_{(\a_1} S_{\a_2 \dots \a_s) \ad(s-1) } =0~.
\label{3.5}
\eea
The equations \eqref{4.2a} and \eqref{3.5} describe the conserved 
current supermultiplet which corresponds to our theory in the gauge \eqref{2.10}.

Taking the sum of \eqref{4.2c} and \eqref{3.4b}
leads to
\bea
\hf \cD^\b J_{\b \a(s-1) \ad(s-1)} 
+\hf \bar \cD^\bd \bar J_{\a(s-1) \bd \ad(s-1)}
+ T_{\a(s-1) \ad(s-1)}-\bar T_{\a(s-1) \ad(s-1)} =0~. 
\label{4.3}
\eea
The equations \eqref{4.2a}, \eqref{4.2b} and \eqref{4.3} describe the conserved 
current supermultiplet which corresponds to our theory in the gauge \eqref{2.12}.
As a consequence of \eqref{4.2b}, the conservation equation \eqref{4.3} 
implies
\bea
\hf \cD_{(\a_1} \left\{\cD^{|\b|} J_{\a_2 \dots \a_s ) \b\ad(s-1)} 
+ \bar \cD^\bd \bar J_{\a_2 \dots \a_s ) \bd \ad(s-1)}\right\}
+\cD_{(\a_1} T_{\a_2 \dots \a_s ) \ad(s-1)} =0~. 
\label{4.4}
\eea

Using our notation introduced in section 3, the transverse linear condition \eqref{4.2a} turns into 
\bea
\bar \cD_{(0,-1)} J_{(s,s-1)} &=& 0~,  \label{4.5a}
\eea
while the longitudinal linear condition \eqref{4.2b} takes the form
\bea
\bar \cD_{(0,1)} T_{(s-1,s-1)} &=& 0~. \label{4.5b}
\eea
The conservation equation \eqref{4.2c} becomes
\bea
-\frac{1}{2s} \cD_{(-1,0)} J_{(s,s-1)} + S_{(s-1,s-1)} + \bar T_{(s-1,s-1)} = 0
\label{4.6}
\eea
and \eqref{4.4} takes the form
\bea
\frac{1}{2s} \cD_{(1,0)} \left\{\cD_{(-1,0)} J_{(s,s-1)} + \bar \cD_{(0,-1)} \bar J_{(s-1,s)}\right\}
+\cD_{(1,0)} T_{(s-1,s-1)} =0~. 
\label{4.7}
\eea

In the flat-superspace limit, the  higher spin supercurrent multiplet
described by eqs.  \eqref{4.2a} and \eqref{3.5} 
reduces to the one proposed in \cite{HK2}.


\subsection{Improvement transformation}

There exist an improvement transformation for the supercurrent multiplet \eqref{3.4}. 
Given a chiral scalar superfield $\O$, we introduce
\begin{subequations}
\bea
\widetilde{J}_{(s,s-1)} &:=& J_{(s,s-1)} 
+ \cD^{s-1}_{(1,1)} \cD_{(1,0)} \O ~, \qquad \bar \cD_\ad \O=0~,\\
\widetilde{\bar T}_{(s-1, s-1)} &:=& \bar T_{(s-1, s-1)} 
+ \frac{s-1}{4s} \cD^{s-1}_{(1,1)} (\cD^2 -4 \bar \m) \O \non \\
&&+ (-1)^s (s-1) \Big(\bar \m + \frac{\m}{s}\Big) \cD^{s-1}_{(1,1)} \bar \O~,\\
\widetilde{S}_{(s-1,s-1)} &:=& S_{(s-1,s-1)} 
+ \m (s-1) \cD^{s-1}_{(1,1)} \O + (-1)^{s-1} \bar \m (s-1)\cD^{s-1}_{(1,1)} \bar \O \non \\
&&+ \bar \m \frac{s-1}{s} \cD^{s-1}_{(1,1)} \O + (-1)^{s-1} \m \frac{s-1}{s} \cD^{s-1}_{(1,1)} \bar \O~.
\eea
\end{subequations}
It may be checked that the operators $\widetilde{J}_{(s,s-1)} $, $\widetilde{\bar T}_{(s-1,s-1)} $ and $\widetilde{S}_{(s-1,s-1)}$ obey the conservation 
equation \eqref{4.6}, as well as \eqref{4.2b} and \eqref{4.5a}.


\section{Higher spin supercurrents for chiral superfields: Half-integer superspin}

In the remainder of this paper we will study explicit realisations
of the higher spin supercurrents introduced above in various supersymmetric 
field theories in AdS. 

\subsection{Superconformal model for a chiral superfield} \label{1chiral}


Let us consider the superconformal theory of a single chiral scalar superfield 
\bea
S = \int \rd^4x \rd^2 \q  \rd^2 \bar \q \,E\, \bar \F \F ~,
\label{chiral}
\eea
where  $\F$ is covariantly chiral, $\bar \cD_\ad \F =0$.
We can define the conformal supercurrent $J_{(s, s)}$ 
in direct analogy with the flat superspace case~\cite{KMT,  HK1}
\bea
J_{(s,s)} &=& \sum_{k=0}^s (-1)^k
\binom{s}{k}
\left\{ \binom{s}{k+1} 
{\cD}^k_{(1,1)}
 \cD_{(1,0)} \F \,\,
{\cD}^{s-k-1}_{(1,1)}
\bar \cD_{(0,1)} 
\bar \F  
\right. \non \\ 
&& \left.
 \qquad \qquad
+ \binom{s}{k} 
{\cD}^k_{(1,1)}
  \F \,\,
{\cD}^{s-k}_{(1,1)}
\bar \F \right\}~.~
\label{7.15}
\eea
Making use of
the massless equations of motion,   $(\cD^2-4\bar \m)\, \F = 0$, 
one may check that $J_{(s,s)}$ satisfies the conservation equation
\bea
\cD_{(-1,0)} J_{(s,s)} = 0 \quad \Longleftrightarrow \quad 
\bar \cD_{(0,-1)} J_{(s,s)} = 0 ~.~
\label{7.9}
\eea
The calculation of \eqref{7.9} in AdS is much more complicated than in flat superspace due to the fact that the algebra of covariant derivatives \eqref{1.2}
is nontrivial. 
Let us sketch the main steps in evaluating the left-hand side of eq.~\eqref{7.9}
with $J_{(s,s)} $ given by \eqref{7.15}.
We start with the obvious relations
\begin{subequations}
\bea
\frac{\pa}{\pa \z^\a} {\cD}_{(1,1)} &=&2\ri {\bar \z}^\ad {\cD}_{\a \ad}~, \\
\frac{\pa}{\pa \z^\a} {\cD}^k_{(1,1)} &=& 
\sum_{n=1}^k\,{\cD}^{n-1}_{(1,1)} \,\,  2\ri \, {\bar \z}^\ad {\cD}_{\a \ad}\,\, {\cD}^{k-n}_{(1,1)} ~, \qquad k>1
~.\label{eq1}
\eea
\end{subequations}
To simplify eq.~\eqref{eq1}, we may push ${\bar \z}^\ad{\cD}_{\a \ad}$, say,  to the left 
provided that we take into account its commutator with ${\cD}_{(1,1)}$:
\bea
[{\bar \z}^\ad {\cD}_{\a \ad}\,, {\cD}_{(1,1)}] = -4\ri \,\bar \m \m \,\z_\a  {\bar \z}^\ad {\bar \z}^\bd {\bar M}_{\ad \bd}~.
\label{555}
\eea
Associated with the Lorentz generators are the operators
\begin{subequations}
\bea
{\bar M}_{(0,2)} &:=& {\bar \z}^\ad {\bar \z}^\bd {\bar M}_{\ad \bd}~,\\
{M}_{(2,0)} &:=& {\z}^\a {\z}^\b {M}_{\a \b}~,
\eea
where ${\bar M}_{(0,2)}$ appears in the right-hand side of \eqref{555}.
These operators annihilate every superfield $U_{(m,n)}(\z, \bar \z) $ of the form 
\eqref{4.100},\footnote{These properties are analogous to those 
that play a fundamental role for the consistent definition of covariant projective supermultiplets
 in 5D $\cN=1$ \cite{KT-M08} and 4D $\cN=2$ \cite{KLRT-M} supergravity theories.}
\bea
{\bar M}_{(0,2)} U_{(m,n)}=0~, \qquad  M_{(2,0)} U_{(m,n)} =0~.
\eea
\end{subequations}
From the above consideration, it follows that
\begin{subequations}
\bea
[{\bar \z}^\ad {\cD}_{\a \ad}\,, {\cD}^k_{(1,1)}]\, U_{(m,n)} &=& 0 ~, \\
\Big(\frac{\pa}{\pa \z^\a} {\cD}^k_{(1,1)}\Big)U_{(m,n)} &=& 2\ri k\, {\bar \z}^\ad {\cD}_{\a \ad}\, {\cD}^{k-1}_{(1,1)}U_{(m,n)}~.
\eea
\end{subequations}
We also state some other properties which we often use throughout our calculations
\begin{subequations}
\bea
{\cD}^2_{(0,1)} &=& -2\bar \m M_{(2,0)} ~,\\
\big[ {\cD}_{(1,0)}\,, {\cD}_{(1,1)} \big] 
&=& 
\big[ \bar \cD_{(0,1)}\,, \cD_{(1,1)} \big] = 0~,\\
\big[ \cD^\a, \cD_{(1,1)} \big] &=& -2 \bar \m \, \z^\a \bar {\cD}_{(0,1)} ~,\\
\big[\cD^\a, \cD^k_{(1,1)}\big] &=& -2 \bar \m \,k \,\z^\a \cD^{k-1}_{(1,1)} \bar {\cD}_{(0,1)}~,\\
\big[\cD^\a, \bar \z^\bd \cD_{\b \bd}\big] &=& \ri \bar \m \d^\a_\b \, \bar \cD_{(0,1)}~.
\eea
\end{subequations}
The above identities suffice to prove that the supercurrent  \eqref{7.15}
does obey the conservation equation \eqref{7.9}.


\subsection{Non-superconformal model for a chiral superfield} \label{subsection5.2}

Let us now add the mass term to~\eqref{chiral} and consider the following action
\bea
S = \int \rd^4x \rd^2 \q  \rd^2 \bar \q \,E\, \bar \F \F
+\Big\{ \hf \int \rd^4x \rd^2 \q \, \cE \, m\F^2 +{\rm c.c.} \Big\}~,
\label{chiral-massive}
\eea
with $m$ a complex mass parameter. 
In the massive case $J_{(s,s)}$ satisfies a more general conservation equation~\eqref{7.9a}
for some superfield $T_{(s-1, s-2)}$.
Making use of the equations of motion
\bea
-\frac{1}{4} (\cD^2-4\bar\m) \F  +\bar m \bar \F =0, \qquad
-\frac{1}{4} (\bar \cD^2-4\m) \bar \F +m \F =0,
\eea
we obtain 
\begin{subequations}
\bea
\bar \cD_{(0,-1)} J_{(s,s)} &=& F_{(s,s-1)}~, \label{7.12a}
\eea
where we have denoted 
\bea
F_{(s,s-1)} &=& 2m(s+1) \sum_{k=0}^s (-1)^{s-1+k} \binom{s}{k} \binom{s}{k+1}
\non \\ 
&& 
\times \left\{1+(-1)^s \frac{k+1}{s-k+1}\right\} 
 {\cD}^k_{(1,1)} \F \,\,{\cD}^{s-k-1}_{(1,1)}
 \cD_{(1,0)} \F ~.
\eea
\end{subequations}

We now  look for a superfield $T_{(s-1, s-2)}$
such that (i) it obeys the transverse linear constraint \eqref{7.8a}; and 
(ii) it satisfies the equation
\bea
F_{(s,s-1)} = \frac{s}{2} A_{(1,1)} T_{(s-1, s-2)}~. 
\eea
Our analysis will be similar to the one performed in~\cite{HK1} in flat superspace. 
We consider a general ansatz
\bea
T_{(s-1, s-2)} = (-1)^s m \sum_{k=0}^{s-2} c_k 
{\cD}^k_{(1,1)} \F\,
{\cD}^{s-k-2}_{(1,1)}
 \cD_{(1,0)} \F 
 \label{T7.15}
\eea
with some coefficients $c_k$ which have to be determined. 
For $k = 1,2,...s-2$, condition (i) implies that 
the coefficients $c_k$ must satisfy
\begin{subequations}\label{7.16}
\begin{align}
kc_k = (s-k-1) c_{s-k-1}~,\label{7.16a}
\end{align}
while (ii) gives the following equation
\begin{align}
c_{s-k-1} + s c_k + (s-1) c_{k-1} &= -4(-1)^k \frac{s+1}{s} \binom{s}{k} \binom{s}{k+1} 
\non \\
& \qquad \qquad \times \left\{ 1+ (-1)^s \frac{k+1}{s-k+1} \right\} ~.\label{7.16b}
\end{align}
Condition (ii) also implies that 
\begin{align}
(s-1) c_{s-2} +c_0 &= 4(-1)^s (s+1)\left\{1+(-1)^s \frac{s}{2}\right\}~, \label{7.16c}\\
c_0 &= -\frac{4}{s}(s+1+(-1)^s)~. \label{7.16d}
\end{align}
\end{subequations}
It turns out that the equations \eqref{7.16} 
lead to a unique expression for $c_k$ given by 
\bea\label{7.17}
c_k &=& -\frac{4(s+1)(s-k-1)}{s(s-1)}
\sum_{l=0}^k \frac{(-1)^k}{s-l} \binom{s}{l} \binom {s}{l+1} \left\{ 1+(-1)^s \frac{l+1}{s-l+1} \right\}  ~,~~~~  \\
&& \qquad \qquad \qquad  k=0,1,\dots s-2~. \non 
\eea

If the parameter $s$ is odd, $s=2n+1$, with  $n=1,2,\dots$, 
one can check that the equations \eqref{7.16a}--\eqref{7.16c} are identically 
satisfied. 
However, if the parameter $s$ is even, $s=2n$, with $n=1,2,\dots$, 
there appears an inconsistency: 
 the right-hand side of \eqref{7.16c} is positive, while the left-hand side 
is negative, $(s-1) c_{s-2} + c_0 < 0$. Therefore, our solution \eqref{7.17} is only consistent for $s=2n+1, n=1,2,\dots$. 

Relations \eqref{7.15}, \eqref{T7.15}, \eqref{7.16d} and \eqref{7.17} determine the non-conformal higher spin supercurrents 
in the massive chiral model \eqref{chiral-massive}.
Unlike the conformal higher spin supercurrents \eqref{7.15},
the non-conformal ones exist only for the odd values of $s$,
$s=2n+1$, with  $n=1,2,\dots$.

In the flat-superspace limit, the above results reduce to those derived in \cite{HK1}
and in a revised version (v3, 26 Oct.) of Ref. \cite{BGK1} (which appeared a few days
before \cite{HK1}).


\subsection{Superconformal model with $N$ chiral superfields}

We now  generalise
the  superconformal model \eqref{chiral}
to the case of $N$ covariantly 
chiral scalar superfields $\F^i$, 
$i=1,\dots N$,
\bea
S = \int \rd^4x \rd^2 \q  \rd^2 \bar \q\,E \,{\bar \F}^i \F^i ~,
\qquad {\bar \cD}_\ad \F^i = 0 ~.~
\label{Nchiral}
\eea
The novel feature of the $N>1$ case is that there exist two different types of
conformal supercurrents, which are:
\bea
J^+_{(s,s)} &=& S^{ij}\sum_{k=0}^s (-1)^k
\binom{s}{k}
\left\{ \binom{s}{k+1} 
{\cD}^k_{(1,1)}
 \cD_{(1,0)} \F^i \,\,
{\cD}^{s-k-1}_{(1,1)}
\bar \cD_{(0,1)} 
\bar \F^j  
\right. \non \\ 
&& \left.
 \qquad \qquad
+ \binom{s}{k} 
{\cD}^k_{(1,1)}
  \F^i \,\,
{\cD}^{s-k}_{(1,1)}
\bar \F^j \right\}~, \qquad S^{ij}= S^{ji} 
\label{A.1}
\eea
and
\bea
J^-_{(s,s)} &=& \ri \, A^{ij}\sum_{k=0}^s (-1)^k
\binom{s}{k}
\left\{ \binom{s}{k+1} 
{\cD}^k_{(1,1)}
 \cD_{(1,0)} \F^i \,\,
{\cD}^{s-k-1}_{(1,1)}
\bar \cD_{(0,1)} 
\bar \F^j  
\right. \non \\ 
&& \left.
 \qquad \qquad
+ \binom{s}{k} 
{\cD}^k_{(1,1)}
  \F^i \,\,
{\cD}^{s-k}_{(1,1)}
\bar \F^j \right\}~, \qquad A^{ij}= -A^{ji} ~.~
\label{A.2}
\eea
Here $S$ and $A$ are arbitrary real symmetric and antisymmetric 
constant matrices, respectively. 
We have put an overall factor  $\sqrt{-1} $ in eq.~\eqref{A.2} in order 
to make $J^-_{(s,s)}$  real. 
One can show that the currents \eqref{A.1} are  \eqref{A.2} are conserved on-shell:
\bea
\cD_{(-1,0)} J^\pm_{(s,s)} = 0 \quad \Longleftrightarrow \quad 
\bar \cD_{(0,-1)} J^\pm_{(s,s)} = 0 ~.~
\label{A.3}
\eea

The above results can be recast in terms of the matrix conformal supercurrent
$J_{(s,s)} =\big(J^{ij}_{(s,s)} \big)$ with components
\bea
J^{ij}_{(s,s)} &:=& \sum_{k=0}^s (-1)^k
\binom{s}{k}
\left\{ \binom{s}{k+1} 
{\cD}^k_{(1,1)}
 \cD_{(1,0)} \F^i \,\,
{\cD}^{s-k-1}_{(1,1)}
\bar \cD_{(0,1)} 
\bar \F^j  
\right. \non \\ 
&& \left.
 \qquad \qquad
+ \binom{s}{k} 
{\cD}^k_{(1,1)}
  \F^i \,\,
{\cD}^{s-k}_{(1,1)}
\bar \F^j \right\}
\label{520}~,
\eea
which is  Hermitian, $J_{(s,s)}{}^\dagger = J_{(s,s)}$. 
The chiral action  \eqref{Nchiral}
possesses rigid ${\rm U}(N)$ symmetry acting on the chiral column-vector $\F = (\F^i$) 
by $\F \to g \F$, with $g \in {\rm U}(N)$, which implies that 
the supercurrent \eqref{520} transforms
as $J_{(s,s)} \to gJ_{(s,s)} g^{-1}$.

\subsection{Massive model with $N$ chiral superfields} 
Now let us consider a  theory of $N$ massive chiral 
multiplets with action
\bea
S = \int \rd^4x \rd^2 \q  \rd^2 \bar \q \,E\, \bar \F^i \F^i
+\Big\{\hf \int \rd^4x \rd^2 \q  \, \cE \, M^{ij} \F^i \F^j +{\rm c.c.} \Big\}~,
\label{Nchiralm}
\eea
where $M^{ij}$ is a constant symmetric $N\times N$ mass matrix.
The corresponding equations of motion are
\bea
-\frac{1}{4} (\cD^2-4\bar\m) \F^i  + {\bar M}^{ij} \bar \F^j =0~, \qquad
-\frac{1}{4} ({\bar \cD}^2-4\m) {\bar \F }^i + M^{ij} \F^j =0~.
\eea

First we will consider the case where $S$ is a real and symmetric matrix. 
Making use of the equations of motion,
we obtain 
\bea
\cD_{(-1,0)} J_{(s,s)} &=& 2(s+1) (S\bar M)^{ji} \sum_{k=0}^s (-1)^{k} \binom{s}{k} \binom{s}{k} 
\non \\ 
&& 
\times  \frac{k}{k+1}
 {\cD}^{k-1}_{(1,1)} \,{\bar \cD}_{(0,1)} {\bar \F}^i \,\,{\cD}^{s-k}_{(1,1)}
 {\bar \F}^j 
 \non \\
 &&
 +2(s+1) (S\bar M)^{ji} \sum_{k=0}^{s-1} (-1)^{k} \binom{s}{k} \binom{s}{k+1} 
\non \\ 
&& 
\times
 {\cD}^{k}_{(1,1)} {\bar \F}^i \,\,{\cD}^{s-k-1}_{(1,1)} {\bar \cD}_{(0,1)} {\bar \F}^j ~.
\label{A.4}
\eea
Now, suppose the product ${S \bar M}$ is symmetric, which implies $[S,\bar M]=0$. Then, \eqref{A.4} becomes
\bea
\cD_{(-1,0)} J_{(s,s)} &=& 2(s+1) (S\bar M)^{ij} \sum_{k=0}^{s-1} (-1)^{k} \binom{s}{k} \binom{s}{k+1} 
\non \\ 
&& 
\times \left\{1+(-1)^s \frac{k+1}{s-k+1}\right\} 
 {\cD}^{k}_{(1,1)} {\bar \F}^i \,\,{\cD}^{s-k-1}_{(1,1)} {\bar \cD}_{(0,1)} {\bar \F}^j ~.
\label{A.4a}
\eea
We now look for a superfield $\bar T_{(s-2, s-1)}$
such that (i) it obeys the transverse antilinear constraint \eqref{7.8b}; and 
(ii) it satisfies the conservation equation \eqref{7.9b}:
\bea
\cD_{(-1,0)} J_{(s,s)} = \frac{s}{2} \bar A_{(1,1)} \bar T_{(s-2, s-1)}~. 
\eea
As in the single field case we consider a general ansatz
\bea
\bar T_{(s-2, s-1)} = (S\bar M)^{ij}\sum_{k=0}^{s-2} c_k 
{\cD}^k_{(1,1)} \bar \F^i\,
{\cD}^{s-k-2}_{(1,1)}
 \bar \cD_{(0,1)} \bar \F^j ~.
\label{A.5}
\eea
Then for $k = 1,2,...s-2$, condition (i) implies that 
the coefficients $c_k$ must satisfy
\begin{subequations}\label{A.6}
\begin{align}
kc_k = (s-k-1) c_{s-k-1}~,\label{A.6a}
\end{align}
while (ii) gives the following equation
\begin{align}
c_{s-k-1} + s c_k + (s-1) c_{k-1} &= -4(-1)^k \frac{s+1}{s} \binom{s}{k} \binom{s}{k+1} 
\non \\
& \qquad \qquad \times \left\{ 1+ (-1)^s \frac{k+1}{s-k+1} \right\} ~.\label{A.6b}
\end{align}
Condition (ii) also implies that 
\begin{align}
(s-1) c_{s-2} +c_0 &= 4(-1)^s (s+1)\left\{1+(-1)^s \frac{s}{2}\right\}~, \label{A.6c}\\
c_0 &= -\frac{4}{s}(s+1+(-1)^s)~. \label{A.6d}
\end{align}
\end{subequations}
The above conditions coincide with eqs.\eqref{7.16a}--\eqref{7.16d} in the case of a single, massive chiral superfield, which are satisfied only for $s= 2n+1, n=1,2,\dots$.
Hence, the solution for the coefficients $c_k$ is given by~\eqref{7.17} for odd values of $s$ and there is no solution for even $s$.

On the other hand, if $S\bar M$ is antisymmetric (which is equivalent to $\{S,\bar M\}=0$), eq.~\eqref{A.4a} is slightly modified
\bea
\cD_{(-1,0)} J_{(s,s)} &=& 2(s+1) (S\bar M)^{ij} \sum_{k=0}^{s-1} (-1)^{k} \binom{s}{k} \binom{s}{k+1} 
\non \\ 
&& 
\times \left\{-1+(-1)^s \frac{k+1}{s-k+1}\right\} 
 {\cD}^{k}_{(1,1)} {\bar \F}^i \,\,{\cD}^{s-k-1}_{(1,1)} {\bar \cD}_{(0,1)} {\bar \F}^j ~.
\eea
Starting with a  general ansatz 
\bea
\bar T_{(s-2, s-1)} = (S\bar M)^{ij}\sum_{k=0}^{s-2} d_k 
{\cD}^k_{(1,1)} \bar \F^i\,
{\cD}^{s-k-2}_{(1,1)}
 \bar \cD_{(0,1)} \bar \F^j 
 \label{TA.7}
\eea
and imposing conditions (i) and (ii) yield the following equations for the coefficients $d_k$
\begin{subequations}\label{A.7}
\begin{align}
kd_k = -(s-k-1) d_{s-k-1}~.\label{A.7a}
\end{align}
\begin{align}
-d_{s-k-1} + s d_k + (s-1) d_{k-1} &= -4(-1)^k \frac{s+1}{s} \binom{s}{k} \binom{s}{k+1} 
\non \\
& \qquad \qquad \times \left\{ -1+ (-1)^s \frac{k+1}{s-k+1} \right\} ~.\label{A.7b}
\end{align}
\begin{align}
(s-1) d_{s-2} -d_0 &= 4(-1)^s (s+1)\left\{-1+(-1)^s \frac{s}{2}\right\}~. \label{A.7c}\\
d_0 &= \frac{4}{s}(s+1+(-1)^{s-1})~. \label{A.7d}
\end{align}
\end{subequations}
The equations \eqref{A.7} 
lead to a unique expression for $d_k$ given by 
\bea\label{A.8}
d_k &=& -\frac{4(s+1)(s-k-1)}{s(s-1)}
\sum_{l=0}^k \frac{(-1)^k}{s-l} \binom{s}{l} \binom {s}{l+1} \left\{ -1+(-1)^s \frac{l+1}{s-l+1} \right\}  ~,~~~~  \\
&& \qquad \qquad \qquad  k=0,1, \dots s-2~. \non 
\eea
If the parameter $s$ is even, $s=2n$, with  $n=1,2,\dots$, 
one can check that the equations \eqref{A.7a}--\eqref{A.7d} are identically 
satisfied. 
However, if the parameter $s$ is odd, $s=2n+1$, with $n=1,2,\dots$, 
there appears an inconsistency: 
 the right-hand side of \eqref{A.7c} is positive, while the left-hand side 
is negative, $(s-1) d_{s-2} - d_0 < 0$. Therefore, our solution \eqref{A.8} is only consistent for $s=2n, n=1,2,\dots$. 

Finally, we consider $A^{ij}=-A^{ji}$ with the corresponding $J_{(s,s)}$ given by~\eqref{A.2}. 
The analysis in this case is similar to the one presented above and we will simply state the results. 
If $s$ is odd the non-conformal higher spin supercurrents exist if  $\{A,\bar M\}=0$. The trace supercurrent $\bar T_{(s-2, s-1)}$ is given by~\eqref{A.5}
with the coefficients $c_k$ given by 
\bea\label{eee1}
c_k &=& \ri \frac{4(s+1)(s-k-1)}{s(s-1)}
\sum_{l=0}^k \frac{(-1)^k}{s-l} \binom{s}{l} \binom {s}{l+1} \left\{ 1+(-1)^s \frac{l+1}{s-l+1} \right\}  ~,~~~~  \\
&& \qquad \qquad \qquad  k=0,1,\dots s-2~. \non 
\eea
If $s$ is even the non-conformal higher spin supercurrents exist if  $[A,\bar M]=0$. The trace supercurrent $\bar T_{(s-2, s-1)}$ is given by~\eqref{TA.7}
with the coefficients $d_k$ given by
\bea\label{eee2}
d_k &=& \ri \frac{4(s+1)(s-k-1)}{s(s-1)}
\sum_{l=0}^k \frac{(-1)^k}{s-l} \binom{s}{l} \binom {s}{l+1} \left\{ -1+(-1)^s \frac{l+1}{s-l+1} \right\}  ~,~~~~  \\
&& \qquad \qquad \qquad  k=0,1, \dots s-2~. \non 
\eea

Note that the coefficients $c_k$ in~\eqref{eee1} differ from similar coefficients in~\eqref{7.17} by a factor of $- \ri$. This means that for odd $s$ we can define a more general supercurrent
\bea
J_{(s,s)} &=& H^{ij}\sum_{k=0}^s (-1)^k
\binom{s}{k}
\left\{ \binom{s}{k+1} 
{\cD}^k_{(1,1)}
 \cD_{(1,0)} \F^i \,\,
{\cD}^{s-k-1}_{(1,1)}
\bar \cD_{(0,1)} 
\bar \F^j  
\right. \non \\ 
&& \left.
 \qquad \qquad
+ \binom{s}{k} 
{\cD}^k_{(1,1)}
  \F^i \,\,
{\cD}^{s-k}_{(1,1)}
\bar \F^j \right\} ~,
\label{eee3}
\eea
where $H^{ij}$ is a generic matrix which can be split into the symmetric and antisymmetric parts $H^{ij} = S^{ij} + \ri A^{ij}$. Here both $S$ and $A$ are real 
and we put an $\ri$ in front of $A$ because $J_{(s,s)} $ must be real. From the above consideration it then follows that the corresponding more general solution for $\bar T_{(s-2, s-1)}$ reads
\bea
\bar T_{(s-2, s-1)} = (\bar H\bar M)^{ij}\sum_{k=0}^{s-2} c_k 
{\cD}^k_{(1,1)} \bar \F^i\,
{\cD}^{s-k-2}_{(1,1)}
 \bar \cD_{(0,1)} \bar \F^j ~,
\label{A.5a}
\eea
where $[S, \bar M]=0$, $\{A, \bar M\}=0$ and $c_k$ are, as before, given by eq.~\eqref{7.17}.
Similarly, the coefficients $d_k$ in~\eqref{eee2} differ from similar coefficients in~\eqref{A.8} by a factor of $- \ri$. This means that for even $s$ we can define a more general 
supercurrent~\eqref{eee3}, where $H^{ij}$ is a generic matrix which we can split as before into the symmetric and antisymmetric parts,  $H^{ij} = S^{ij} + \ri A^{ij}$. 
From the above consideration it then follows that the corresponding more general solution for $\bar T_{(s-2, s-1)}$ reads
\bea
\bar T_{(s-2, s-1)} = (\bar H\bar M)^{ij}\sum_{k=0}^{s-2} d_k 
{\cD}^k_{(1,1)} \bar \F^i\,
{\cD}^{s-k-2}_{(1,1)}
 \bar \cD_{(0,1)} \bar \F^j ~,
\label{A.5b}
\eea
where $\{S, \bar M\}=0$, $[A, \bar M]=0$ and $d_k$ are given by eq.~\eqref{A.8}.


\section{Higher spin supercurrents for chiral superfields: Integer superspin}

In this section we provide explicit realisations for the fermionic higher spin supercurrents 
(integer superspin) in models described by chiral scalar superfields. 


\subsection{Massive hypermultiplet model}

Consider a free massive hypermultiplet in AdS\footnote{This model possesses
off-shell $\cN=2$ AdS supersymmetry \cite{BKsigma,KT-M-ads}.}
\bea
S = \int \rd^4x \rd^2 \q  \rd^2 \bar \q \,E\, \Big( \bar \J_+ \J_+
+\bar \J_- \J_-\Big)
+\Big\{ {m} \int \rd^4x \rd^2 \q  \,\cE\, \J_+ \J_- +{\rm c.c.} \Big\}~,
\label{hyper1}
\eea
where the superfields $\J_\pm$ are covariantly chiral, $\bar \cD_\ad \J_\pm =0$ and $m$ is a complex mass parameter.
By a change of variables it is possible to make $m$ real. 
Let us introduce another set of fields $\F_\pm$, $\bar \cD_\ad \F_\pm =0$,  related to $\J_\pm$ by the following transformations
\bea \label{phase}
\F_\pm = e^{\ri \a/2} \J_\pm~, \qquad
m = M e^{\ri \a}~.
\eea
Under the transformations \eqref{phase}, the action \eqref{hyper1} turns into
\bea
S = \int \rd^4x \rd^2 \q  \rd^2 \bar \q \,E \, \Big( \bar \F_+ \F_+
+\bar \F_- \F_-\Big)
+\Big\{ {M} \int \rd^4x \rd^2 \q \,\cE \, \F_+ \F_- +{\rm c.c.} \Big\}~,
\label{hyper2}
\eea
where the mass parameter $M$ is now real.
In the massless case, $M=0$, 
the conserved fermionic supercurrent $J_{\a(s) \ad(s-1)}$ was constructed in~\cite{KMT} and 
is given by
\bea
J_{(s,s-1)} &=& \sum_{k=0}^{s-1} (-1)^k
\binom{s-1}{k}
\left\{ \binom{s}{k+1} 
{\cD}^k_{(1,1)}
 \cD_{(1,0)} \F_{+} \,\,
{\cD}^{s-k-1}_{(1,1)}
 \F_{-}  
\right. \non \\ 
&& \left.
 \qquad \qquad
- \binom{s}{k} 
{\cD}^k_{(1,1)}
  \F_{+} \,\,
{\cD}^{s-k-1}_{(1,1)}
\cD_{(1,0)} \F_{-} \right\}~.
\label{4.8}
\eea
Making use of
the massless equations of motion,  $-\frac{1}{4}(\cD^2-4\bar \m)\, \F_\pm = 0$, 
one may check that $J_{(s,s-1)}$ obeys, 
for $s > 1$,  the conservation equations
\bea
\cD_{(-1,0)} J_{(s,s-1)} = 0, \qquad
\bar \cD_{(0,-1)} J_{(s,s-1)} = 0 ~.~
\label{4.9}
\eea

We will now construct fermionic higher spin supercurrents 
corresponding to the massive model \eqref{hyper2}.
Making use of the massive equations of motion 
\bea
-\frac{1}{4} (\cD^2-4\bar\m) \F_+ +M \bar \F_{-} =0, \qquad
-\frac{1}{4} (\cD^2-4\bar\m)\F_- +M \bar \F_{+} =0,
\eea
we obtain 
\bea
\cD_{(-1,0)} J_{(s,s-1)} &=& 2M (s+1) \sum_{k=0}^{s-1} (-1)^{k+1} \binom{s-1}{k} \binom{s}{k} \non \\
&&\qquad \times \left\{ -\frac{s-k}{k+1} {\cD}^k_{(1,1)} \bar \F_- \,{\cD}^{s-k-1}_{(1,1)} \F_-
+ {\cD}^k_{(1,1)} \F_+ \,{\cD}^{s-k-1}_{(1,1)} \bar \F_+ \right\}\non \\
&&+ 2M(s+1) \sum_{k=1}^{s-1} (-1)^{k+1} \binom{s-1}{k} \binom{s}{k} \frac{k}{k+1} 
\non \\
&& \qquad \times {\cD}^{k-1}_{(1,1)} \bar \cD_{(0,1)} \bar \F_- \,\,{\cD}^{s-k-1}_{(1,1)} \cD_{(1,0)} \F_- 
\non \\
&&+ 2M(s+1)\sum_{k=0}^{s-2} (-1)^{k+1} \binom{s-1}{k} \binom{s}{k} \frac{s-1-k}{k+1} 
\non \\
&& \qquad \times {\cD}^{k}_{(1,1)} \cD_{(1,0)} \F_+ \,\,{\cD}^{s-k-2}_{(1,1)} \bar \cD_{(0,1)} \bar \F_+ ~.~ \label{4.10}
\eea
It can be shown that the massive supercurrent $J_{(s,s-1)}$ also obeys \eqref{4.5a}. 

We now look for a superfield $T_{(s-1,s-1)}$ such that (i) it obeys the longitudinal linear constraint \eqref{4.5b}; and 
(ii) it satisfies \eqref{4.7}, which is a consequence of the conservation equation \eqref{4.6}. 
For this we consider a general ansatz 
\bea
T_{(s-1, s-1)} &=& 
\sum_{k=0}^{s-1} c_k \,{\cD}^k_{(1,1)} \F_-\,\, {\cD}^{s-k-1}_{(1,1)} \bar \F_-  \non \\
&&+ \sum_{k=0}^{s-1} d_k \,{\cD}^k_{(1,1)} \F_+ \,\,
{\cD}^{s-k-1}_{(1,1)} \bar \F_+  \non \\
&&+ \sum_{k=1}^{s-1} f_k \,{\cD}^{k-1}_{(1,1)} \cD_{(1,0)} \F_-\,\, {\cD}^{s-k-1}_{(1,1)} \bar \cD_{(0,1)} \bar \F_-  \non \\
&&+ \sum_{k=1}^{s-1} g_k \,{\cD}^{k-1}_{(1,1)}  \cD_{(1,0)} \F_+\,\, {\cD}^{s-k-1}_{(1,1)} \bar \cD_{(0,1)} \bar \F_+ ~.
\label{T4.11}
\eea
Condition (i) implies that the coefficients must be related by
\begin{subequations} 
\bea
c_0 = d_0 = 0~, \qquad f_k = c_k~, \qquad g_k = d_k~, 
\label{qqq1}
\eea
while for $k=1,2, \dots s-2$, condition  (ii) gives the following recurrence relations:
\bea 
c_k + c_{k+1} &=& 
\frac{M(s+1)}{s} (-1)^{s+k} \binom{s-1}{k} \binom{s}{k} 
\non \\
&& \times \frac{1}{(k+2)(k+1)} \Big\{(2k+2-s)(s+1)-k-2\Big\}~, \\
d_k + d_{k+1} &=& \frac{M(s+1)}{s} (-1)^{k} \binom{s-1}{k} \binom{s}{k} 
\non \\
&& \times \frac{1}{(k+2)(k+1)} \Big\{(2k+2-s)(s+1)-k-2\Big\}~.
\eea
Condition (ii) also implies that
\bea
c_1 = -(-1)^s \frac{M(s^2-1)}{2}~, \qquad c_{s-1} &=&- \frac{M(s^2-1)}{s}~;\\
d_1 =- \frac{M(s^2-1)}{2}~, \qquad d_{s-1} &=& -(-1)^s \frac{M(s^2-1)}{s}~.
\eea
\end{subequations}
The above conditions lead to simple expressions for $c_k$ and $d_k$:
\begin{subequations}
\bea
d_k &=& \frac{M(s+1)}{s} \frac{k}{k+1} (-1)^{k} \binom{s-1}{k} \binom{s}{k}~,\\
c_k &=& (-1)^s d_k ~,
\eea
\label{qqq2}
\end{subequations}
where $ k=1,2,\dots s-1$.


\subsection{Superconformal model with $N$ chiral superfields}


In this subsection we will generalise the above results for $N$ chiral superfields $\Phi^i$, $i=1, \dots N$. We first consider the superconformal model \eqref{Nchiral}.
Let us construct the following fermionic supercurrent
\bea
J_{(s,s-1)} &=& C^{ij}\sum_{k=0}^{s-1} (-1)^k
\binom{s-1}{k}
\left\{ \binom{s}{k+1} 
{\cD}^k_{(1,1)}
 \cD_{(1,0)} \F^i \,\,
{\cD}^{s-k-1}_{(1,1)} \F^j  
\right. \non \\ 
&& \left.
 \qquad \qquad
- \binom{s}{k} 
{\cD}^k_{(1,1)}
  \F^i \,\,
{\cD}^{s-k-1}_{(1,1)} \cD_{(1,0)} \F^j \right\}~, 
\label{B.1}
\eea
where $C^{ij}$ is a constant complex matrix. By changing the summation index it is not hard to show that $J_{(s,s-1)} =0$ if  (i) $s$ is odd and 
$C^{ij}$ is symmetric; and 
 (ii) 
 $s$ is even and $C^{ij}$ is antisymmetric, that is
 \begin{subequations}
 \bea
 C^{ij} =C^{ji} ~, \quad s=1,3,\dots  \quad & \Longrightarrow \quad J_{(s,s-1)}=0~; \\
 C^{ij} =-C^{ji} ~, \quad s=2,4,\dots  \quad & \Longrightarrow \quad J_{(s,s-1)}=0~.
 \eea
 \end{subequations}
This means that we have to consider the two separate cases: the case of even $s$ with symmetric $C$, and the case of odd $s$ with antisymmetric $C$. 
Using the massless equation of motion,  $-\frac{1}{4}(\cD^2-4\bar \m)\, \F^i = 0$, one may check that $J_{(s,s-1)}$ satisfies the conservation equations \eqref{4.9}
\bea
\cD_{(-1,0)} J_{(s,s-1)} = 0~, \qquad
\bar \cD_{(0,-1)} J_{(s,s-1)} = 0~.
\label{B.2}
\eea

In the case of a single chiral superfield, the supercurrent \eqref{B.1} exists 
for even $s$, 
\bea
J_{(s,s-1)} &=& \sum_{k=0}^{s-1} (-1)^k
\binom{s-1}{k}
\left\{ \binom{s}{k+1} 
{\cD}^k_{(1,1)}
 \cD_{(1,0)} \F \,
{\cD}^{s-k-1}_{(1,1)} \F
\right. \non \\ 
&& \left.
 \qquad \qquad
- \binom{s}{k} 
{\cD}^k_{(1,1)}
  \F \,
{\cD}^{s-k-1}_{(1,1)} \cD_{(1,0)} \F \right\}~, \qquad  s=2,4,\dots  
\label{6.14}
\eea
The flat-superspace version of \eqref{6.14} can be extracted from the results 
 of \cite{HK2,KMT}.


\subsection{Massive model with $N$ chiral superfields}

Now we move to the massive model \eqref{Nchiralm}. As was discussed in previous subsection, to construct the conserved currents we first have to 
calculate ${\cD}_{(-1,0)} J_{(s,s-1)} $ using the equations of motion in the massive theory. The calculation depends on whether $C^{ij}$ is symmetric or antisymmetric. 

\subsubsection{Symmetric $C$}

If $C^{ij}$ is a symmetric matrix, using the massive equation of motion, we obtain
\bea
{\cD}_{(-1,0)} J_{(s,s-1)} &=& -2(s+1)(C\bar M)^{ji}\sum_{k=0}^{s-1} (-1)^{k+1} \binom{s-1}{k} \binom{s}{k} \frac{s-k}{k+1} \non \\
&&\qquad \times {\cD}^k_{(1,1)} \bar \F^i \,{\cD}^{s-k-1}_{(1,1)} \F^j \non\\
&&+ 2(s+1)(C\bar M)^{ij}\sum_{k=0}^{s-1} (-1)^{k+1} \binom{s-1}{k} \binom{s}{k} \non \\
&& \qquad \times {\cD}^k_{(1,1)} \F^i \,{\cD}^{s-k-1}_{(1,1)} \bar \F^j
\non \\
&&+ 2(s+1) (C\bar M)^{ji} \sum_{k=1}^{s-1} (-1)^{k+1} \binom{s-1}{k} \binom{s}{k} \frac{k}{k+1}
\non \\
&& \qquad \times {\cD}^{k-1}_{(1,1)} \bar \cD_{(0,1)} \bar \F^i \,\,{\cD}^{s-k-1}_{(1,1)} \cD_{(1,0)} \F^j
\non \\
&&+ 2(s+1) (C\bar M)^{ij}\sum_{k=0}^{s-2} (-1)^{k+1} \binom{s-1}{k} \binom{s}{k} \frac{s-1-k}{k+1} 
\non \\
&& \qquad \times {\cD}^{k}_{(1,1)} \cD_{(1,0)} \F^i \,\,{\cD}^{s-k-2}_{(1,1)} \bar \cD_{(0,1)} \bar \F^j  ~.~ 
\label{B.3}
\eea

Here we have two cases to consider:
\begin{enumerate}
\item $C \bar M$ is symmetric $\Longleftrightarrow [C,\bar M]=0, \,\, s$ even.
\item $C \bar M$ is antisymmetric $\Longleftrightarrow \{C,\bar M\}=0, \,\,s$ even.
\end{enumerate}
\textbf{Case 1:} Eq. \eqref{B.3} can be simplified to yield
\bea
{\cD}_{(-1,0)} J_{(s,s-1)} &=& 4(s+1)(C\bar M)^{ij}\sum_{k=0}^{s-1} (-1)^{k+1} \binom{s-1}{k} \binom{s}{k} \non \\
&&\qquad \times {\cD}^k_{(1,1)} \F^i \,{\cD}^{s-k-1}_{(1,1)} \bar \F^j \non\\
&&+ 4(s+1) (C\bar M)^{ij} \sum_{k=1}^{s-1} (-1)^{k+1} \binom{s-1}{k} \binom{s}{k} \frac{k}{k+1}
\non \\
&& \qquad \times {\cD}^{k-1}_{(1,1)} \bar \cD_{(0,1)}\bar \F^i \,\,{\cD}^{s-k-1}_{(1,1)} \cD_{(1,0)} \F^j ~.~ 
\label{B.4}
\eea
We now look for a superfield $T_{(s-1,s-1)}$ such that (i) it obeys the longitudinal linear constraint \eqref{4.5b}; and (ii) it satisfies \eqref{4.7}, 
which is a consequence of the conservation equation \eqref{4.6}.
The precise form of eq.~\eqref{4.7} in the present case is
\bea
&&\frac{1}{2s} \cD_{(1,0)} \left\{\cD_{(-1,0)} J_{(s,s-1)} + \bar \cD_{(0,-1)} \bar J_{(s-1,s)}\right\}\non\\
&&= \frac{2}{s+1} {\cD}_{(1,0)} \sum_{k=0}^{s-1}(-1)^{k+1} \binom{s-1}{k} \binom{s}{k} \non \\
&&\qquad \times \left\{\frac{s}{k+1} (C \bar M)^{ij}-\frac{(s+1)(s-k)}{(k+1)(k+2)} (\bar C M)^{ij} \right\} \non\\
&& \qquad \times {\cD}^k_{(1,1)} \F^i \,{\cD}^{s-k-1}_{(1,1)} \bar \F^j \non \\
&& =-\cD_{(1,0)} T_{(s-1,s-1)} ~. \label{B.4a}
\eea
To find $T_{(s-1,s-1)}$ we  consider a general ansatz 
\bea
T_{(s-1, s-1)} &=& 
\sum_{k=0}^{s-1} (c_k)^{ij} \,{\cD}^k_{(1,1)} \F^i\,\, {\cD}^{s-k-1}_{(1,1)} \bar \F^j  \non \\
&&+ \sum_{k=1}^{s-1} (d_k)^{ij} \,{\cD}^{k-1}_{(1,1)} \cD_{(1,0)} \F^j\,\, {\cD}^{s-k-1}_{(1,1)} \bar \cD_{(0,1)} \bar \F^j ~.
\label{B.5}
\eea
It is possible to show that no solution for $T_{(s-1, s-1)} $ can be found unless we impose\footnote{Since $C$ and ${\bar M}$ commute 
we can take them both to be diagonal, $C= {\rm diag} (c_1, \dots, c_N)$, $M= {\rm diag} (m_1, \dots, m_N)$. 
Then the condition~\eqref{e1} means that ${\rm arg} (c_i) - {\rm arg} (m_i) =n_i \pi$ for some integers $n_i$.} 
\be 
C \bar M = \bar C M~.
\label{e1}
\ee
Furthermore,  condition (i) implies that the coefficients must be related by
\begin{subequations} 
\bea
(c_0)^{ij} = 0~, \qquad (c_k)^{ij} = (d_k)^{ij}~,
\eea
while for $k=1,2, \dots s-2$, while condition (ii) and eq.~\eqref{e1} gives the following recurrence relations 
\bea 
(d_k)^{ij} + (d_{k+1})^{ij} &=& -2\frac{(s+1)}{s} (C\bar M)^{ij} (-1)^{k+1} \binom{s-1}{k} \binom{s}{k} 
\non \\
&& \times \frac{1}{k+1} \Big\{s-\frac{(s+1)(s-k)}{k+2}\Big\}~.
\eea
Condition (ii) also implies that
\bea
(d_1)^{ij} = (1-s^2) (C \bar M)^{ij}~, \qquad (d_{s-1})^{ij} &=& \frac{2(1-s^2)}{s} (C \bar M)^{ij}~.
\eea
\label{ee1}
\end{subequations}
The above conditions lead to simple expressions for $d_k$:
\bea
(d_k)^{ij} &=& \frac{2(s+1)}{s} (C \bar M)^{ij} \frac{k}{k+1} (-1)^{k} \binom{s-1}{k} \binom{s}{k}~,
\eea
where $ k=1,2,\dots s-1$ and $s$ is even.\\
\textbf{Case 2:}
If we take $C \bar M$ to be antisymmetric, a similar analysis shows that no solution for $T_{(s-1, s-1)}$ exists for even values of $s$. 

\subsubsection{Antisymmetric $C$}
If $C^{ij}$ is antisymmetric we get: 
\bea 
{\cD}_{(-1,0)} J_{(s,s-1)} &=& 2(s+1)(C\bar M)^{ji}\sum_{k=0}^{s-1} (-1)^{k+1} \binom{s-1}{k} \binom{s}{k} \frac{s-k}{k+1} \non \\
&&\qquad \times {\cD}^k_{(1,1)} \bar \F^i \,{\cD}^{s-k-1}_{(1,1)} \F^j \non\\
&&+ 2(s+1)(C\bar M)^{ij}\sum_{k=0}^{s-1} (-1)^{k+1} \binom{s-1}{k} \binom{s}{k} \non \\
&& \qquad \times {\cD}^k_{(1,1)} \F^i \,{\cD}^{s-k-1}_{(1,1)} \bar \F^j
\non \\
&&- 2(s+1) (C\bar M)^{ji} \sum_{k=1}^{s-1} (-1)^{k+1} \binom{s-1}{k} \binom{s}{k} \frac{k}{k+1}
\non \\
&& \qquad \times {\cD}^{k-1}_{(1,1)} \bar \cD_{(0,1)} \bar \F^i \,\,{\cD}^{s-k-1}_{(1,1)} \cD_{(1,0)} \F^j
\non \\
&&+ 2(s+1) (C\bar M)^{ij}\sum_{k=0}^{s-2} (-1)^{k+1} \binom{s-1}{k} \binom{s}{k} \frac{s-1-k}{k+1} 
\non \\
&& \qquad \times {\cD}^{k}_{(1,1)} \cD_{(1,0)} \F^i \,\,{\cD}^{s-k-2}_{(1,1)} \bar \cD_{(0,1)} \bar \F^j  ~.~ 
\label{B.3.new}
\eea
As in the symmetric $C$ case, there are also two cases to consider:
\begin{enumerate}
\item $C \bar M$ is symmetric $\Longleftrightarrow \{C,\bar M\}=0, \,\,s$ odd.
\item $C \bar M$ is antisymmetric $\Longleftrightarrow [C,\bar M]=0, \,\,s$ odd.
\end{enumerate}
\textbf{Case 1:}
Using eq.~\eqref{B.3.new} and keeping in mind that $s$ is odd, we obtain
\bea
{\cD}_{(-1,0)} J_{(s,s-1)} &=& 4(s+1)(C\bar M)^{ij}\sum_{k=0}^{s-1} (-1)^{k+1} \binom{s-1}{k} \binom{s}{k} \non \\
&&\qquad \times {\cD}^k_{(1,1)} \F^i \,{\cD}^{s-k-1}_{(1,1)} \bar \F^j \non\\
&&- 4(s+1) (C\bar M)^{ij} \sum_{k=1}^{s-1} (-1)^{k+1} \binom{s-1}{k} \binom{s}{k} \frac{k}{k+1}
\non \\
&& \qquad \times {\cD}^{k-1}_{(1,1)} \bar \cD_{(0,1)}\bar \F^i \,\,{\cD}^{s-k-1}_{(1,1)} \cD_{(1,0)} \F^j ~.~ 
\label{B.6}
\eea
Then it follows that eq.~\eqref{4.7} becomes
\bea
&&\frac{1}{2s} \cD_{(1,0)} \left\{\cD_{(-1,0)} J_{(s,s-1)} + \bar \cD_{(0,-1)} \bar J_{(s-1,s)}\right\}\non\\
&&= \frac{2}{s+1} {\cD}_{(1,0)} \sum_{k=0}^{s-1}(-1)^{k+1} \binom{s-1}{k} \binom{s}{k} \non \\
&&\qquad \times \left\{\frac{s}{k+1} (C \bar M)^{ij}-\frac{(s+1)(s-k)}{(k+1)(k+2)} (\bar C M)^{ij} \right\} \non\\
&& \qquad \times {\cD}^k_{(1,1)} \F^i \,{\cD}^{s-k-1}_{(1,1)} \bar \F^j \non \\
&& =-\cD_{(1,0)} T_{(s-1,s-1)} ~. \label{B.6a}
\eea
Note that it is the equation same as eq.~\eqref{B.4a} which means that the solution for $T_{(s-1,s-1)}$ is the same as in \textbf{Case 1}. That is, 
the matrices $C$ and $M$ must satisfy $C \bar M = \bar C M$, $T_{(s-1,s-1)}$ is given by eq.~\eqref{B.5} and the coefficients 
$(c_k)^{ij}, (d_k)^{ij}$ are given by eqs.~\eqref{ee1}.
\textbf{Case 2:}
If we take $C \bar M$ to be antisymmetric, a similar analysis shows that no solution for $T_{(s-1, s-1)}$ exists for odd values of $s$. 

\subsubsection{Massive hypermultiplet model revisited}

As a consistency check of our general method, let us reconsider the case of a hypermultiplet studied previously. 
For this we will take $N=2$, the mass matrix in the form 
\be
M=
\begin{pmatrix}
0 & m \\
m \,& 0 
\end{pmatrix} ~,
\label{ee2}
\ee
and denote $\Phi^i = (\Phi_+, \Phi_-)$. If $s$ is even we will take $C$ in the form
\be
C=
\begin{pmatrix}
0 & c \\
c \,& 0 
\end{pmatrix} ~.
\label{ee3}
\ee
Note that $C$ commutes with $M$. The condition $C \bar M= \bar C M$ is equivalent to ${\rm arg} (c)= {\rm arg} (m) +n \pi$. For simplicity, let us choose both $c$ and $m$ 
to be real. Under these conditions eq.~\eqref{B.1} for $J_{(s, s-1)}$ becomes
\bea
J_{(s,s-1)} &=& c\sum_{k=0}^{s-1} (-1)^k
\binom{s-1}{k} \binom{s}{k+1} 
\left\{{\cD}^k_{(1,1)}
 \cD_{(1,0)} \F_+ \,\,
{\cD}^{s-k-1}_{(1,1)} \F_- \right.
\non \\
&& \left. \qquad +  {\cD}^k_{(1,1)} \cD_{(1,0)} \F_- \,\,
{\cD}^{s-k-1}_{(1,1)} \F_+ \right\} 
\non \\ 
&&+ c\sum_{k=0}^{s-1} (-1)^{k+1}
\binom{s-1}{k} \binom{s}{k} 
\left\{\cD^k_{(1,1)} \F_+ \,\,
{\cD}^{s-k-1}_{(1,1)} {\cD}_{(1,0)} \F_- \right.
\non \\
&& \left. \qquad  +\cD^k_{(1,1)} \F_- \,\,
{\cD}^{s-k-1}_{(1,1)} {\cD}_{(1,0)} \F_+ \right\}~.
\label{ee4}
\eea
Introducing a new summation variable $k'= s-1-k$ for the second and fourth terms, we obtain
\bea
J_{(s,s-1)} &=& c\sum_{k=0}^{s-1} (-1)^k
\binom{s-1}{k} \binom{s}{k+1} 
\Big[(1+(-1)^s\Big]
{\cD}^k_{(1,1)}
 \cD_{(1,0)} \F_+ \,\,
{\cD}^{s-k-1}_{(1,1)} \F_- \non\\
&&- c\sum_{k=0}^{s-1} (-1)^{k}
\binom{s-1}{k} \binom{s}{k} 
\Big[(1+(-1)^s\Big]
{\cD}^{k}_{(1,1)} \F_+ {\cD}^{s-k-1}_{(1,1)} {\cD}_{(1,0)} \F_- ~.
\label{J.even}
\eea
We see that for even $s$ it coincides with the hypermultiplet supercurrent given by~\eqref{4.8} up to an overall coefficient $2c$. 
If $s$ is odd we have to choose $C$ to be antisymmetric 
\be
C=
\begin{pmatrix}
0 & c \\
-c \,& 0 
\end{pmatrix} ~.
\label{ee5}
\ee
Note that $C$ now anticommutes with $M$. For simplicity, we again choose $c$ and $m$ to be real. Now the expression~\eqref{B.1} for $J_{(s, s-1)}$ becomes
\bea
J_{(s,s-1)} &=& c\sum_{k=0}^{s-1} (-1)^k
\binom{s-1}{k} \binom{s}{k+1} 
\Big[(1- (-1)^s\Big]
{\cD}^k_{(1,1)}
 \cD_{(1,0)} \F_+ \,\,
{\cD}^{s-k-1}_{(1,1)} \F_- \non\\
&&- c\sum_{k=0}^{s-1} (-1)^{k}
\binom{s-1}{k} \binom{s}{k} 
\Big[(1- (-1)^s\Big]
{\cD}^{k}_{(1,1)} \F_+ {\cD}^{s-k-1}_{(1,1)} {\cD}_{(1,0)} \F_- ~.
\label{J.odd}
\eea
We see that for odd $s$ it coincides with the hypermultiplet supercurrent given by~\eqref{4.8} up to an overall coefficient $2c$. 
To summarise, we reproduced the hypermultiplet supercurrent~\eqref{4.8} for both even and odd values of $s$. However, for even $s$ it came 
from a symmetric matrix~\eqref{ee3} and for odd $s$ it came from an antisymmetric matrix~\eqref{ee5}.

Let us now consider $T_{(s-1, s-1)}$. First, we will note that the product $C \bar M$ is given by 
\be
C \bar M= c m 
\begin{pmatrix}
1 & 0 \\
0\,& (-1)^s 
\end{pmatrix} ~.
\label{ee6}
\ee
This means that $T_{(s-1, s-1)}$ is given by the following expression valid for all values of  $s$
\be
T_{(s-1, s-1)} = 
\sum_{k=0}^{s-1} (d_k)^{ij} \, \Big[ {\cD}^k_{(1,1)} \F^i\,\, {\cD}^{s-k-1}_{(1,1)} \bar \F^j  
+  {\cD}^{k-1}_{(1,1)} \cD_{(1,0)} \F^j\,\, {\cD}^{s-k-1}_{(1,1)} \bar \cD_{(0,1)} \bar \F^j  \Big]~, 
\label{ee7}
\ee
where the matrix $(d_k)^{ij}$ is given by 
\be
(d_k)^{ij} = 2 c m \frac{s+1}{s} \frac{k}{k+1} (-1)^k \binom{s-1}{k} \binom{s}{k} 
\begin{pmatrix}
1 & 0 \\
0\,& (-1)^s 
\end{pmatrix} ~.
\label{ee8}
\ee
It is easy to see that this expression for $T_{(s-1, s-1)}$ coincides with the one obtained for the hypermultiplet in the previous subsections 
in eqs.~\eqref{T4.11}, \eqref{qqq1}, \eqref{qqq2} up to an overall factor $2 c$.



\section{Summary and applications}

In this paper, we have proposed higher spin conserved supercurrents 
for ${\cal N}=1$ supersymmetric theories in four-dimensional anti-de Sitter space. 
We have explicitly constructed such supercurrents 
 in the case of $N$ chiral scalar superfields with an arbitrary mass matrix $M$.  
The structure of the supercurrents depends on whether the superspin is integer or half-integer, as well as on the value of the superspin, and the mass matrix. 
Let us summarise our results. 

In the case of half-integer superspin $s+ 1/2$, the supercurrent has the structure $J_{(s,s)}= H^{ij} J^{ij}_{(s, s)}$, 
where $i, j =1, \dots N$ and $H^{ij}$ is a Hermitian matrix. The precise form of  $ J^{ij}_{(s, s)}$ was discussed in Section 5. In massless theory it is conserved for all values of $s$. 
In massive theory, the conservation 
equation involves an additional complex multiplet $T_{(s-1, s-2)}$ whose existence depends on the value of $s$ and the mass matrix. 
For odd values of $s$, it exists provided $[S, \bar M]=0$, $\{A, \bar M\}=0$, where $S$ and $A$  are the symmetric and antisymmetric parts of $H$, respectively. When $s$ is even, it exists provided $\{S, \bar M\}=0$, $[A, \bar M]=0$.

In the case of integer superspin $s$, the fermionic supercurrent was discussed in Section 6. It has the form $J_{(s, s-1)}= C^{ij} J^{ij}_{(s, s-1)}$. In massless 
theory it exists for even values of $s$ if $C$ is symmetric and for odd values of $s$ if $C$ is antisymmetric. In massive theory the conservation equation involves 
an additional complex multiplet $T_{(s-1 , s-1)}$ and a real multiplet $S_{(s-1, s-1)}$. Their existence also depends on the value of $s$. 
For $s$ even they exist provided $C \bar M = \bar C M$, $[C, \bar M]=0$ and for $s$ odd provided $C \bar M = \bar C M$, $\{ C, \bar M\}=0$.

In the rest of this section, we will discuss several applications of the results obtained in the paper.

\subsection{Higher spin supercurrents for a massive chiral 
multiplet: Integer superspin} \label{subsection7.1}

Let us return to the model \eqref{chiral-massive} describing the dynamics of a single 
massive chiral multiplet in AdS. It proves to possess conserved fermionic higher spin supercurrents.  For even integer superspin, $s=2,4, \dots$,
the supercurrent $J_{(s,s-1)}$ is given by \eqref{6.14}. The corresponding 
trace multiplet is 
\bea
T_{(s-1,s-1)} &=& \sum_{k=0}^{s-1} c_k {\cD}^k_{(1,1)} \F \,\, {\cD}^{s-k-1}_{(1,1)} \bar \F \non \\
&&  + \sum_{k=1}^{s-1} d_k {\cD}^{k-1}_{(1,1)} {\cD}_{(1,0)} \F \,\, \cD^{s-k-1}_{(1,1)}  \bar {\cD}_{(0,1)} \bar \F ~.
\eea
where the coefficients $c_k$ and $d_k$ are given by \eqref{qqq2}. 
As an example, for $s=2$ we obtain
\begin{subequations}
\bea
J_{(2,1)} &=& 4 {\cD}_{(1,1)} \F \,\, {\cD}_{(1,0)} \F -2 \F \,\,{\cD}_{(1,1)} {\cD}_{(1,0)} \F ~,\\
T_{(1,1)} &=& -3 \bar m \Big( \bar \F {\cD}_{(1,1)} \F + {\cD}_{(1,0)} \F \,\, \bar \cD_{(0,1)} \bar \F \Big)~.
\eea
\end{subequations}

It was claimed in \cite{BGK1} that the chiral model in Minkowski superspace
\bea
S_{\rm massive} = \int \rd^4x \rd^2 \q  \rd^2 \bar \q \, \bar \F \F
+\Big\{ \frac{m}{2} \int \rd^4x \rd^2 \q\, \F^2 +{\rm c.c.} \Big\}~, \qquad 
\bar D_\ad \F=0
\eea
does not possess any conserved fermionic supercurrents $J_{(s,s-1)}$, 
for any value of the mass parameter $m$.
Here we have demonstrated that they, in fact,  do exist when $s$ is even.

There is a simple explanation for why the conserved fermionic supercurrents 
were overlooked in the analysis of  \cite{BGK1}. 
The point is that the authors of  \cite{BGK1}
considered only a particular ansatz for the Noether procedure to construct 
cubic vertices, $\d_g \F = \cA \F$, 
where  $\cA$ is a higher-derivative operator containing 
infinitely many local parameters. However, 
in order to generate  the conserved fermionic supercurrents we constructed, 
it is necessary to deal with a more general ansatz
$\d_g \F = \cA \F + \bar D^2 \cB\bar \F$, with $\cB$ another higher-derivative 
operator.\footnote{We thank Konstantinos Koutrolikos for clarifying comments.}


\subsection{Higher spin supercurrents for a 
tensor multiplet}

Let us consider a special case of the non-superconformal chiral model 
\eqref{chiral-massive} with the mass parameter $m=\m$, 
\bea
S [\F, \bar \F]= \hf \int \rd^4x \rd^2 \q  \rd^2 \bar \q \,E\, ( \F +\bar \F)^2~,
\qquad \bar \cD_\ad \F=0~.
\label{7.111}
\eea
This theory is known to be dual to a tensor multiplet model \cite{Siegel79}
\bea
S [L]=- \hf \int \rd^4x \rd^2 \q  \rd^2 \bar \q \,E\, L^2~,
\label{tensor}
\eea
which is realised in terms of a real linear superfield $L=\bar L$, 
constrained by $(\bar \cD^2 -4\m) L =0$,
which is the gauge-invariant field strength
of a chiral spinor superfield 
\bea
L= \cD^\a \eta_\a +\bar \cD_\ad \bar \eta^\ad~, \qquad \bar \cD_\bd \eta_\a =0~.
\eea
We recall that the duality between \eqref{7.111} and \eqref{tensor} follows, e.g., 
from the fact the off-shell constraint
\begin{subequations}
\bea
 (\bar \cD^2 -4\m) \cD_\a (\F +\bar \F)=0
 \eea
 and the equation of motion for $\F$ 
 \bea
  (\bar \cD^2 -4\m) (\F +\bar \F) =0
  \eea
  \end{subequations}
are equivalent to the equation of motion for $\eta_\a$ 
\begin{subequations}
\bea
 (\bar \cD^2 -4\m) \cD_\a L=0
 \eea
 and the off-shell constraint
 \bea
  (\bar \cD^2 -4\m) L =0~,
  \eea
  \end{subequations}
respectively. 

Higher spin supercurrents for the tensor model \eqref{tensor} 
can be obtained from the results derived in section \ref{subsection5.2} in conjunction
with an improvement transformation of the type \eqref{improvement4.20}
with  $\O = -\hf \F^2$.
Given an odd $s=3, 5 \dots$, for the supercurrent we get
\bea
{J}_{(s,s)} &=& -L\,\, {\cD}^{s-1}_{(1,1)} \, [{\cD}_{(1,0)},{\bar \cD}_{(0,1)}]L
\non \\
&&+ \sum_{k=0}^{s-1} (-1)^k
\binom{s}{k}
\binom{s}{k+1} 
{\cD}^k_{(1,1)}
 \cD_{(1,0)} L \,\,
{\cD}^{s-k-1}_{(1,1)}
\bar \cD_{(0,1)} L  
\non \\ 
&&+ \hf \sum_{k=1}^{s-1} \left\{-1+ (-1)^k
\binom{s}{k}\right\}
\binom{s}{k} 
{\cD}^{k-1}_{(1,1)}\, [\cD_{(1,0)}, {\bar \cD}_{(0,1)}] L \,\,\,
{\cD}^{s-k}_{(1,1)} L ~.
\label{currentL}
\eea
The corresponding trace multiplet proves to be
\bea
{T}_{(s-1, s-2)} &=& -\frac{4\m}{s} L \,\, {\cD}^{s-2}_{(1,1)} {\cD}_{(1,0)}L 
+ 4\m \frac{s+1}{s} {\cD}_{(1,0)}L \,\, {\cD}^{s-3}_{(1,1)} {\bar \cD}_{(0,1)} {\cD}_{(1,0)}L \non \\
&&-\frac{2}{s} {\cD}^{s-2}_{(1,1)} \left\{{\cD}_{(1,0)} {\bar \cD}_\ad L \,\,\, {\bar \cD}^\ad L \right\} \non \\
&&+ \m \sum_{k=1}^{s-2} c_k {\cD}^{k-1}_{(1,1)} {\bar \cD}_{(0,1)} {\cD}_{(1,0)}L \,\, {\cD}^{s-k-2}_{(1,1)} {\cD}_{(1,0)}L \non \\
&&+ \frac{4\m}{s} \sum_{k=1}^{s-2} \binom{s-2}{k} {\cD}^{k-1}_{(1,1)} {\cD}_{(1,0)} {\bar \cD}_{(0,1)}L \,\, {\cD}^{s-k-2}_{(1,1)} {\cD}_{(1,0)}L \non \\
&&+ 2\m \frac{s+1}{s}\sum_{k=1}^{s-3} \binom{s-2}{k} \left\{ {\cD}^{k}_{(1,1)} {\cD}_{(1,0)}L \,\, {\cD}^{s-k-3}_{(1,1)} {\bar \cD}_{(0,1)} {\cD}_{(1,0)}L \right. \non \\
&& \left. \qquad \qquad + {\cD}^{k-1}_{(1,1)} {\bar \cD}_{(0,1)} {\cD}_{(1,0)}L \,\, {\cD}^{s-k-2}_{(1,1)} {\cD}_{(1,0)}L \right\} ~. 
\label{traceL}
\eea
The coefficient $c_k$ is given by eq.~\eqref{7.17}, $s$ is odd. 
The Ferrara-Zumino supercurrent ($s=1$) for the model \eqref{tensor}
in an arbitrary supergravity background
 was derived in section 6.3 of \cite{BK}. Modulo normalisation, 
 the AdS supercurrent is 
 \begin{subequations}
 \bea
 J_{\a\ad} = \bar \cD_\ad L \cD_\a L+ L \big[\cD_\a, \bar \cD_\ad \big] L~,
 \eea
 and the corresponding trace multiplet is 
 \bea
 T=\frac 14 (\bar \cD^2 -4\m) L^2~.
 \eea
 \end{subequations}
 The supercurrent obeys the conservation equation  \eqref{FZsupercurrent}.


\subsection{Higher spin supercurrents for a complex linear multiplet}

The superconformal non-minimal scalar multiplet in AdS is described by the action 
\bea
S [\G, \bar \G]= - \int \rd^4x \rd^2 \q  \rd^2 \bar \q \,E\, \bar \G  \G ~,
\label{7.88}
\eea
where $\G$ is a complex linear scalar, $(\bar \cD^2 -4\m) \G =0$. 
This is a dual formulation for the superconformal chiral model \eqref{chiral}.
As is well known, the duality between \eqref{chiral} and \eqref{7.88} 
follows from the fact that the off-shell constraint 
\begin{subequations}
\bea
(\cD^2 -4 \bar \m) \bar \G =0
\eea
and the equation of motion for $\G$ 
\bea
\bar \cD_\ad \bar \G=0
\eea
\end{subequations}
are equivalent to the equation of motion for $\bar \F$, $(\cD^2 -4 \bar \m) \F=0$, 
and the off-shell constraint $\bar \cD_\ad \F=0$, respectively.
In other words, on the mass shell we can identify $\bar \G$ with $\F$.  

The higher spin supercurrents, $J_{(s,s)}$ and $J_{(s,s-1)}$, 
 for the model \eqref{7.88} are obtained from \eqref{7.15} and \eqref{6.14}, 
 respectively, by replacing $\F$ with $\bar \G$. The fermionic supercurrent 
 $J_{(s,s-1)}$ exists for even values of $s$. In the flat-superspace limit, 
 the expression for  $J_{(s,s)}$ obtained coincides with the main result of 
 \cite{KKvU}.\footnote{Actually the higher spin supercurrents derived in \cite{KKvU} 
 are obtained from eq. (5.6) 
 in \cite{KMT} by replacing  $\F$ with $\bar \G$. }
 It was claimed in \cite{KKvU} that the flat-superspace model 
 \bea
S [\G, \bar \G]= - \int \rd^4x \rd^2 \q  \rd^2 \bar \q  \bar \G  \G ~,
\qquad \bar D^2 \G=0
\eea
does not possess any conserved fermionic supercurrents $J_{(s,s-1)}$. 
Here we have demonstrated that they, in fact, do exist when $s$ is even.
Just like in the case of a massive chiral multiplet, 
the fermionic supercurrents were overlooked in  \cite{KKvU} because 
only a particular ansatz for the Noether procedure was studied in  \cite{KKvU}.


\subsection{Gauge higher spin multiplets and conserved supercurrents}

For each of the two off-shell formulations for 
the massless multiplet of half-integer superspin $s+1/2$, with $s=2, 3, \ldots$, 
which we reviewed in section \ref{subsection4.1}, it was shown
in \cite{KS94} that there exists a  gauge-invariant field strength 
$W_{\a(2s+1)} $ which is covariantly chiral, $\cDB_\bd W_{\a(2s+1)} = 0$, 
and is given by the expression
\bea
W_{\a (2s + 1)} &=& -\frac 14(\cDB^2 - 4 \mu) \cD_{(\a_1}{}^{\bd_1} \cdots \cD_{(\a_s}{}^{\bd_s} \cD_{\a_{s+1}} H_{\a_{s+2} \cdots \a_{2s+1}) \bd_1 \cdots \bd_s} ~.
 \label{7.115}
\eea
It was also shown in \cite{KS94} that 
on the mass shell it holds that  (i) $W_{\a(2s+1)} $ and its conjugate $\bar W_{\ad(2s+1)} $ 
are the only  independent gauge-invariant  field strengths; and (ii) $W_{\a(2s+1)} $
obeys the irreducibility condition 
\bea
\cD^\b W_{\b \a(2s)}=0~.
\label{7.116}
\eea
The relations \eqref{7.115} and \eqref{7.116} also hold for the cases 
$s=0$ and $s=1$, which correspond to the vector multiplet and linearised
supergravity, respectively. 
In terms of $W_{\a(2s+1)} $ and  $\bar W_{\ad(2s+1)} $, 
we can define the following 
higher spin supercurrent 
\bea
J_{\a(2s+1) \ad(2s+1)} =  W_{\a(2s+1)} \bar W_{\ad(2s+1)} ~,\qquad 
s=0,1,\dots~, 
\eea
which obeys the conservation equation
\bea
 \bar \cD_{(0,-1)} J_{(2s+1,2s+1)} =0 \quad \Longleftrightarrow \quad
 \cD_{(-1,0)} J_{(2s+1,2s+1)} =0 ~.
\eea

In the case of the longitudinal formulation for 
the massless multiplet of integer superspin $s$, with $s=2, 3, \ldots$, 
which we described in section \ref{section3}, it was shown
in \cite{KS94} that there exists a  gauge-invariant field strength 
$W_{\a(2s)} $ which is covariantly chiral, $\cDB_\bd W_{\a(2s)} = 0$, 
and is given by the expression\footnote{The flat-superspace version 
of \eqref{7.199} is given in section 6.9 of \cite{BK}.}
\bea
W_{\a (2s )} &=& -\frac 14(\cDB^2 - 4 \mu) \cD_{(\a_1}{}^{\bd_1} \cdots 
\cD_{(\a_{s-1}}{}^{\bd_{s-1}} \cD_{\a_{s}} \J_{\a_{s+1} \cdots \a_{2s}) 
\bd_1 \cdots \bd_{s-1}} ~.
\label{7.199}
\eea
As demonstrated  in \cite{KS94},
on the mass shell it holds that  (i) $W_{\a(2s)} $ and its conjugate $\bar W_{\ad(2s) }$ 
are the only  independent gauge-invariant  field strengths; and (ii) $W_{\a(2s)} $
obeys the irreducibility condition 
\bea
\cD^\b W_{\b \a(2s-1) }=0~.
\label{7.120}
\eea
The relations \eqref{7.199} and \eqref{7.120} also hold for the case $s=1$, 
which corresponds to the gravitino  multiplet.
In terms of $W_{\a(2s)} $ and  $\bar W_{\ad(2s)} $, 
we can define the higher spin supercurrent 
\bea
J_{\a (2s) \ad(2s) } =  W_{\a(2s)} \bar W_{\ad(2s)} ~, \qquad s=1,2,\dots~,
\eea
which obeys the  conservation equation
\bea
 \bar \cD_{(0,-1)} J_{(2s,2s)} =0 \quad \Longleftrightarrow \quad
 \cD_{(-1,0)} J_{(2s,2s)} =0 ~.
\eea
The conserved supercurrents $J_{\a (n) \ad(n) } =  W_{\a(n)} \bar W_{\ad(n)}$,
with $n=1,2,\dots$, are the AdS extensions of those introduced many years
ago by Howe, Stelle and Townsend \cite{HST}.

Now, for any positive integer $n>0$,
 we can try to generalise the higher spin supercurrent \eqref{7.15}
as follows:
\bea
{\mathfrak J}_{(s+n,s+n)} &=& \sum_{k=0}^s (-1)^{k} \frac{\binom{s}{k} \binom{s+n}{k}}{\binom{n+k}{n}}
\left\{ (-1)^n \frac{s-k}{n+k+1}
{\cD}^k_{(1,1)}
 \cD_{(1,0)} W_{(n,0)} \,\,
{\cD}^{s-k-1}_{(1,1)} \bar {\cD}_{(0,1)}
\bar W_{(0,n)} 
\right. \non \\ 
&& \left.
 \qquad 
+{\cD}^k_{(1,1)}
  W_{(n,0)} \,\,
{\cD}^{s-k}_{(1,1)}
\bar W_{(0,n)} \right\}~.~
\label{currentW}
\eea
Making use of
the on-shell condition   
\bea
\cD_{(-1,0)} W_{(n,0)} = 0 \quad \Longleftrightarrow \quad ({\cD}^2- 2(n+2) \bar \m)W_{(n,0)} = 0 ~,
\eea
one may check that 
\bea
\cD_{(-1,0)} {\mathfrak J}_{(s+n,s+n)} &=& 2n \bar \m \sum_{k=0}^{s-1} (-1)^{n+k} \frac{s-k}{n+k+1} \,\, \frac{\binom{s}{k} \binom{s+n}{k}}{\binom{n+k}{n}}
\non \\
&&\qquad \quad \times 
{\cD}^k_{(1,1)} W_{(n,0)} \,\,
{\cD}^{s-k-1}_{(1,1)}
\bar {\cD}_{(0,1)} \bar W_{(0,n)} ~.~
\label{eqW}
\eea
This demonstrates that ${\mathfrak J}_{(s+n,s+n)} $ is not conserved in AdS${}^{4|4}$.

In the flat-superspace limit, $ \m\to 0$, the right-hand side of \eqref{eqW}
vanishes and ${\mathfrak J}_{(s+n,s+n)} $ becomes conserved. 
In Minkowski superspace, the conserved supercurrent 
${\mathfrak J}_{(s+n,s+n)} $ was recently constructed in \cite{BGK3}
as an extension of the non-supersymmetric approach \cite{GSV}.

As a generalisation of the  conserved supercurrents $J_{\a (n) \ad(n) } =  W_{\a(n)} \bar W_{\ad(n)}$, one can introduce 
\bea
J_{\a (n) \ad(m) } =  W_{\a(n)} \bar W_{\ad(m)}~,
\label{7.266}
\eea
with $n\neq m$. They obey the conservation equations
\bea
 \bar \cD_{(0,-1)} J_{(n,m)} =0 ~, \qquad 
 \cD_{(-1,0)} J_{(n,m)} =0 
\eea
and can be viewed as Noether currents for the generalised superconformal 
higher spin multiplets introduced in \cite{KMT}.
Starting from the conserved supercurrents \eqref{7.266}, one can construct a generalisation of \eqref{currentW}.  We will not elaborate on a construction 
here.
\\


\noindent
{\bf Acknowledgements:}\\
SMK is grateful to Ruben Manvelyan for pointing out  important references.
The work of JH is supported by an Australian Government Research Training Program (RTP) Scholarship.
The work of SMK is supported in part by the Australian 
Research Council, project No. DP160103633.


\appendix


\section{AdS supercurrents}

There are only two irreducible AdS supercurrents, 
with $(12+12)$ and $(20+20)$ degrees of 
freedom \cite{BK11}.\footnote{A supercurrent multiplet is called irreducible 
if it is associated with an off-shell formulation for pure supergravity.}
The former is associated with minimal AdS supergravity (see, e.g., \cite{BK,GGRS} for reviews) and the corresponding conservation equation is 
\bea
\bar \cD^\ad J_{\a\ad} = \cD_\a T~, \qquad \bar \cD_\ad T =0~.
\label{FZsupercurrent}
\eea
The latter corresponds to non-minimal AdS supergravity \cite{BK12}, 
and the conservation equation is 
\bea
\label{nonminCurrent}
\bar \cD^\ad {\mathbb J}_{\alpha \ad} = -\frac{1}{4} \bar \cD^2 \zeta_\alpha~,\qquad
\cD_{(\beta} \zeta_{\alpha)} = 0~.
\eea
The vector superfields $J_a$ and ${\mathbb J}_a$ are real.

The non-minimal supercurrent \eqref{nonminCurrent} is equivalent to the Ferrara-Zumino multiplet \eqref{FZsupercurrent} in the sense that there always exists 
a well-defined improvement 
transformation that turns  \eqref{nonminCurrent} into \eqref{FZsupercurrent}, 
as demonstrated in \cite{BK12}. In AdS superspace, the constraint on the 
longitudinal linear compensator $\z_\a$ is equivalent to 
\bea
\zeta_\alpha 
	= \cD_\alpha (V + \ri \,U)~,
\eea
for well-defined real operators $V$ and $U$.
If we now introduce 
\bea
J_{\alpha \ad} := {\mathbb J}_{\alpha \ad} 
+ \frac{1}{6} [\cD_\alpha, \bar \cD_\ad] V
	- \cD_{\alpha \ad} U~,
	\qquad
T := \frac{1}{12} (\bar \cD^2 - 4 \mu) (V - 3 \ri U)~,
\eea
then the operators $J_{\a\ad} $ and $T$ prove to satisfy the conservation equation 
 \eqref{FZsupercurrent}.
 
 For the Ferrara-Zumino supercurrent  \eqref{FZsupercurrent}, 
 there exists an improvement transformation that is 
 generated by a chiral scalar operator $\O$.
 Specifically, using the operator $\O$ allows one to  
 introduce new supercurrent $\widetilde J_{\a\ad}$ 
 and chiral trace multiplet $\widetilde T$ defined by 
 \begin{subequations}\label{A.55}
 \bea
 \widetilde J_{\a\ad} &=& J_{\a\ad} +\ri \cD_{\a\ad} \big( \O -\bar \O \big)~, 
 \qquad \bar \cD_\ad \O=0~,
 \\
 \widetilde T &=& T +2\m \O +\frac 14 (\bar \cD^2 -4\m) \bar \O~.
 \eea
 \end{subequations}
 The operators  $\widetilde J_{\a\ad}$ and $\widetilde T$ obey the conservation equation  \eqref{FZsupercurrent} for arbitrary $\O$.\footnote{Extension 
 of the improvement transformation \eqref{A.55} to the case of supergravity is discussed
in section 6.3 of \cite{BK}.}

\section{Conserved currents for free real scalars}
\label{AppendixC}


In this appendix we will consider higher spin currents in free scalar field theory in flat space. Similar analysis for free fermions will be done in the next appendix.

Given an integer $s\geq 2$, 
the massless spin-$s$ field  \cite{Fronsdal} is described   by
real potentials
$h_{\a(s) \ad(s)}$ and $h_{\a(s-2) \ad(s-2)}$
with the gauge freedom\footnote{We follow the description of Fronsdal's
theory \cite{Fronsdal} given in section 
6.9 of \cite{BK}.}
\begin{subequations} \label{gauge1}
\bea
 \d h_{\a_1 \dots \a_{s} \ad_1 \dots \ad_{s} } 
&=& \pa_{(\a_1 (\ad_1} \l_{\a_2\dots \a_{s}) \ad_2 \dots \ad_{s})}~, \\
\d h_{\a_1 \dots \a_{s-2} \ad_1 \dots \ad_{s-2} } 
&=& \frac{s-1}{s^2} \pa^{\b \bd} \l_{\b \a_1\dots \a_{s-2} \bd \ad_1 \dots \ad_{s-2}}~,
\eea
\end{subequations}
for an arbitrary real gauge parameter 
$\l_{\a(s-1) \ad(s-1)}$. 
The field $h_{\a(s) \ad(s)}$ may be interpreted as a conformal spin-$s$ field \cite{FT,FL}.

To construct non-conformal higher spin currents, we couple $h_{\a(s) \ad(s)}$ and $h_{\a(s-2) \ad(s-2)}$ to external sources
\bea
S^{(s)}_{\rm source} &=& \int \rd^4x \, \Big\{ 
h^{ \a (s) \ad (s) } j_{ \a (s) \ad (s) }
+h^{ \a (s-2) \ad (s-2) } t_{ \a (s-2) \ad (s-2) } \Big\}~.
\label{source}
\eea
Requiring that $S^{(s)}_{\rm source}$ be invariant under the $\l$-transformation 
in \eqref{gauge1} gives the conservation equation
\bea
{\pa}^{\b \bd} j_{\b \a_1\dots \a_{s-1} \bd \ad_1 \dots \ad_{s-1}} 
+ \frac{s-1}{s^2} \pa_{(\a_1 (\ad_1} t_{\a_2 \dots \a_{s-1}) \ad_2 \dots \ad_{s-1})}= 0 ~.~
\label{cons-eq1}
\eea
Our derivation of \eqref{cons-eq1} is analogous to that given in \cite{Anselmi}.

Let us introduce the following operators 
\begin{subequations} \label{notation}
\bea
{\pa}_{(1,1)} &:=& 2\ri \z^\a \bar \z^\ad \pa_{\a\ad}~, \\
{\pa}_{(-1,-1)} &:=& 2\ri {\pa}^{\a \ad} \frac{\pa}{\pa \z^\a} \frac{\pa}{\pa \bar \z^\ad}~.
\eea
\end{subequations}
The conservation equation \eqref{cons-eq1} then becomes
\bea
{\pa}_{(-1,-1)} j_{(s,s)} + (s-1) {\pa}_{(1,1)} t_{(s-2,s-2)} = 0
\label{cons-eq2}
\eea
Note that both $j_{(s,s)}$ and $t_{(s-2,s-2)}$ are real.

Let us now consider the model for $N$ massless real scalar fields $\f^i$, with $i=1,\dots N$,
in Minkowski space
\bea
S = - \hf  \int \rd^4x \, \pa_\m \f^i \pa^\m \f^i ~,
\label{Nreal}
\eea
which admits conserved higher spin currents of the form
\bea
j_{(s,s)} = {\rm i}^s \,C^{ij}\sum_{k=0}^s (-1)^k
\binom{s}{k} \binom{s}{k} 
{\pa}^k_{(1,1)} \f^i \,\,
{\pa}^{s-k}_{(1,1)} \f^j ~,
\label{C.1}
\eea
where $C^{ij}$ is a constant matrix. It can be shown that $j_{(s,s)}=0$ if $s$ is odd and $C^{ij}$ is symmetric. Similarly, $j_{(s,s)}=0$ if $s$ is even and $C^{ij}$ is antisymmetric. Thus, we have to consider two separate cases: the case of even $s$ with symmetric $C$ and, the case of odd $s$ with antisymmetric $C$. Using the massless equation of motion $\Box \f^i =0 ~,$ one may show that $j_{(s,s)}$ satisfies the conservation equation
\bea
{\pa_{(-1,-1)}}j_{(s,s)}=0 ~.
\eea

We now turn to the massive model
\bea
S = -\hf \int \rd^4x \, \Big\{  \pa_\m \f^i \pa^\m \f^i+(M^2)^{ij} \f^i \f^j \Big\}~,
\label{Nrealm}
\eea
where $M =(M^{ij})$ is a real, symmetric $N\times N$ mass matrix. In the massive theory, the conservation equation is described by \eqref{cons-eq2} and so we first need to compute $\pa_{(-1,-1)} j_{(s,s)}$ using the massive equations of motion
\bea
\Box \f^i  -(M^2)^{ij}\f^j=0~.
\eea
For symmetric $C$, we obtain 
\bea
{\pa}_{(-1,-1)} j_{(s,s)} &=&- 8(s+1)^2 (C M^2)^{ij} \sum_{k=0}^{s-1} (-1)^{k} \binom{s}{k} \binom{s}{k} 
\non\\
&&
\times \frac{(s-k)^2}{(k+1)(k+2)}
 {\pa}^{k}_{(1,1)} \,{\f}^j \,\,{\pa}^{s-k-1}_{(1,1)}{\f}^i~.
\label{C.4}
\eea
If $C^{ij}$ is antisymmetric, we get
\bea
{\pa}_{(-1,-1)} j_{(s,s)} &=& 8\ri(s+1)^2 (C M^2)^{ij} \sum_{k=0}^{s-1} (-1)^{k} \binom{s}{k} \binom{s}{k} 
\non\\
&&
\times \frac{(s-k)^2}{(k+1)(k+2)}
 {\pa}^{k}_{(1,1)} \,{\f}^j \,\,{\pa}^{s-k-1}_{(1,1)}{\f}^i ~.
\label{C.4.anti}
\eea
Thus, in the massive real scalars there are four cases to consider:
\begin{enumerate}
\item Both $C$ and $CM^2$ are symmetric $\Longleftrightarrow [C,M^2]=0, \,\, s$ even.
\item $C$ is symmetric; $CM^2$ is antisymmetric $\Longleftrightarrow \{C,M^2\}=0, \,\,s$ even.
\item $C$ is antisymmetric; $CM^2$ is symmetric $\Longleftrightarrow \{C,M^2\}=0, \,\,s$ odd.
\item Both $C$ and $CM^2$ are antisymmetric $\Longleftrightarrow [C,M^2]=0, \,\,s$ odd.
\end{enumerate}

\textbf{Case 1:} Eq. \eqref{C.4} is equivalent to 
\bea
{\pa}_{(-1,-1)} j_{(s,s)} &=& -4(s+1)^2 (C M^2)^{ij} \sum_{k=0}^{s-1} (-1)^{k} \binom{s}{k} \binom{s}{k} (s-k)
\non \\ 
&& 
\times \left\{\frac{s-k}{(k+1)(k+2)}+(-1)^{s-1} \frac{1}{s-k+1}\right\} 
 {\pa}^{k}_{(1,1)} {\f}^j \,\,{\pa}^{s-k-1}_{(1,1)} {\f}^i ~.
\label{C.4a}
\eea
We look for $t_{(s-2, s-2)}$ such that (i) it is real; and (ii) it satisfies the conservation equation \eqref{cons-eq2}. We consider a general ansatz
\bea
t_{(s-2, s-2)} = -(C M^2)^{ij}\sum_{k=0}^{s-2} d_k \,
{\pa}^k_{(1,1)} \f^j\,
{\pa}^{s-k-2}_{(1,1)} \f^i ~.
\label{C.5}
\eea
For $k = 1,2,...s-2$, condition (ii) gives
\begin{subequations}\label{C.6}
\begin{align}
d_{k-1} + d_k &= -4\frac{(s+1)^2}{s-1} (-1)^k\binom{s}{k} \binom{s}{k} (s-k)\label{C.6a}
\non \\
& \qquad \qquad \times \left\{ \frac{s-k}{(k+1)(k+2)} + (-1)^{s-1} \frac{1}{s-k+1} \right\} ~.
\end{align}
Condition (ii) also implies that 
\begin{align}
d_{s-2} + d_0 &= -4s (s+1)(s+2)~,\label{C.6b}
\end{align}
\end{subequations}
Equations \eqref{C.6} lead to the following expression for $d_k,\, k=1,2,\dots s-2$ 
\begin{subequations}\label{C.7}
\begin{align}
d_k &= (-1)^k d_0 - \frac{4(s+1)^2}{s-1}
\sum_{l=1}^k (-1)^k \binom{s}{l} \binom {s}{l}(s-l)\left\{ \frac{s-l}{(l+1)(l+2)}- \frac{1}{s-l+1} \right\}  ~, 
\end{align}
\begin{align}
d_0 &= d_{s-2} = -2s(s+1)(s+2)~.
\end{align}
\end{subequations}
One can check that the equations \eqref{C.6a}--\eqref{C.6b} are identically 
satisfied if $s$ is even. 

\textbf{Case 2:} If we take $C M^2$ to be antisymmetric, a similar analysis shows that no solution for $t_{(s-2,s-2)}$ exists for even $s$.

\textbf{Case 3:} Now we consider the case where $C$ is antisymmetric and $CM^2$ symmetric. Again, similar consideration shows that no solution for $t_{(s-2,s-2)}$ exists for odd $s$.

\textbf{Case 4:} Eq. \eqref{C.4.anti} is equivalent to
\bea
{\pa}_{(-1,-1)} j_{(s,s)} &=& 4\ri(s+1)^2 (C M^2)^{ij} \sum_{k=0}^{s-1} (-1)^{k} \binom{s}{k} \binom{s}{k} (s-k)
\non \\ 
&& 
\times \left\{\frac{s-k}{(k+1)(k+2)}- \frac{1}{s-k+1}\right\} 
 {\pa}^{k}_{(1,1)} {\f}^j \,\,{\pa}^{s-k-1}_{(1,1)} {\f}^i ~.
\eea
We consider a general ansatz
\bea
t_{(s-2, s-2)} = -\ri (C M^2)^{ij}\sum_{k=0}^{s-2} d_k \,
{\pa}^k_{(1,1)} \f^j\,
{\pa}^{s-k-2}_{(1,1)} \f^i ~.
\label{C.5.anti}
\eea
Imposing (i) and (ii) and keeping in mind that $s$ is odd, we obtain the following conditions for $d_k$:
\begin{subequations}\label{C.6.anti}
\begin{align}
d_{k-1} + d_k &= 4\frac{(s+1)^2}{s-1} (-1)^k\binom{s}{k} \binom{s}{k} (s-k)\label{C.6a.anti}
\non \\
& \qquad \qquad \times \left\{ \frac{s-k}{(k+1)(k+2)} - \frac{1}{s-k+1} \right\} ~.
\end{align}
Condition (ii) also implies that 
\begin{align}
d_{s-2} - d_0 &= -4s (s+1)(s+2)~,\label{C.6b.anti}
\end{align}
\end{subequations}
Equations \eqref{C.6.anti} lead to the following expression for $d_k,\, k=1,2,\dots s-2$ 
\begin{subequations}\label{C.7.anti}
\begin{align}
d_k &= (-1)^k d_0 + \frac{4(s+1)^2}{s-1}
\sum_{l=1}^k (-1)^k \binom{s}{l} \binom {s}{l}\left\{ \frac{(s-l)^2}{(l+1)(l+2)}- \frac{s-l}{s-l+1} \right\}  ~, 
\end{align}
\begin{align}
d_0 &=-d_{s-2} = 2s(s+1)(s+2)~.
\end{align}
\end{subequations}
One can check that the equations \eqref{C.6a.anti}--\eqref{C.6b.anti} are identically 
satisfied if $s$ is odd.



\section{Conserved currents for free Majorana fermions}
\label{AppendixD}


Let us now consider $N$ free massless Majorana fermions
\bea
S = -\ri \,\int \rd^4x \,\, {\j}^{\a i} {\pa}_{\a \ad} {\bar \j}^{\ad i}~,
\label{Nfermions}
\eea
with the equation of motion 
\bea
{\pa}_{\a \ad} {\bar \j}^{\ad i} \Longrightarrow \Box {\bar \j}_\ad^i =0 ~, \qquad i=1,\dots N ~.~
\label{eom}
\eea
We can construct the following higher spin currents
\bea
j_{(s,s)} &=& C^{ij}\sum_{k=0}^{s-1} (-1)^k
\binom{s}{k} \binom{s}{k+1}\, 
{\pa}^k_{(1,1)}
 \z^\a \psi^{i}_\a \,\,
{\pa}^{s-k-1}_{(1,1)}
{\bar \z}^\ad {\bar \psi}^{j}_\ad ~, \qquad C^{ij}= C^{ji} ~,~
\label{D.1} \\
j_{(s,s)} &=& \ri\, C^{ij}\sum_{k=0}^{s-1} (-1)^k
\binom{s}{k} \binom{s}{k+1}\, 
{\pa}^k_{(1,1)}
 \z^\a \psi^{i}_\a \,\,
{\pa}^{s-k-1}_{(1,1)}
{\bar \z}^\ad {\bar \psi}^{j}_\ad ~, \qquad C^{ij}= -C^{ji} ~,~
\label{D.2}
\eea
where we put an extra $\ri$ in eq. \eqref{D.2} since $j_{(s,s)}$ has to be real. Using the equation of motion \eqref{eom}, it can be shown that the currents \eqref{D.1}, \eqref{D.2} are conserved
\bea
{\pa}_{(-1,-1)} j_{(s,s)} = 0~.~
\label{D.3}
\eea

We now look at the massive model
\bea
S = -\int \rd^4x \,\, \Big\{\ri {\j}^{\a i} {\pa}_{\a \ad} {\bar \j}^{\ad i}
+\Big(\hf M^{ij} {\j}^{\a i} \j_\a^j 
+ \hf \bar M^{ij} {\bar \j}_\ad^ i {\bar \j}^{\ad j}\Big)\Big\}~,
\label{fermions-massive}
\eea
where $M^{ij}$ is a constant symmetric $N\times N$ mass matrix.
To construct the conserved currents, we compute $\pa_{(-1,-1)} j_{(s,s)}$ using the massive equations of motion ($i=1,\dots,N$)
\begin{subequations}
\bea
\ri {\pa}_{\a \ad} {\bar \j}^{\ad i} + M^{ij}\j_\a^j &=&0 
\quad\Longrightarrow  \quad \Box {\bar \j}_\ad^i = (M \bar M)^{ij} \bar \j_\ad^j~,\\
-\ri {\pa}_{\a \ad} {\j}^{\a i} + \bar M^{ij} \bar \j_\ad^j &=&0 
\quad \Longrightarrow \quad \Box {\j}_\a^i = (\bar M M)^{ij} \j_\a^j~.
\eea
\end{subequations}
If $C^{ij}$ is a real symmetric matrix, we find
\bea 
{\pa}_{(-1,-1)} j_{(s,s)} &=& -2(s+1)\sum_{k=0}^{s-1}\frac{k+1}{s-k+1} (-1)^k \binom{s}{k} \binom{s}{k+1} 
\non \\
&& 
\times \left\{(CM)^{ij} {\pa}^{k}_{(1,1)} {\j}^{\a i} \,\,{\pa}^{s-k-1}_{(1,1)} {\j}_\a^j +(-1)^s (C\bar M)^{ij} {\pa}^{k}_{(1,1)} {\bar \j}_\ad^i \,\,{\pa}^{s-k-1}_{(1,1)} {\bar \j}^{\ad j}\right\}
\non \\ 
&&+ 4(s+1)(s+2)\sum_{k=1}^{s-1} k (-1)^k \binom{s}{k} \binom{s}{k+1}
\non \\
&&\times \left\{ \frac{1}{k+2} (M \bar M C)^{ij} -\frac{k+1}{(s-k+2)(s-k+1)} (C M \bar M)^{ij} \right\} 
\non \\
&&\times {\pa}^{k-1}_{(1,1)} \z^\a {\j}_\a^ i \,\,{\pa}^{s-k-1}_{(1,1)} \bar \z^\ad {\bar \j}_\ad^j ~.~
\label{D.4}
\eea 
If $C^{ij}$ is antisymmetric, we have
\bea 
{\pa}_{(-1,-1)} j_{(s,s)} &=& -2\ri(s+1)\sum_{k=0}^{s-1}\frac{k+1}{s-k+1} (-1)^k \binom{s}{k} \binom{s}{k+1} 
\non \\
&& 
\times \left\{(CM)^{ij} {\pa}^{k}_{(1,1)} {\j}^{\a i} \,\,{\pa}^{s-k-1}_{(1,1)} {\j}_\a^j +(-1)^{s-1}(C\bar M)^{ij} {\pa}^{k}_{(1,1)} {\bar \j}_\ad^i \,\,{\pa}^{s-k-1}_{(1,1)} {\bar \j}^{\ad j}\right\}
\non \\ 
&&+ 4\ri(s+1)(s+2)\sum_{k=1}^{s-1} k (-1)^k \binom{s}{k} \binom{s}{k+1}
\non \\
&&\times \left\{ \frac{1}{k+2} (M \bar M C)^{ij} -\frac{k+1}{(s-k+2)(s-k+1)} (C M \bar M)^{ij} \right\} 
\non \\
&&\times {\pa}^{k-1}_{(1,1)} \z^\a {\j}_\a^ i \,\,{\pa}^{s-k-1}_{(1,1)} \bar \z^\ad {\bar \j}_\ad^j ~.
\label{D.4a}
\eea

There are four cases to consider:
\begin{enumerate}
\item $C, CM, CM \bar M$ are symmetric $\Longleftrightarrow [C,M]=[C,\bar M]=0, [M,\bar M]=0$.
\item $C, C M \bar M$ symmetric; $C M$ antisymmetric  $\Longleftrightarrow \{C,M\}=\{C,\bar M\}=0, [M,\bar M]=0$.
\item $C, CM \bar M$ antisymmetric; $CM$ symmetric $\Longleftrightarrow \{C,M\}=\{C,\bar M\}=0, [M,\bar M]=0$.
\item $C, CM, CM \bar M$ are antisymmetric $\Longleftrightarrow [C,M]=[C,\bar M]=0, [M,\bar M]=0$.
\end{enumerate}

\textbf{Case 1:} Eq. \eqref{D.4} becomes
\bea
{\pa}_{(-1,-1)} j_{(s,s)} &=& -(s+1)\sum_{k=0}^{s-1}(-1)^k \binom{s}{k} \binom{s}{k+1} 
\non \\
&&\times \left\{\frac{k+1}{s-k+1} +(-1)^{s-1} \frac{s-k}{k+2}\right\}(CM)^{ij}\,\, {\pa}^{k}_{(1,1)} {\j}^{\a i} \,\,{\pa}^{s-k-1}_{(1,1)} {\j}_\a^j 
\non \\
&&+ (-1)^{s-1}(s+1)\sum_{k=0}^{s-1}(-1)^k \binom{s}{k} \binom{s}{k+1} 
\non \\
&& 
\times \left\{\frac{k+1}{s-k+1} +(-1)^{s-1} \frac{s-k}{k+2}\right\}(C\bar M)^{ij}\,\, {\pa}^{k}_{(1,1)} {\bar \j}_{\ad}^ i \,\,{\pa}^{s-k-1}_{(1,1)} {\bar \j}^{\ad j} 
\non \\ 
&&+ 4(s+1)(s+2)\sum_{k=1}^{s-1} k (-1)^k \binom{s}{k} \binom{s}{k+1}
\non \\
&& \times \left\{ \frac{1}{k+2} -\frac{k+1}{(s-k+2)(s-k+1)}\right\}
\non \\
&&\times (C M \bar M)^{ij} \,\, {\pa}^{k-1}_{(1,1)} \z^\a {\j}_\a^ i \,\,{\pa}^{s-k-1}_{(1,1)} \bar \z^\ad {\bar \j}_\ad^j ~.
\label{D.4new}
\eea
We look for $t_{(s-2, s-2)}$ such that (i) it is real; and (ii) it satisfies the conservation equation \eqref{cons-eq2}:
\bea
{\pa}_{(-1,-1)} j_{(s,s)} = -(s-1) {\pa}_{(1,1)} t_{(s-2, s-2)}~. 
\eea
Consider a general ansatz
\bea
t_{(s-2, s-2)} &=& (C M)^{ij}\,\sum_{k=0}^{s-2} c_k \,
{\pa}^k_{(1,1)} \j^{\a i}\,\,{\pa}^{s-k-2}_{(1,1)} \j^j_\a
\non \\
&&+(-1)^s(C \bar M)^{ij}\,\sum_{k=0}^{s-2} c_k \,
{\pa}^k_{(1,1)} {\bar \j}^i_\ad \,\,{\pa}^{s-k-2}_{(1,1)} {\bar \j}^{\ad j}
\non \\
&&+ (C M \bar M)^{ij} \, \sum_{k=1}^{s-2} g_k \,{\pa}^{k-1}_{(1,1)} \z^\a {\j}_\a^ i \,\,{\pa}^{s-k-1}_{(1,1)} \bar \z^\ad {\bar \j}_\ad^j  ~.
\label{D.5}
\eea
For $k = 1,2,...s-2$, condition (i) gives
\begin{subequations}\label{D.6}
\begin{align}
g_k &= (-1)^{s-1} g_{s-1-k}~, \label{D.6a}
\end{align}
while condition (ii) gives
\begin{align}
c_{k-1} + c_k &= \frac{s+1}{s-1} (-1)^k\binom{s}{k} \binom{s}{k+1} \left\{\frac{k+1}{s-k+1}+(-1)^{s-1} \frac{s-k}{k+2} \right\}~,\label{D.6b}
\end{align}
\begin{align}
g_{k-1} + g_k &= -4\frac{(s+1)(s+2)}{s-1} (-1)^k\binom{s}{k} \binom{s}{k+1} k\left\{\frac{1}{k+2}- \frac{k+1}{(s-k+2)(s-k+1)} \right\}~.\label{D.6c}
\end{align}
Condition (ii) also implies that 
\begin{align}
c_{s-2} + c_0 &= \frac{1}{s-1} \left\{2s + (-1)^{s-1} s^2(s+1)\right\}~,\label{D.6d}
\end{align}
\begin{align}
g_1 &= \frac{2s(s-2)}{3} (s^2+5s+6)~,\label{D.6e}
\end{align}
\begin{align}
g_{s-2} &= (-1)^{s-1} \frac{2s(s-2)}{3} (s^2+5s+6)~~.\label{D.6f}
\end{align}
\end{subequations}
The above conditions lead to the following expressions for $c_k$ and $g_k$ ($k=1,2,\dots s-2$) 
\begin{subequations}\label{D.7}
\begin{align}
c_k &= (-1)^k c_0 + \frac{s+1}{s-1} 
\sum_{l=1}^k (-1)^k \binom{s}{l} \binom {s}{l+1}\left\{ \frac{l+1}{s-l+1}+ (-1)^{s-1} \frac{s-l}{l+2} \right\}  ~, \label{D.7a}
\end{align}
\begin{align}
g_k &= 4(-1)^k \frac{(s+1)(s+2)}{s-1} 
\sum_{l=1}^k \binom{s}{l} \binom {s}{l+1}\left\{ \frac{l(l+1)}{(s-l+1)(s-l+2)}- \frac{l}{l+2} \right\}  ~. \label{D.7b}
\end{align}
If the parameter $s$ is even, \eqref{D.7a} gives 
\begin{align}
c_{s-2} &= c_0 = -\hf s(s+2)\label{D.7c}
\end{align}
\end{subequations}
and \eqref{D.6a}-\eqref{D.6f} are identically satisfied. However, when $s$ is odd, there appears an inconsistency: 
the right-hand side of \eqref{D.6d} is positive, while the left-hand side 
is negative, $c_{s-2} + c_0 < 0$. Therefore, our solution \eqref{D.7} is only consistent for $s=2n, n=1,2,\dots$. 

\textbf{Case 2:} If $C M$ is antisymmetric while $C M \bar M$ symmetric, eq.~\eqref{D.4} is slightly modified
\bea
{\pa}_{(-1,-1)} j_{(s,s)} &=& -(s+1)\sum_{k=0}^{s-1}(-1)^k \binom{s}{k} \binom{s}{k+1} 
\non \\
&&\times \left\{\frac{k+1}{s-k+1} +(-1)^{s} \frac{s-k}{k+2}\right\}(CM)^{ij}\,\, {\pa}^{k}_{(1,1)} {\j}^{\a i} \,\,{\pa}^{s-k-1}_{(1,1)} {\j}_\a^j 
\non \\
&&+ (-1)^{s-1}(s+1)\sum_{k=0}^{s-1}(-1)^k \binom{s}{k} \binom{s}{k+1} 
\non \\
&& 
\times \left\{\frac{k+1}{s-k+1} +(-1)^{s} \frac{s-k}{k+2}\right\}(C\bar M)^{ij}\,\, {\pa}^{k}_{(1,1)} {\bar \j}_{\ad}^ i \,\,{\pa}^{s-k-1}_{(1,1)} {\bar \j}^{\ad j} 
\non \\ 
&&+ 4(s+1)(s+2)\sum_{k=1}^{s-1} k (-1)^k \binom{s}{k} \binom{s}{k+1}
\non \\
&& \times \left\{ \frac{1}{k+2} -\frac{k+1}{(s-k+2)(s-k+1)}\right\}
\non \\
&&\times (C M \bar M)^{ij} \,\, {\pa}^{k-1}_{(1,1)} \z^\a {\j}_\a^ i \,\,{\pa}^{s-k-1}_{(1,1)} \bar \z^\ad {\bar \j}_\ad^j ~.
\label{D.4b}
\eea
Starting with a general ansatz
\bea
t_{(s-2, s-2)} &=& (C M)^{ij}\,\sum_{k=0}^{s-2} d_k \,
{\pa}^k_{(1,1)} \j^{\a i}\,\,{\pa}^{s-k-2}_{(1,1)} \j^j_\a
\non \\
&&+(-1)^s(C \bar M)^{ij}\,\sum_{k=0}^{s-2} d_k \,
{\pa}^k_{(1,1)} {\bar \j}^i_\ad \,\,{\pa}^{s-k-2}_{(1,1)} {\bar \j}^{\ad j}
\non \\
&&+ (C M \bar M)^{ij} \, \sum_{k=1}^{s-2} g_k \,{\pa}^{k-1}_{(1,1)} \z^\a {\j}_\a^ i \,\,{\pa}^{s-k-1}_{(1,1)} \bar \z^\ad {\bar \j}_\ad^j  
\label{D.8}
\eea
and imposing conditions (i) and (ii) yield 
\begin{subequations}\label{D.9}
\begin{align}
g_k &= (-1)^{s-1} g_{s-1-k}~, \label{D.9a}
\end{align}
\begin{align}
d_{k-1} + d_k &= \frac{s+1}{s-1} (-1)^k\binom{s}{k} \binom{s}{k+1} \left\{\frac{k+1}{s-k+1}-(-1)^{s-1} \frac{s-k}{k+2} \right\}~,\label{D.9b}
\end{align}
\begin{align}
g_{k-1} + g_k &= -4\frac{(s+1)(s+2)}{s-1} (-1)^k\binom{s}{k} \binom{s}{k+1} k\left\{\frac{1}{k+2}- \frac{k+1}{(s-k+2)(s-k+1)} \right\}~,\label{D.9c}
\end{align}
\begin{align}
d_{0} - d_{s-2} &= \frac{1}{s-1} \left\{2s + (-1)^{s} s^2(s+1)\right\}~,\label{D.9d}
\end{align}
\begin{align}
g_1 &= \frac{2s(s-2)}{3} (s^2+5s+6)~,\label{D.9e}
\end{align}
\begin{align}
g_{s-2} &= (-1)^{s-1} \frac{2s(s-2)}{3} (s^2+5s+6)~~.\label{D.9f}
\end{align}
\end{subequations}
As a result, the coefficients $d_k$ and $g_k$ are given by ($k=1,2,\dots s-2$) 
\begin{subequations}\label{D.10}
\begin{align}
d_k &= (-1)^k d_0 + \frac{s+1}{s-1} 
\sum_{l=1}^k (-1)^k \binom{s}{l} \binom {s}{l+1}\left\{ \frac{l+1}{s-l+1}- (-1)^{s-1} \frac{s-l}{l+2} \right\}  ~, \label{D.10a}
\end{align}
\begin{align}
g_k &= 4(-1)^k \frac{(s+1)(s+2)}{s-1} 
\sum_{l=1}^k \binom{s}{l} \binom {s}{l+1}\left\{ \frac{l(l+1)}{(s-l+1)(s-l+2)}- \frac{l}{l+2} \right\}  ~. \label{D.10b}
\end{align}
When the parameter $s$ is odd, \eqref{D.10a} gives 
\begin{align} \label{D.10c}
d_{s-2} &= -d_0 = \hf s(s+2)
\end{align}
\end{subequations}
and \eqref{D.9a}-\eqref{D.9f} are identically satisfied. However, when $s$ is even, there appears an inconsistency: 
the right-hand side of \eqref{D.9d} is positive, while the left-hand side 
is negative, $d_{0} - d_{s-2} < 0$. Therefore, our solution \eqref{D.10} is only consistent for $s=2n+1, n=1,2,\dots$. 

Finally, we consider $C^{ij}=-C^{ji}$ with the corresponding $j_{(s,s)}$ given by \eqref{D.2}. Similar considerations show that in \textbf{Case 3}, the non-conformal currents exist only if $s$ is even. The trace $t_{(s-2,s-2)}$ is given by \eqref{D.5} with the coefficients $c_k$ and $g_k$ given by
\begin{subequations}
\begin{align}
c_k &= \ri(-1)^k c_0 + \ri \frac{s+1}{s-1} 
\sum_{l=1}^k (-1)^k \binom{s}{l} \binom {s}{l+1}\left\{ \frac{l+1}{s-l+1}+ (-1)^{s-1} \frac{s-l}{l+2} \right\}  ~, \label{ck.a}
\end{align}
\begin{align}
g_k &=4 \ri  \, (-1)^k \frac{(s+1)(s+2)}{s-1} 
\sum_{l=1}^k \binom{s}{l} \binom {s}{l+1}\left\{ \frac{l(l+1)}{(s-l+1)(s-l+2)}- \frac{l}{l+2} \right\}  ~. \label{gk.a}
\end{align}
\end{subequations}
In \textbf{Case 4}, the non-conformal currents exist only for odd values of $s$. The trace $t_{(s-2,s-2)}$ is given by \eqref{D.8} with the coefficients $d_k$ and $g_k$ given by
\begin{subequations}
\begin{align}
d_k &= \ri (-1)^k d_0 + \ri \frac{s+1}{s-1} 
\sum_{l=1}^k (-1)^k \binom{s}{l} \binom {s}{l+1}\left\{ \frac{l+1}{s-l+1}- (-1)^{s-1} \frac{s-l}{l+2} \right\}  ~, \label{dk.b}
\end{align}
\begin{align}
g_k &= 4\ri \, (-1)^k \frac{(s+1)(s+2)}{s-1} 
\sum_{l=1}^k \binom{s}{l} \binom {s}{l+1}\left\{ \frac{l(l+1)}{(s-l+1)(s-l+2)}- \frac{l}{l+2} \right\}  ~. \label{gk.b}
\end{align}
\end{subequations}
We observe that the coefficients $c_k$ and $g_k$ in eq.~\eqref{ck.a} and \eqref{gk.a}, respectively differ from similar coefficients in \eqref{D.7a} and \eqref{D.7b} by a factor of $\ri$. Hence, for even $s$ we may define a more general supercurrent
\bea
j_{(s,s)} = C^{ij}\sum_{k=0}^{s-1} (-1)^k
\binom{s}{k} \binom{s}{k+1}\, 
{\pa}^k_{(1,1)}
 \z^\a \psi^{i}_\a \,\,
{\pa}^{s-k-1}_{(1,1)}
{\bar \z}^\ad {\bar \psi}^{j}_\ad ~,
\label{j.even}
\eea
where $C^{ij}$ is a generic matrix which can be split into the symmetric and antisymmetric parts: $C^{ij} =S^{ij}+\ri A^{ij}$. Here both $S$ and $A$ are real and we put an $\ri$ in front of $A$ because $j_{(s,s)} $ must be real. From the above consideration it then follows that the corresponding more general solution for $t_{(s-2, s-2)}$ reads
\bea
t_{(s-2, s-2)} &=& (C M)^{ij}\,\sum_{k=0}^{s-2} c_k \,
{\pa}^k_{(1,1)} \j^{\a i}\,\,{\pa}^{s-k-2}_{(1,1)} \j^j_\a
\non \\
&&+(-1)^s(\bar C \bar M)^{ij}\,\sum_{k=0}^{s-2} c_k \,
{\pa}^k_{(1,1)} {\bar \j}^i_\ad \,\,{\pa}^{s-k-2}_{(1,1)} {\bar \j}^{\ad j}
\non \\
&&+ (C M \bar M)^{ij} \, \sum_{k=1}^{s-2} g_k \,{\pa}^{k-1}_{(1,1)} \z^\a {\j}_\a^ i \,\,{\pa}^{s-k-1}_{(1,1)} \bar \z^\ad {\bar \j}_\ad^j  ~,
\label{t.even}
\eea
where $[S, M]=[S, \bar M]=0$, $\{A, M\}=\{A, \bar M\}=0$ and $[M, \bar M]=0$. The coefficients $c_k$ and $g_k$ are given by eqs.~\eqref{D.7a} and \eqref{D.7b}, respectively.
Similarly, the coefficients $d_k$ and $g_k$ in~\eqref{dk.b} and \eqref{gk.b} differ from similar coefficients in~\eqref{D.10a} and \eqref{D.10b} by a factor of $\ri$. This means that for odd $s$ we can define a more general supercurrent~\eqref{j.even}, where $C^{ij}$ is a generic matrix which we can split as before into the symmetric and antisymmetric parts,  $C^{ij} = S^{ij} + \ri A^{ij}$. 
From the above consideration it then follows that the corresponding more general solution for $t_{(s-2, s-1)}$ reads
\bea
t_{(s-2, s-1)} &=& (C M)^{ij}\,\sum_{k=0}^{s-2} d_k \,
{\pa}^k_{(1,1)} \j^{\a i}\,\,{\pa}^{s-k-2}_{(1,1)} \j^j_\a
\non \\
&&+(-1)^s(\bar C \bar M)^{ij}\,\sum_{k=0}^{s-2} d_k \,
{\pa}^k_{(1,1)} {\bar \j}^i_\ad \,\,{\pa}^{s-k-2}_{(1,1)} {\bar \j}^{\ad j}
\non \\
&&+ (C M \bar M)^{ij} \, \sum_{k=1}^{s-2} g_k \,{\pa}^{k-1}_{(1,1)} \z^\a {\j}_\a^ i \,\,{\pa}^{s-k-1}_{(1,1)} \bar \z^\ad {\bar \j}_\ad^j  ~,
\label{t.odd}
\eea
where $\{S, M\}=\{S, \bar M\}=0$, $[A, M]=[A, \bar M]=0$ and $[M, \bar M]=0$. The coefficients $d_k$ and $g_k$ are given by eqs.~\eqref{D.10a} and \eqref{D.10b}, respectively.


\begin{footnotesize}

\end{footnotesize}

\end{document}